\PassOptionsToPackage{unicode}{hyperref}
\PassOptionsToPackage{hyphens}{url}
\PassOptionsToPackage{dvipsnames,svgnames,x11names}{xcolor}
\documentclass[
  11pt,
  letterpaper,
]{article}
\usepackage{xcolor}
\usepackage[margin=1in]{geometry}
\usepackage{amsmath,amssymb}
\usepackage[T1]{fontenc}
\usepackage[utf8]{inputenc}
\usepackage{textcomp} 
\usepackage{lmodern}
\IfFileExists{upquote.sty}{\usepackage{upquote}}{}
\IfFileExists{microtype.sty}{
  \usepackage[]{microtype}
  \UseMicrotypeSet[protrusion]{basicmath} 
}{}
\makeatletter
\@ifundefined{KOMAClassName}{
  \IfFileExists{parskip.sty}{%
    \usepackage{parskip}
  }{
    \setlength{\parindent}{0pt}
    \setlength{\parskip}{6pt plus 2pt minus 1pt}}
}{
  \KOMAoptions{parskip=half}}
\makeatother
\makeatletter
\ifx\paragraph\undefined\else
  \let\oldparagraph\paragraph
  \renewcommand{\paragraph}{
    \@ifstar
      \xxxParagraphStar
      \xxxParagraphNoStar
  }
  \newcommand{\xxxParagraphStar}[1]{\oldparagraph*{#1}\mbox{}}
  \newcommand{\xxxParagraphNoStar}[1]{\oldparagraph{#1}\mbox{}}
\fi
\ifx\subparagraph\undefined\else
  \let\oldsubparagraph\subparagraph
  \renewcommand{\subparagraph}{
    \@ifstar
      \xxxSubParagraphStar
      \xxxSubParagraphNoStar
  }
  \newcommand{\xxxSubParagraphStar}[1]{\oldsubparagraph*{#1}\mbox{}}
  \newcommand{\xxxSubParagraphNoStar}[1]{\oldsubparagraph{#1}\mbox{}}
\fi
\makeatother

\usepackage{color}
\usepackage{fancyvrb}

\DefineVerbatimEnvironment{Highlighting}{Verbatim}{commandchars=\\\{\}}
\usepackage{framed}
\definecolor{shadecolor}{RGB}{241,243,245}
\newenvironment{Shaded}{\begin{snugshade}}{\end{snugshade}}

\newcommand{\CommentTok}[1]{\textcolor[rgb]{0.37,0.37,0.37}{#1}}

\newcommand{\ControlFlowTok}[1]{\textcolor[rgb]{0.00,0.23,0.31}{\textbf{#1}}}
\newcommand{\DataTypeTok}[1]{\textcolor[rgb]{0.68,0.00,0.00}{#1}}
\newcommand{\DecValTok}[1]{\textcolor[rgb]{0.68,0.00,0.00}{#1}}

\newcommand{\FloatTok}[1]{\textcolor[rgb]{0.68,0.00,0.00}{#1}}

\newcommand{\KeywordTok}[1]{\textcolor[rgb]{0.00,0.23,0.31}{\textbf{#1}}}
\newcommand{\NormalTok}[1]{\textcolor[rgb]{0.00,0.23,0.31}{#1}}

\newcommand{\StringTok}[1]{\textcolor[rgb]{0.13,0.47,0.30}{#1}}

\usepackage{longtable,booktabs,array}
\usepackage{calc} 
\usepackage{etoolbox}
\makeatletter
\patchcmd\longtable{\par}{\if@noskipsec\mbox{}\fi\par}{}{}
\makeatother
\IfFileExists{footnotehyper.sty}{\usepackage{footnotehyper}}{\usepackage{footnote}}
\makesavenoteenv{longtable}
\usepackage{graphicx}
\makeatletter
\newsavebox\pandoc@box
\newcommand*\pandocbounded[1]{
  \sbox\pandoc@box{#1}%
  \Gscale@div\@tempa{\textheight}{\dimexpr\ht\pandoc@box+\dp\pandoc@box\relax}%
  \Gscale@div\@tempb{\linewidth}{\wd\pandoc@box}%
  \ifdim\@tempb\p@<\@tempa\p@\let\@tempa\@tempb\fi
  \ifdim\@tempa\p@<\p@\scalebox{\@tempa}{\usebox\pandoc@box}%
  \else\usebox{\pandoc@box}%
  \fi%
}
\def\fps@figure{htbp}
\makeatother

\NewDocumentCommand\citeproctext{}{}

\makeatletter
 \let\@cite@ofmt\@firstofone
 \def\@biblabel#1{}
 \def\@cite#1#2{{#1\if@tempswa , #2\fi}}
\makeatother
\newlength{\cslhangindent}
\newlength{\csllabelwidth}
\newenvironment{CSLReferences}[2] 
 {\begin{list}{}{%
  \setlength{\itemindent}{0pt}
  \setlength{\leftmargin}{0pt}
  \setlength{\parsep}{0pt}
  \ifodd #1
   \setlength{\leftmargin}{\cslhangindent}
   \setlength{\itemindent}{-1\cslhangindent}
  \fi
  \setlength{\itemsep}{#2\baselineskip}}}
 {\end{list}}
\usepackage{calc}

\providecommand{\tightlist}{%
  \setlength{\itemsep}{0pt}\setlength{\parskip}{0pt}}

\usepackage{amsmath,amssymb,amsthm}
\usepackage{bm}
\usepackage[htt]{hyphenat}
\usepackage{placeins}

\theoremstyle{definition}

\theoremstyle{remark}

\makeatletter
\@ifpackageloaded{tcolorbox}{}{\usepackage[skins,breakable]{tcolorbox}}
\@ifpackageloaded{fontawesome5}{}{\usepackage{fontawesome5}}
\definecolor{quarto-callout-color}{HTML}{909090}
\definecolor{quarto-callout-note-color}{HTML}{0758E5}
\definecolor{quarto-callout-important-color}{HTML}{CC1914}
\definecolor{quarto-callout-warning-color}{HTML}{EB9113}
\definecolor{quarto-callout-tip-color}{HTML}{00A047}
\definecolor{quarto-callout-caution-color}{HTML}{FC5300}
\definecolor{quarto-callout-color-frame}{HTML}{acacac}
\definecolor{quarto-callout-note-color-frame}{HTML}{4582ec}
\definecolor{quarto-callout-important-color-frame}{HTML}{d9534f}
\definecolor{quarto-callout-warning-color-frame}{HTML}{f0ad4e}
\definecolor{quarto-callout-tip-color-frame}{HTML}{02b875}
\definecolor{quarto-callout-caution-color-frame}{HTML}{fd7e14}
\makeatother
\makeatletter
\@ifpackageloaded{caption}{}{\usepackage{caption}}
\AtBeginDocument{%
\ifdefined\contentsname
  \renewcommand*\contentsname{Table of contents}
\else
  \newcommand\contentsname{Table of contents}
\fi
\ifdefined\listfigurename
  \renewcommand*\listfigurename{List of Figures}
\else
  \newcommand\listfigurename{List of Figures}
\fi
\ifdefined\listtablename
  \renewcommand*\listtablename{List of Tables}
\else
  \newcommand\listtablename{List of Tables}
\fi
\ifdefined\figurename
  \renewcommand*\figurename{Figure}
\else
  \newcommand\figurename{Figure}
\fi
\ifdefined\tablename
  \renewcommand*\tablename{Table}
\else
  \newcommand\tablename{Table}
\fi
}
\@ifpackageloaded{float}{}{\usepackage{float}}
\floatstyle{ruled}
\@ifundefined{c@chapter}{\newfloat{codelisting}{h}{lop}}{\newfloat{codelisting}{h}{lop}[chapter]}
\floatname{codelisting}{Listing}

\makeatother
\makeatletter
\@ifpackageloaded{caption}{}{\usepackage{caption}}
\@ifpackageloaded{subcaption}{}{\usepackage{subcaption}}
\makeatother
\usepackage{bookmark}
\IfFileExists{xurl.sty}{\usepackage{xurl}}{} 
\hypersetup{
  pdftitle={Sensitivity to Subjective Expected Utility Maximization: A Methodological Study, with an Illustrative Application to LLM Decision-Making},
  pdfauthor={Jeff Helzner},
  pdfkeywords={subjective expected
utility, identifiability, simulation-based calibration, Bayesian
workflow, softmax choice, LLM decision-making},
  colorlinks=true,
  linkcolor={blue},
  filecolor={Maroon},
  citecolor={Blue},
  urlcolor={Blue},
  pdfcreator={LaTeX via pandoc}}

\title{Sensitivity to Subjective Expected Utility Maximization: A
Methodological Study, with an Illustrative Application to LLM
Decision-Making}
\author{Jeff Helzner}
\date{7 July 2026}
\begin{document}
\maketitle
\begin{abstract}
Evaluating the quality of decisions made under uncertainty is hard when
labeled outcomes are scarce, costly, or confounded with luck. We treat
subjective expected utility (SEU) maximization as a \emph{stated
standard} and define a graded measure --- \emph{SEU sensitivity} --- of
an agent's conformity to it. The formal vehicle is a softmax choice
model with a sensitivity parameter \(\alpha\) applied to SEU-valued
alternatives; the methodological contribution is a sequence of
identifiability results for \(\alpha\) and for the belief and utility
parameters \((\beta, \delta)\), validated in Stan via prior predictive
checks, parameter recovery, and simulation-based calibration (SBC). The
findings are reported with their finite-sample caveats intact. In the
uncertain-choice-only model \texttt{m\_0}, \(\alpha\) is identifiable
\emph{given} the expected-utility vector \(\eta\) --- and, under the
study's prior and design, sharply recovered --- while
\((\beta, \delta)\) are only weakly informed by uncertain choices: the
joint posterior barely contracts and concentrates on a
\(\beta\)--\(\delta\) trade-off at realistic sample sizes. In the
extended model \texttt{m\_1}, \(\delta\) becomes identifiable in
principle (hence estimable in the limit) via a \(\beta\)-free risky
block, but its practical recovery gain at realistic sample sizes is
negligible (a matched-count CI-width reduction under 1\%); and at
matched choice count the \(\beta\)-free risky block yields no
\emph{detected} \(\alpha\)-precision gain in this design either, so
finite-sample \(\alpha\) precision here tracks data \emph{quantity}
rather than the \emph{type} of choice. These are two distinct phenomena:
for \(\delta\), identifiability does not imply precise estimability at
realistic \(n\); for \(\alpha\), identifiability in principle is silent
about what governs finite-\(n\) precision. Marginal SBC passes for both
models even where the joint (belief, utility) posterior is only weakly
informed --- a \emph{marginal-SBC demarcation} we make precise. A
two-by-two illustrative application (GPT-4o and Claude 3.5 Sonnet, each
on insurance-claims triage and Ellsberg-style urns, with sampling
temperature as the lever) shows the workflow runs end-to-end on real LLM
choice data, detecting a structured comparative \(\alpha\) effect in two
of four cells and declining to support one in the other two.
\end{abstract}

\renewcommand*\contentsname{Table of contents}
{
\hypersetup{linkcolor=}
\setcounter{tocdepth}{3}
\tableofcontents
}

\section*{Motivation}\label{sec-motivation}
\addcontentsline{toc}{section}{Motivation}

\subsection*{1.1 Two approaches to assessing a decision
maker}\label{sec-two-questions}
\addcontentsline{toc}{subsection}{1.1 Two approaches to assessing a
decision maker}

There are two broad approaches to assessing a decision maker from a body
of choices. The first is \emph{external}, or label-based: do the agent's
choices agree with a labeled set of decisions held to be correct --- a
key that records, for each decision problem, which alternative an
informed judge would select; would such a judge have forwarded the same
claim, taken the same gamble, made the same call? The criterion is
agreement with a curated answer key, whoever supplies it. The second is
\emph{internal}: are the choices coherent with one another in the way a
stated standard of consistency requires --- responsive to the
probabilities, the stakes, and the tradeoffs in the way the standard
prescribes? We assess this second kind of conformity from the agent's
choices as observed, without attributing to the agent any particular
internal computation or deliberative procedure: ``internal'' qualifies
the \emph{standard} --- coherence among the agent's own choices --- not
the agent's cognition. The standard constrains behavior, and the
question is whether the behavior conforms to it.

The two assessments answer to different things and can come apart. The
consistency standard constrains how an agent's choices hang together,
not which alternative an external key marks as correct: an agent whose
beliefs and values differ from those the key encodes can be fully
consistent yet disagree with it, and an agent can match the key on a
given decision while its choices violate the standard elsewhere. The
internal-consistency assessment is the one this paper develops, and it
is especially relevant where a correct-choice key is unavailable or
costly to construct --- in assessing a human analyst, an institution,
or, increasingly, a large language model deployed to make or recommend
choices.

Both approaches have a long precedent in the study of decision makers.
One strand of internal-consistency critique assesses choices against
expected utility theory as a standard for decision making under risk:
the Allais paradox (Allais 1953) exhibits a pattern of preferences
standardly read as violating the independence axiom (roughly, that a
component common to two options should not affect the preference between
them),\footnote{The diagnosis is analysis-relative rather than forced.
  Levi (1986) accounts for the same pattern within a generalization of
  subjective expected utility theory that admits indeterminacy in both
  probabilities and utilities: relaxing the assumption that any two
  alternatives are comparable in terms of preference accommodates the
  Allais choices with no violation of independence. Unlike the
  descriptive theories built on the pattern, Levi's theory is not
  reducible to a binary preference relation; we return to it in
  §\hyperref[sec-seu-standard]{1.4}.} a tension Kahneman and Tversky
took up in developing prospect theory (Kahneman and Tversky 1979;
Tversky and Kahneman 1992). These demonstrations, and the large-scale
replication that subsequently confirmed their patterns (Ruggeri et al.
2020), were built on \emph{pairwise} choices. Such choices reveal a
pattern inconsistent with subjective expected utility maximization, but
they do not by themselves fix \emph{which} assumption of the standard is
responsible: the same Allais pattern reads as a violation of the
independence axiom or, on Levi's analysis, as a relaxation of the
ordering assumption with no independence violation (see the footnote
above and §\hyperref[sec-seu-standard]{1.4}). The instrument developed
here, by contrast, places no restriction on the number of alternatives
in a decision problem, which opens the possibility of probing how
sensitivity itself varies with menu size (an extension we return to in
§\hyperref[sec-disc-limitations]{8.5}). A second strand --- the
heuristics-and-biases program --- assesses judgments about uncertainty
against the probability axioms, construed as a coherence standard on
degrees of belief in the Dutch-book tradition (Ramsey 1926; Finetti
1937), on which a set of degrees of belief is faulted if it would
sanction a combination of bets guaranteeing a sure loss: the conjunction
fallacy of the ``Linda problem'' (Tversky and Kahneman 1983) exhibits
credences that violate that calculus, evidence used to motivate the
representativeness heuristic. In each case the decision maker is faulted
not against a labeled key but against a coherence requirement that its
own choices or judgments fail to meet. The contrasting tradition scores
choices against an externally curated set of good options: the choice
architecture of Thaler and Sunstein (2008), for instance, assesses
diners against a menu whose ``healthy'' items have been labeled by
dieticians. The instrument developed here belongs to the
internal-consistency tradition --- it measures conformity to an internal
standard --- with subjective expected utility maximization, which
couples a coherence requirement on degrees of belief with one on
preferences, in the role the axioms of rational choice play in those
precedents.

\subsection*{1.2 Challenges for label-based assessment of decision
quality}\label{sec-why-labels-fail}
\addcontentsline{toc}{subsection}{1.2 Challenges for label-based
assessment of decision quality}

The label-based approach scores choices against a key, drawn either from
an adjudicated set of correct choices or from the realized outcomes of
the decisions. Two difficulties limit what such a key can tell us about
the quality of the decisions: a key of correct choices is often
impractical to obtain, and a key of outcomes measures the wrong thing.

The first concerns obtaining the labels at all. A key of correct choices
records, decision by decision, which alternative is the right one to
select, and assembling one is often impractical. The insurance-claims
study we report in §\hyperref[sec-application]{7} is a case in point:
there is no off-the-shelf key that says which choice was correct, and
constructing one would mean asking subject-matter experts what they
would choose across the decision problems --- an exercise that is costly
and, because expert judgment varies, sensitive to which experts are
consulted. We could not produce such a key credibly under realistic
experimental conditions.

The second difficulty is more basic, and remains even where a key
\emph{is} at hand --- in particular where it records realized
\emph{outcomes} rather than adjudicated choices. A good outcome is not
the same as a good decision: the outcome of a choice typically depends
on how the relevant uncertainty happens to resolve, which lies outside
the decision maker's control. A score computed purely from outcomes also
ignores the way the context in which a decision is made informs the
beliefs and desires of the decision maker. Treating outcomes as a direct
measure of decision quality is defensible only under strong assumptions
--- for instance, that the decision maker is properly incentivized and
has enough information to learn the ``objective'' probabilities of the
relevant states. Those assumptions are reasonable in some controlled
settings, but we would not want to impose them in a study like the
insurance experiment of §\hyperref[sec-application]{7}, where neither
condition can be taken for granted. A decision that is well structured
given the agent's information may still be scored poorly by an outcome
key that was generated under different information, and vice versa.

\subsection*{1.3 The need for consistency-based
evaluation}\label{sec-normative}
\addcontentsline{toc}{subsection}{1.3 The need for consistency-based
evaluation}

What remains available when neither kind of key can be had is the
\emph{structure} of the choices themselves. If we are willing to state a
standard of good decision making explicitly, we can ask whether an
agent's choices are responsive to probabilities, utilities, and
tradeoffs in the way that standard prescribes --- without a labeled set
of correct choices, and without reference to which outcome happened to
be realized. Evaluation of this internal-consistency kind replaces a
claim about outcomes with a claim about conformity to a stated standard.
It does not require any labeled set of decisions; it requires that the
standard be named explicitly and --- as the rest of the paper makes
concrete --- that a statistical model operationalizing conformity to it
be specified, since any graded measure of conformity is read off a
fitted model and is therefore conditional on that model's assumptions,
prior, and design (§\hyperref[sec-disc-alpha]{8.4}). In return it yields
an evaluation that is well defined even where no key of correct choices
and no outcome record exist, and that does not depend on which outcome
happened to be realized.

\subsection*{1.4 SEU as the stated standard}\label{sec-seu-standard}
\addcontentsline{toc}{subsection}{1.4 SEU as the stated standard}

We take \emph{subjective expected utility (SEU) maximization} as the
reference standard. An SEU-committed agent holds a subjective
probability over the states of the world and a utility over
consequences, and prefers an alternative whose expected utility is
maximal (Savage 1954; Anscombe and Aumann 1963; Neumann and Morgenstern
1947). The standard does not require a unique maximizer: when several
alternatives attain the maximal expected utility they are all
admissible, and \emph{picking} among them lies beyond what the standard
prescribes (Ullmann-Margalit and Morgenbesser 1977). We do \emph{not}
argue that SEU is the uniquely correct theory of rational choice, and
nothing below depends on that stronger claim. The argument is
conditional: \emph{if} SEU is the reference standard, then a graded,
statistically rigorous measure of conformity to it is methodologically
useful. SEU is the natural first case for such an instrument: it is the
most widely shared normative benchmark in decision theory, so a
conformity measure built on it reaches the largest audience. The
identifiability questions that occupy
§§\hyperref[sec-m0-identifiability]{3} and
\hyperref[sec-m1-identifiability]{5} are then taken up for that
instrument in turn.

Two qualifications bound the sense in which other standards could be
substituted for SEU. First, several theories often mentioned alongside
SEU are not normative standards of decision quality at all but
descriptive accounts of how people actually choose; cumulative prospect
theory (Kahneman and Tversky 1979; Tversky and Kahneman 1992) is the
clearest example. Substituting such a theory is therefore more than a
change of functional form: the quantity being measured changes from
conformity to a norm to fit to a behavioral model, and the
interpretation of the instrument changes with it. Put plainly: a
descriptive substitute would still return a number, but that number
would measure fit to observed behavior, not conformity to a standard of
good decision making. Maxmin expected utility (Gilboa and Schmeidler
1989) is more plausibly read as a normative alternative, and it supplies
an admissible value function for the softmax construction; but because
that value function is a \emph{minimum} over a set of priors, it is not
smooth, and substituting it into the gradient-based (Hamiltonian Monte
Carlo) implementation of §§\hyperref[sec-m0-implementation]{4} and
\hyperref[sec-m1-implementation]{6} is not a drop-in replacement --- the
kinks of the min would require either a smoothing that introduces a
precision parameter whose finite-sample estimate would be entangled with
\(\alpha\) (the two are separately defined, but a smoothed-min sharpness
and a choice sensitivity press on overlapping features of the data), or
a non-gradient sampler. Smooth ambiguity models (Klibanoff, Marinacci,
and Mukerji 2005) avoid the non-smoothness by construction and would
substitute more readily. Cumulative prospect theory, by contrast, could
not be substituted at all without abandoning the normative reading.
Second, the template applies only to standards that rank alternatives by
a single real-valued evaluation. Standards that decline to reduce choice
to such an evaluation --- for instance the decision theory of Levi
(1980, 1986), in which admissibility does not reduce to maximizing a
preference ordering: two admissible options may be incomparable rather
than equally preferred --- fall outside the framework entirely, not
merely outside its current scope. Parallel instruments for the
real-valued non-SEU standards are possible by the same template and are
out of scope here.

\subsection*{1.5 Why a graded measure rather than a binary
verdict}\label{sec-graded}
\addcontentsline{toc}{subsection}{1.5 Why a graded measure rather than a
binary verdict}

Real decision makers, human or machine, only ever \emph{approximately}
satisfy any rationality axiom (Simon 1955; Lieder and Griffiths 2020). A
binary verdict --- ``rational'' or ``not'' --- discards exactly the
information that an evaluation should preserve, namely \emph{how
closely} and \emph{how reliably} an agent's choices track the standard.
We therefore measure conformity on a continuum. The formal vehicle,
developed in §\hyperref[sec-abstract-model]{2}, is a softmax
(Luce--McFadden) choice model (Luce 1959; McFadden 1974; Train 2009)
carrying a single scalar \emph{sensitivity} parameter \(\alpha \geq 0\)
applied to SEU-valued alternatives. As \(\alpha\) ranges from \(0\) to
\(\infty\), the implied behavior runs from uniform-random choice
(insensitive to expected utility) to deterministic SEU maximization
(perfectly sensitive to it). The empirically interesting agents live
strictly between these poles, and \(\alpha\) is the coordinate that
locates them there. We call \(\alpha\) the agent's \textbf{SEU
sensitivity}: the disposition of an agent committed to SEU maximization
to act in accordance with that commitment, held separate from both the
\emph{content} of the commitment (the agent's beliefs and utilities) and
the \emph{outcomes} it happens to realize.

\subsection*{1.6 Scope}\label{sec-scope}
\addcontentsline{toc}{subsection}{1.6 Scope}

This paper is methodological. Its focus is a calibrated measurement
instrument and the identifiability analysis that licenses it; the
decision-theoretic and philosophical material frames that analysis.
§§\hyperref[sec-abstract-model]{2}--\hyperref[sec-m1-implementation]{6}
specify, motivate, and validate the instrument: the abstract choice
model and its three characterizing properties
(§\hyperref[sec-abstract-model]{2}), the identifiability of \(\alpha\)
together with the finding that the belief and utility parameters
\((\beta, \delta)\) are only \emph{weakly informed} in the finite-sample
Bayesian workflow for an uncertain-choice-only model
(§\hyperref[sec-m0-identifiability]{3}), a Stan implementation validated
by prior predictive checks, parameter recovery, and simulation-based
calibration (SBC) (§\hyperref[sec-m0-implementation]{4}), and the same
arc for an extended model that adds risky choices
(§§\hyperref[sec-m1-identifiability]{5}--\hyperref[sec-m1-implementation]{6}).
§\hyperref[sec-application]{7} then runs the full workflow end-to-end on
real LLM choice data: a \(2\times2\) illustrative application crossing
GPT-4o and Claude 3.5 Sonnet with insurance-claims triage and
Ellsberg-style urn gambles (Ellsberg 1961), using sampling temperature
as the experimental lever. The application is included to show that the
methodology does real evaluative work on real choice data (Binz and
Schulz 2023; Hagendorff, Fabi, and Kosinski 2023); it is \emph{not} a
substantive contribution to LLM behavioral science, and
§\hyperref[sec-app-limits]{7.6} draws that line explicitly.

Several things are deliberately out of scope, and are flagged once at
their natural locations rather than revisited: an information-optimal
lottery design for sharpening the utility parameter (future work,
§\hyperref[sec-disc-limitations]{8.5}); the hierarchical extension
\texttt{h\_m01} and the multi-LLM alignment study it was built for (a
planned companion paper, §\hyperref[sec-app-followup]{7.7});
generalized-sensitivity models with block-specific or context-specific
\(\alpha\); a test for ambiguity aversion in the Ellsberg setting (the
SEU-plus-softmax model is, by assumption, of the wrong form for that,
§\hyperref[sec-app-limits]{7.6}); and any absolute-rationality ranking
\emph{between} LLMs (only within-design comparative claims about
\(\alpha\) are supported).

\subsection*{1.7 What is new}\label{sec-contribution}
\addcontentsline{toc}{subsection}{1.7 What is new}

Multinomial logit is McFadden's (McFadden 1974); Stan, SBC, and the
prior-predictive/recovery/SBC workflow have been standard practice for a
decade (Carpenter et al. 2017; Talts et al. 2018; Gelman et al. 2020).
The contribution here is not any one of these tools. It is, first, a
\emph{conceptual} contribution --- item (a) below: a principled
explication of what it means for a decision maker to be \emph{sensitive
to}, or \emph{aligned with}, the requirements of SEU maximization across
a range of decision contexts --- and, resting on that explication, a
four-part \emph{methodological} package, items (b)--(e), that makes the
explication measurable and validates the measurement.

\textbf{(a) A principled explication of SEU sensitivity.} Within the
softmax/Luce--McFadden specification, three characterizing properties of
the choice rule (§\hyperref[sec-three-properties]{2.3}, proved in
Appendix A) license reading its single scalar parameter \(\alpha\) as
the degree to which an agent's choices track SEU maximization, with
\(\alpha = 0\) at indifference to expected utility and
\(\alpha \to \infty\) at deterministic maximization. This gives a
graded, well-defined meaning to ``alignment with expected-utility
reasoning'' across the family of decision problems the estimable model
accommodates (the experiments expressible in the data block of the Stan
models, e.g., \texttt{m\_01}), and separates that disposition from the
\emph{content} of the agent's commitments (its beliefs and utilities)
and from the \emph{outcomes} it realizes
(§\hyperref[sec-conceptual-payoff]{2.5}). The generalization across many
agents and contexts via a hierarchical model is left to a companion
paper (§\hyperref[sec-app-followup]{7.7}). \emph{In plain terms:}
\(\alpha\) assigns a precise number to how closely an agent's choices
follow expected-utility reasoning, holding apart what the agent
believes, what it wants, and how things happen to turn out.

\textbf{(b) A clean identifiability decomposition.} We separate the
identifiability of \(\alpha\) \emph{from the expected-utility vector}
\(\eta\) --- stated as a proposition under an explicit genericity
condition (§\hyperref[sec-alpha-from-eta]{3.3}, Appendix B.1) --- from
the question of whether the data also pin down the ingredients of
\(\eta\), namely beliefs (\(\beta\)) and utilities (\(\delta\)). At
realistic sample sizes they do not: uncertain choices leave
\((\beta, \delta)\) only \emph{weakly informed}, as the Bayesian
workflow shows directly --- wide, correlated \((\beta, \delta)\)
posteriors against tight \(\alpha\) intervals
(§\hyperref[sec-bd-weak]{3.4}, §\hyperref[sec-m0-recovery]{4.3}). The
two are one structural picture (§\hyperref[sec-m0-link]{3.5}): given the
\(\bm{\upsilon}\)-endpoint convention that fixes the utility scale, the
within-menu choice-probability contrasts inform \(\alpha\) through the
products \(\alpha(\eta_r - \eta_s)\), while beliefs and utilities enter
only \emph{through} \(\eta\) and trade off against each other in a way
that uncertain-choice data barely resolve. Because the data inform
\(\alpha\) only through those products, what ``recovered'' means is
always relative to the utility spreads a particular design and prior
induce --- a design-conditionality the application makes quantitative
(§\hyperref[sec-app-validation]{7.4}). \emph{In plain terms:} under a
given design and prior, the alignment number \(\alpha\) can be recovered
from choice data, but the agent's underlying beliefs and utilities often
cannot be told apart from one another by that same data.

\textbf{(c) A worked demonstration that in-principle identifiability
does not predict finite-sample estimability.} The
\texttt{m\_0}/\texttt{m\_1} contrast makes the gap concrete in two
distinct ways. For the utility parameter \(\delta\), adding the risky
block supplies a \(\beta\)-free route that secures its identifiability
in principle (§\hyperref[sec-delta-id]{5.5}) yet buys almost no
precision at the design sample size --- a matched-count
credible-interval-width reduction under 1\%
(§\hyperref[sec-m1-recovery]{6.4}). For the sensitivity parameter
\(\alpha\), a matched recovery study shows that finite-sample precision,
\emph{in this design}, is governed by data \emph{quantity} rather than
the \emph{type} of choice that an in-principle argument might privilege.
\emph{In plain terms:} a quantity that is recoverable in theory may
still not be recoverable from a realistic amount of data.

\textbf{(d) The marginal-SBC demarcation.} Marginal rank uniformity is
\emph{necessary but not sufficient} for joint posterior calibration.
Separately, the correlated, barely-contracted joint \((\beta, \delta)\)
posterior --- which can itself be perfectly calibrated --- reflects weak
joint \emph{informativeness}, and marginal SBC does not reveal it:
per-parameter ranks pass for both models even where beliefs and
utilities are jointly only weakly resolved
(§§\hyperref[sec-m0-sbc]{4.4}, \hyperref[sec-m1-sbc]{6.5},
\hyperref[sec-discussion]{8.3}). \emph{In plain terms:} a standard
calibration check can look healthy one parameter at a time while still
concealing that two parameters are jointly unresolved.

\textbf{(e) An illustrative full-pipeline application} of the instrument
to LLM choice data, in which the \(\alpha\) inference and the
posterior-predictive summaries come in clean --- the only flagged
diagnostics being confined to the weakly-informed nuisance parameters
(§\hyperref[sec-app-validation]{7.4}) --- and the framework reports
posterior support for a structured comparative effect in two of four
LLM\(\times\)task cells while withholding it in the other two, which it
reads as inconclusive on their own diagnostics --- with the design's
resolution quantified explicitly for one of them
(§\hyperref[sec-application]{7}). \emph{In plain terms:} run end-to-end
on real language-model choices, the instrument's sensitivity estimates
behave as designed and report an effect in two of four settings, with
the non-detections reflecting limited resolution rather than established
absence.

\subsection*{1.8 Roadmap}\label{sec-roadmap}
\addcontentsline{toc}{subsection}{1.8 Roadmap}

The paper alternates between \emph{abstract/identifiability} sections
and their \emph{computational} realizations, then applies the result.
§\hyperref[sec-abstract-model]{2} fixes the abstract model.
§\hyperref[sec-m0-identifiability]{3} establishes the identifiability
picture for the uncertain-choice-only model \texttt{m\_0}, and
§\hyperref[sec-m0-implementation]{4} realizes and validates it in Stan.
§\hyperref[sec-m1-identifiability]{5} extends the model with risky
choices and revisits identifiability, and
§\hyperref[sec-m1-implementation]{6} realizes and validates the
extension --- inheriting both the design and the identifiability
questions from the earlier pair. §\hyperref[sec-application]{7} runs the
validated workflow on real LLM data, and
§\hyperref[sec-discussion-top]{8} draws the methodological lessons
together. The formal core --- the proofs that close
§§\hyperref[sec-m0-identifiability]{3} and
\hyperref[sec-m1-identifiability]{5} --- is in Appendix B; the body
states results and points to it. Concretely, the dependency structure is

\[
\underbrace{\S2}_{\text{model}}
\;\longrightarrow\;
\underbrace{\S3 \to \S4}_{\texttt{m\_0}}
\;\longrightarrow\;
\underbrace{\S5 \to \S6}_{\texttt{m\_1}}
\;\longrightarrow\;
\underbrace{\S7}_{\text{application}}
\;\longrightarrow\;
\underbrace{\S8}_{\text{discussion}},
\]

with §\hyperref[sec-m0-identifiability]{3} feeding \emph{both}
§\hyperref[sec-m0-implementation]{4} and
§\hyperref[sec-m1-identifiability]{5}, the two implementation sections
(§§\hyperref[sec-m0-implementation]{4},
\hyperref[sec-m1-implementation]{6}) converging on the application, and
Appendix B supplying the proofs that close the two identifiability
sections.

\textbf{How to read this paper.} Readers approaching from statistics or
the Bayesian workflow may treat
§§\hyperref[sec-abstract-model]{2}--\hyperref[sec-m0-identifiability]{3}
and \hyperref[sec-m1-identifiability]{5} as the modeling setup and
concentrate on the validation and application
(§§\hyperref[sec-m0-implementation]{4},
\hyperref[sec-m1-implementation]{6}, \hyperref[sec-application]{7});
readers approaching from decision theory or formal epistemology will
find the conceptual claims in
§§\hyperref[sec-motivation]{1}--\hyperref[sec-abstract-model]{2} and
\hyperref[sec-discussion-top]{8} and can take the identifiability
propositions (§§\hyperref[sec-m0-identifiability]{3},
\hyperref[sec-m1-identifiability]{5}) on their statements, with proofs
in Appendix B. Terms of art from each field are glossed at first use for
the others.

\section*{The Abstract Model: Softmax Choice and
Sensitivity}\label{sec-abstract-model}
\addcontentsline{toc}{section}{The Abstract Model: Softmax Choice and
Sensitivity}

This section specifies the choice model abstractly, before any
commitment to where values come from. The payoff of the abstraction is
that the three properties that license reading \(\alpha\) as a
\emph{sensitivity} parameter (§\hyperref[sec-three-properties]{2.3})
hold for \emph{any} value function; the subjective expected utility
(SEU) specialization of §\hyperref[sec-seu-specialization]{2.4} then
inherits them as corollaries.

\subsection*{2.1 Alternatives, values, and the sensitivity
parameter}\label{sec-alternatives}
\addcontentsline{toc}{subsection}{2.1 Alternatives, values, and the
sensitivity parameter}

Let \(\mathcal{R} = \{1, \dots, R\}\) be a finite set of distinct
\emph{alternatives}. A \emph{value function}
\(V : \mathcal{R} \to \mathbb{R}\) assigns a real number \(V(r)\) to
each alternative. A decision problem \(m\) presents a subset of
alternatives, encoded by availability indicators
\(I_{m,r} \in \{0,1\}\). A \emph{sensitivity parameter}
\(\alpha \geq 0\) governs how sharply choices track value differences.

\subsection*{2.2 The softmax choice rule}\label{sec-softmax-rule}
\addcontentsline{toc}{subsection}{2.2 The softmax choice rule}

The probability of selecting alternative \(r\) from the available set in
problem \(m\) is the \textbf{softmax} (Luce--McFadden--Boltzmann) rule
\begin{equation}\phantomsection\label{eq-softmax}{
P(\text{choose } r \mid \alpha, V)
  = \frac{\exp\!\big(\alpha\, V(r)\big)}{\sum_{j : I_{m,j}=1} \exp\!\big(\alpha\, V(j)\big)}.
}\end{equation} The functional form has independent precedents: Luce's
(1959) ratio-scale choice axiom, McFadden's (1974) random-utility
derivation of multinomial logit, and the Boltzmann distribution of
statistical mechanics (where \(\alpha\) is inverse temperature). These
three routes reach the same functional form --- a choice likelihood
exponential in value differences --- from independent starting points,
which is part of what recommends it here. We use \(\alpha\) rather than
a temperature \(T = 1/\alpha\) because \emph{higher} \(\alpha\)
corresponds to \emph{higher} sensitivity, the more intuitive direction
for our purposes. (This model temperature \(T = 1/\alpha\) is distinct
from the LLM \emph{sampling} temperature that appears as an experimental
covariate in §\hyperref[sec-application]{7}; the application asks, in
part, how the estimated \(\alpha\) responds to that sampling
temperature.)

\emph{Why softmax rather than probit or another stochastic-choice
wrapping?} Three features recommend it. Luce's (1959)
independence-of-irrelevant-alternatives axiom yields a ratio form
\(P(r \mid \mathcal{A}) = w(r)/\sum_{j \in \mathcal{A}} w(j)\) over
positive weights; specifying an exponential link
\(w(r) = \exp(\alpha V(r))\) from the real-valued value scale to those
weights --- equivalently, log-odds linearity in value differences ---
selects the softmax (Equation~\ref{eq-softmax}). Given that link,
softmax embeds the optimization and uniform-choice limits as the two
endpoints of a single scalar \(\alpha\) (Properties 2 and 3 of
§\hyperref[sec-three-properties]{2.3}, proved as Theorems A.2 and A.3 in
Appendix A), and its log-likelihood is concave in the index \(\alpha V\)
--- a standard multinomial-logit property (McFadden 1974; Train 2009)
--- so that, with the value scale \(V\) held fixed, the
\(\alpha\)-likelihood is unimodal and free of spurious local optima.
(The concavity is in the index \(\alpha V\), i.e.~the surface in
\(\alpha\) with \(V\) fixed; it is \emph{not} joint concavity in
\(\alpha\) together with the value parameters, because \(\alpha V\) is
bilinear in the two. Nor does concavity guarantee a \emph{finite}
maximizer: if the observed choices happen to select a value-maximal
alternative in every problem --- the analogue of complete separation in
logistic regression --- the likelihood increases in \(\alpha\) without
bound, and it is the prior on \(\alpha\) that restores a proper
posterior. The Bayesian workflow of §\hyperref[sec-m0-implementation]{4}
never relies on a maximum-likelihood point estimate.) These are two
distinct virtues, and we keep them apart: the log-odds linearity and
single-scalar limit structure above make \(\alpha\)
\emph{interpretable}, while the concavity is what makes it
\emph{well-behaved to estimate}. Probit, by contrast, replaces the Luce
ratio with Gaussian latent errors: a probit scale parameter carries the
same two limits (uniform choice as it grows, deterministic optimization
as it vanishes), but probit offers no closed-form IIA/log-odds structure
and no comparably tractable single-scalar link, so it is the softmax's
closed-form log-odds linearity that singles it out here (Train 2009).

\emph{Related uses of a softmax precision parameter.} The scalar
\(\alpha\) has close relatives across several literatures. In
experimental game theory it is the precision parameter \(\lambda\) of
quantal response equilibrium (McKelvey and Palfrey 1995), which plays
for strategic choice the role \(\alpha\) plays here for single-agent
choice. In the econometrics of risky choice, Hey and Orme (1994)
estimate expected-utility and generalized-EU models wrapped in
stochastic choice, initiating a literature on how the noise
specification shapes inference about the deterministic core. In machine
learning, the same exponential-in-value rule appears as \emph{Boltzmann
rationality}, the standard observation model in inverse reinforcement
learning (Ziebart et al. 2008). Two cautions from the econometric branch
carry over. Wilcox (2011) argues that a logit-style precision is not
comparable across contexts unless the utility scale is normalized per
context (``contextual utility''), and Apesteguia and Ballester (2018)
show that fixed-precision random-utility wrappings can order risk
attitudes non-monotonically. We take both seriously rather than claiming
immunity: the \(\bm{\upsilon}\)-endpoint convention of
§\hyperref[sec-notation]{2.6} fixes the utility scale (best consequence
\(= 1\), worst \(= 0\)) uniformly across problems. This is a
\emph{global} normalization over a fixed consequence space, not the
per-context renormalization Wilcox's contextual utility prescribes; the
two coincide here only because every problem in a given design draws its
consequences from the same \(K\)-element space, so the within-context
utility range is the same fixed range in every problem. Within such
fixed-consequence-space designs the convention supplies the scale
comparability the critique asks for; across designs with different
consequence spaces it does not, which is one reason the paper licenses
only \emph{within-design comparative} readings of \(\alpha\)
(§\hyperref[sec-app-limits]{7.6}, §\hyperref[sec-disc-alpha]{8.4}), not
cross-instrument cardinal ones.

\subsection*{2.3 Three characterizing properties of
softmax}\label{sec-three-properties}
\addcontentsline{toc}{subsection}{2.3 Three characterizing properties of
softmax}

The following hold for \emph{any} value function \(V\), within each
decision problem \(m\) and relative to its available set
\(\mathcal{A}_m = \{r : I_{m,r} = 1\}\). Write
\(V^\ast_m = \max_{r \in \mathcal{A}_m} V(r)\) for the maximal available
value,
\(\mathcal{R}^\ast_m = \{r \in \mathcal{A}_m : V(r) = V^\ast_m\}\) for
the value-maximizing set, and
\(\mathcal{R}^-_m = \mathcal{A}_m \setminus
\mathcal{R}^\ast_m\). Proofs are in Appendix A.

\begin{tcolorbox}[enhanced jigsaw, rightrule=.15mm, breakable, colback=white, arc=.35mm, leftrule=.75mm, colframe=quarto-callout-note-color-frame, toprule=.15mm, bottomrule=.15mm, left=2mm, opacityback=0]
\begin{minipage}[t]{5.5mm}
\textcolor{quarto-callout-note-color}{\faInfo}
\end{minipage}%
\begin{minipage}[t]{\textwidth - 5.5mm}

\textbf{Property 1 (Monotonicity in sensitivity).} Holding \(V\) fixed,
for any two available alternatives \(r, s \in \mathcal{A}_m\) the
log-odds are linear in \(\alpha\), \[
\frac{\partial}{\partial \alpha}
  \log\frac{P(\text{choose } r \mid \alpha, V)}{P(\text{choose } s \mid \alpha, V)}
  = V(r) - V(s),
\] so the odds always shift monotonically toward the higher-valued
alternative as \(\alpha\) increases. Consequently, \emph{provided the
available values are not all equal}, the \emph{total} probability on the
value-maximizing set \(\mathcal{R}^\ast_m\) is strictly increasing in
\(\alpha\); in particular each value-maximizing alternative
\(r \in \mathcal{R}^\ast_m\) gains and each value-minimizing alternative
loses probability. (If all available values coincide, every alternative
is value-maximizing and all choice probabilities are constant in
\(\alpha\).) One subtlety is worth flagging. Since the \emph{total}
probability on the value-maximizing set rises monotonically in
\(\alpha\), one might expect the probability of every individual
\emph{non-maximal} alternative to \emph{fall} monotonically in
\(\alpha\). This is not so. As shown in Appendix A (Theorem A.1(iii)),
the \(\alpha\)-derivative of a non-maximal alternative's choice
probability has the sign of \(V(r) -
\mathbb{E}_\alpha[V]\), where \(\mathbb{E}_\alpha[V]\) is the current
choice-weighted mean --- the average of the available values weighted by
their current choice probabilities. So while a non-maximal alternative's
value still exceeds this mean its probability \emph{rises}, and only
once the mean overtakes it does its probability begin to fall. Such an
alternative's probability therefore rises before it falls rather than
declining monotonically; individual non-maximal probabilities are not
monotone one-by-one.

\end{minipage}%
\end{tcolorbox}

\begin{tcolorbox}[enhanced jigsaw, rightrule=.15mm, breakable, colback=white, arc=.35mm, leftrule=.75mm, colframe=quarto-callout-note-color-frame, toprule=.15mm, bottomrule=.15mm, left=2mm, opacityback=0]
\begin{minipage}[t]{5.5mm}
\textcolor{quarto-callout-note-color}{\faInfo}
\end{minipage}%
\begin{minipage}[t]{\textwidth - 5.5mm}

\textbf{Property 2 (Optimization limit).} As \(\alpha \to \infty\),
choice concentrates on the value-maximizing set:
\(P(\text{choose } r \mid \alpha, V) \to
1/|\mathcal{R}^\ast_m|\) for \(r \in \mathcal{R}^\ast_m\) and \(\to 0\)
otherwise. With a unique maximizer (the generic case under continuous
priors) choice becomes deterministic.

\end{minipage}%
\end{tcolorbox}

\begin{tcolorbox}[enhanced jigsaw, rightrule=.15mm, breakable, colback=white, arc=.35mm, leftrule=.75mm, colframe=quarto-callout-note-color-frame, toprule=.15mm, bottomrule=.15mm, left=2mm, opacityback=0]
\begin{minipage}[t]{5.5mm}
\textcolor{quarto-callout-note-color}{\faInfo}
\end{minipage}%
\begin{minipage}[t]{\textwidth - 5.5mm}

\textbf{Property 3 (Uniform-choice limit).} As \(\alpha \to 0\), choice
approaches the uniform distribution over available alternatives,
independent of \(V\).

\end{minipage}%
\end{tcolorbox}

Together the three properties give exactly the conceptual content needed
to read \(\alpha\) as ``sensitivity to value maximization,'' with no
reference yet to where values come from: \(\alpha\) interpolates
monotonically (Property 1) between indifference to value (Property 3)
and exclusive pursuit of the maximum (Property 2).

\subsection*{2.4 SEU specialization}\label{sec-seu-specialization}
\addcontentsline{toc}{subsection}{2.4 SEU specialization}

Specialize the value function to a \emph{subjective expected utility}.
Fix \(K\) consequences with an (ordered) utility vector
\(\bm{\upsilon} \in \mathbb{R}^K\). Each alternative \(r\) carries a
subjective probability distribution \(\bm{\psi}_r \in \Delta^{K-1}\)
over consequences, and its value is its expected utility
\begin{equation}\phantomsection\label{eq-eu}{
V(r) = \eta_r = \bm{\psi}_r^\top \bm{\upsilon} = \sum_{k=1}^{K} \psi_{r,k}\,\upsilon_k .
}\end{equation} Because \(\eta_r\) is just a particular value function,
Properties 1--3 apply verbatim, and \(\alpha\) now measures sensitivity
to \emph{SEU maximization} specifically: the disposition to choose the
expected-utility-maximizing alternative.

\subsection*{2.5 The conceptual payoff}\label{sec-conceptual-payoff}
\addcontentsline{toc}{subsection}{2.5 The conceptual payoff}

Sensitivity to SEU maximization is the disposition of an SEU-committed
agent to \emph{act in accordance with its commitments}. The contrast
between an agent's \emph{commitment} to a standard and its
\emph{performance} relative to that standard is Levi's (Levi 1980, chap.
1); we return to it, and to the sense in which \(\alpha\) measures a
tendency to perform, in §\hyperref[sec-disc-meaning]{8.1}. The
disposition is distinct from (a) the \emph{content} of those commitments
--- the agent's beliefs \(\bm{\psi}\) and utilities \(\bm{\upsilon}\)
--- and (b) the \emph{outcomes} the agent happens to realize. The SEU
construction thus decomposes choice behavior conceptually into beliefs,
utilities, and sensitivity. The three pieces are \emph{conceptually}
separable in this way, but they are not equally separable
\emph{empirically}: as §\hyperref[sec-bd-weak]{3.4} and
§\hyperref[sec-delta-id]{5.5} show, \(\alpha\) separates cleanly from
the rest, whereas \(\beta\) and \(\delta\) are only weakly informed by
uncertain choices and are recovered jointly rather than each in
isolation.

\begin{tcolorbox}[enhanced jigsaw, rightrule=.15mm, breakable, colback=white, arc=.35mm, leftrule=.75mm, colframe=quarto-callout-warning-color-frame, toprule=.15mm, bottomrule=.15mm, left=2mm, opacityback=0]
\begin{minipage}[t]{5.5mm}
\textcolor{quarto-callout-warning-color}{\faExclamationTriangle}
\end{minipage}%
\begin{minipage}[t]{\textwidth - 5.5mm}

\textbf{Caveat on uniqueness of the decomposition.} The
\((\alpha, \bm{\psi},
\bm{\upsilon})\) decomposition is not the only way to organize the same
observable choice data: rank-dependent utility, cumulative prospect
theory (Tversky and Kahneman 1992), and maxmin expected utility (Gilboa
and Schmeidler 1989) organize it differently. These alternatives are not
interchangeable substitutes for SEU, and they differ among themselves in
kind --- some are descriptive accounts rather than normative standards,
as §\hyperref[sec-seu-standard]{1.4} discusses. Our claim is only that
\emph{under} an SEU reference standard the (sensitivity, beliefs,
utilities) decomposition is well-defined and --- with the
identifiability caveats of §\hyperref[sec-m0-identifiability]{3} and
§\hyperref[sec-m1-identifiability]{5} --- measurable. It is not a claim
that this decomposition is uniquely correct for all theoretical
purposes.

\end{minipage}%
\end{tcolorbox}

\subsection*{\texorpdfstring{2.6 Notation, the utility-scale convention,
and the \(\beta\)
gauge}{2.6 Notation, the utility-scale convention, and the \textbackslash beta gauge}}\label{sec-notation}
\addcontentsline{toc}{subsection}{2.6 Notation, the utility-scale
convention, and the \(\beta\) gauge}

For the parameterized models of §\hyperref[sec-m0-identifiability]{3}
and §\hyperref[sec-m1-identifiability]{5} we generate subjective
probabilities from \(D\)-dimensional feature vectors \(w_r\) via a
linear-softmax belief map
\(\bm{\psi}_r = \mathrm{softmax}(\beta\, w_r)\) with
\(\beta \in \mathbb{R}^{K
\times D}\), in the multinomial-logit tradition (McFadden 1974; Train
2009). The utility vector is built from increments
\(\delta \in \Delta^{K-2}\) via cumulative sums, with endpoints fixed by
convention. (The \(K-1\) increments \(\delta_j =
\upsilon_{j+1} - \upsilon_j\) are nonnegative and sum to
\(\upsilon_K - \upsilon_1
= 1\), so \(\delta\) lies in the \((K-2)\)-simplex \(\Delta^{K-2}\),
matching the Stan declaration \texttt{simplex{[}K\ -\ 1{]}\ delta}.)

Two facts about this parameterization recur throughout and are stated
once here, then referenced rather than redefined. The first fixes the
zero and unit of the utility scale, which is what makes \(\alpha\) a
well-posed target; the second records an indeterminacy of the belief
map.

\begin{tcolorbox}[enhanced jigsaw, rightrule=.15mm, breakable, colback=white, arc=.35mm, leftrule=.75mm, colframe=quarto-callout-important-color-frame, toprule=.15mm, bottomrule=.15mm, left=2mm, opacityback=0]
\begin{minipage}[t]{5.5mm}
\textcolor{quarto-callout-important-color}{\faExclamation}
\end{minipage}%
\begin{minipage}[t]{\textwidth - 5.5mm}

\textbf{The \(\bm{\upsilon}\)-endpoint convention.} vNM utilities are
unique only up to a positive affine transformation
\(\tilde{\upsilon}_k = a\,\upsilon_k + b\) with \(a > 0\), and rescaling
by \(a\) is observationally equivalent to rescaling \(\alpha\) by
\(1/a\) (Theorem A.4). Fixing the endpoints \[
0 = \upsilon_1 \leq \upsilon_2 \leq \cdots \leq \upsilon_K = 1
\] fixes the zero and unit of the utility scale, removing this affine
indeterminacy and making \(\alpha\) a well-posed sensitivity target on a
standardized one-unit utility scale; its identifiability additionally
requires nonconstant expected utilities on some menu
(§\hyperref[sec-alpha-from-eta]{3.3}).

\end{minipage}%
\end{tcolorbox}

\begin{tcolorbox}[enhanced jigsaw, rightrule=.15mm, breakable, colback=white, arc=.35mm, leftrule=.75mm, colframe=quarto-callout-important-color-frame, toprule=.15mm, bottomrule=.15mm, left=2mm, opacityback=0]
\begin{minipage}[t]{5.5mm}
\textcolor{quarto-callout-important-color}{\faExclamation}
\end{minipage}%
\begin{minipage}[t]{\textwidth - 5.5mm}

\textbf{The \(\bm{\beta}\) gauge (additive row-shift).} Adding the same
vector \(\gamma \in \mathbb{R}^D\) to every row of \(\beta\) --- that
is, \(\beta_{k\cdot} \mapsto \beta_{k\cdot} + \gamma^\top\) for all
\(k\) --- shifts every entry \((\beta\,w_r)_k\) by the common amount
\(\gamma^\top w_r\) and so leaves \(\mathrm{softmax}(\beta\,w_r)\)
unchanged. We rely on no other invariance of the belief map; we refer to
it as the \textbf{\(\beta\) gauge} and reference it in
§\hyperref[sec-alpha-from-eta]{3.3},
§\hyperref[sec-m0-implementation]{4},
§\hyperref[sec-m1-implementation]{6}, and Appendix B.4.

\end{minipage}%
\end{tcolorbox}

The consolidated glossary is collected once here
(Table~\ref{tbl-notation}) and referenced, not redefined, elsewhere.

\begin{longtable}[]{@{}
  >{\raggedright\arraybackslash}p{(\linewidth - 4\tabcolsep) * \real{0.3333}}
  >{\raggedright\arraybackslash}p{(\linewidth - 4\tabcolsep) * \real{0.3333}}
  >{\raggedright\arraybackslash}p{(\linewidth - 4\tabcolsep) * \real{0.3333}}@{}}
\caption{Consolidated notation.}\label{tbl-notation}\tabularnewline
\toprule\noalign{}
\begin{minipage}[b]{\linewidth}\raggedright
Symbol
\end{minipage} & \begin{minipage}[b]{\linewidth}\raggedright
Meaning
\end{minipage} & \begin{minipage}[b]{\linewidth}\raggedright
First used
\end{minipage} \\
\midrule\noalign{}
\endfirsthead
\toprule\noalign{}
\begin{minipage}[b]{\linewidth}\raggedright
Symbol
\end{minipage} & \begin{minipage}[b]{\linewidth}\raggedright
Meaning
\end{minipage} & \begin{minipage}[b]{\linewidth}\raggedright
First used
\end{minipage} \\
\midrule\noalign{}
\endhead
\bottomrule\noalign{}
\endlastfoot
\(\alpha \geq 0\) & sensitivity to value maximization (primary quantity
of interest) & §2.1 \\
\(V(r),\ \eta_r\) & value / expected utility of \(r\),
\(\eta_r = \bm{\psi}_r^\top \bm{\upsilon}\) & §2.1, §2.4 \\
\(\bm{\psi}_r\) & subjective probabilities for uncertain alternative
\(r\) (from \(\beta\)) & §2.4 \\
\(\beta \in \mathbb{R}^{K\times D}\) & belief-formation weights;
\(\bm{\psi}_r = \mathrm{softmax}(\beta w_r)\) & §2.6 \\
\(w_r\) & \(D\)-dimensional feature vector of alternative \(r\) &
§2.6 \\
\(\delta \in \Delta^{K-2}\) & utility increments (\(K-1\) entries; Stan
\texttt{simplex{[}K\ -\ 1{]}}) & §2.6 \\
\(\bm{\upsilon}\) & ordered utilities,
\(\upsilon_1 = 0 \le \cdots \le \upsilon_K = 1\) & §2.6 \\
\(\pi_s\) & objective lottery simplex for risky alternative \(s\) (Stan
data: \texttt{x}) & §5.2 \\
\(\beta\) gauge & additive row-shift indeterminacy of the belief softmax
& §2.6 \\
\(\bm{\upsilon}\)-endpoint convention & fixes the zero and unit of the
vNM utility scale & §2.6 \\
\end{longtable}

\subsection*{2.7 The identifiability
question}\label{sec-identifiability-preview}
\addcontentsline{toc}{subsection}{2.7 The identifiability question}

The choice-probability function is determined by \((\alpha, \bm{\psi},
\bm{\upsilon})\). Whether the \emph{parameterizations} of \(\bm{\psi}\)
and \(\bm{\upsilon}\) --- namely \((\beta, \delta)\) --- are recoverable
from observed choices is the question
§\hyperref[sec-m0-identifiability]{3} and
§\hyperref[sec-m1-identifiability]{5} take up.

\section*{\texorpdfstring{Choice Under Uncertainty Alone:
Identifiability of \(\alpha\), and the Weakly-Informed
\((\beta,\delta)\)}{Choice Under Uncertainty Alone: Identifiability of \textbackslash alpha, and the Weakly-Informed (\textbackslash beta,\textbackslash delta)}}\label{sec-m0-identifiability}
\addcontentsline{toc}{section}{Choice Under Uncertainty Alone:
Identifiability of \(\alpha\), and the Weakly-Informed
\((\beta,\delta)\)}

\subsection*{3.1 The parameterization}\label{sec-m0-param}
\addcontentsline{toc}{subsection}{3.1 The parameterization}

Recall from §\hyperref[sec-notation]{2.6} the uncertain-choice
(``\texttt{m\_0}'') model. Subjective probabilities are generated from
\(D\)-dimensional features \(w_r\) via
\(\bm{\psi}_r = \mathrm{softmax}(\beta\, w_r)\) with
\(\beta \in \mathbb{R}^{K\times
D}\). Utilities are constructed from increments \(\delta\) by cumulative
summation, yielding \(0 = \upsilon_1 \le \upsilon_2 \le \cdots \le
\upsilon_K = 1\). There are \(K-1\) nonnegative increments, summing to
one, so \(\delta\) ranges over the \((K-2)\)-dimensional simplex
\(\Delta^{K-2}\) --- the sum-to-one constraint removes one degree of
freedom --- and Stan declares it \texttt{simplex{[}K\ -\ 1{]}\ delta},
counting the \(K-1\) entries. Expected utilities are
\(\eta_r = \bm{\psi}_r^\top \bm{\upsilon}\), and choices follow
Equation~\ref{eq-softmax} with \(V(r) = \eta_r\). The \(\beta\) gauge
(§\hyperref[sec-notation]{2.6}) is the only \(\beta\) indeterminacy we
use.

\subsection*{3.2 Why this parameterization}\label{sec-m0-why}
\addcontentsline{toc}{subsection}{3.2 Why this parameterization}

Linear-softmax for \(\bm{\psi}\) aligns with the multinomial-logit /
discrete-choice tradition (McFadden 1974; Train 2009); ordered utilities
with endpoints fixed by convention remove the affine scale-and-shift
indeterminacy of vNM utility (Theorem A.4). The two conventions together
leave \(\alpha\) as a scalar with the limit interpretation of
§\hyperref[sec-three-properties]{2.3}.

\subsection*{\texorpdfstring{3.3 Identifiability of \(\alpha\) from
\(\eta\)}{3.3 Identifiability of \textbackslash alpha from \textbackslash eta}}\label{sec-alpha-from-eta}
\addcontentsline{toc}{subsection}{3.3 Identifiability of \(\alpha\) from
\(\eta\)}

\begin{tcolorbox}[enhanced jigsaw, rightrule=.15mm, breakable, colback=white, arc=.35mm, leftrule=.75mm, colframe=quarto-callout-note-color-frame, toprule=.15mm, bottomrule=.15mm, left=2mm, opacityback=0]
\begin{minipage}[t]{5.5mm}
\textcolor{quarto-callout-note-color}{\faInfo}
\end{minipage}%
\begin{minipage}[t]{\textwidth - 5.5mm}

\textbf{Proposition 3.1 (Identifiability of \(\alpha\) from \(\eta\)).}
Fix the expected-utility vector \(\eta = (\eta_r)_{r \in \mathcal{R}}\).
If there is a menu \(\mathcal{M} \subseteq \mathcal{R}\) of positive
design probability on which the values
\(\{\eta_r : r \in \mathcal{M}\}\) are not all equal, then
\(\alpha > 0\) is globally identifiable from the choice-probability
function on \(\mathcal{M}\).

\end{minipage}%
\end{tcolorbox}

\emph{Sketch (full proof in Appendix B.1).} On any menu with
non-constant \(\eta\) the log-odds between two alternatives with
\(\eta_r \neq \eta_s\) is \(\log[P(r)/P(s)] =
\alpha(\eta_r - \eta_s)\), strictly monotone in \(\alpha\); the softmax
choice-probability map is therefore injective in \(\alpha\) on such a
menu.

\begin{tcolorbox}[enhanced jigsaw, rightrule=.15mm, breakable, colback=white, arc=.35mm, leftrule=.75mm, colframe=quarto-callout-note-color-frame, toprule=.15mm, bottomrule=.15mm, left=2mm, opacityback=0]
\begin{minipage}[t]{5.5mm}
\textcolor{quarto-callout-note-color}{\faInfo}
\end{minipage}%
\begin{minipage}[t]{\textwidth - 5.5mm}

\textbf{Boundary remark (\(\alpha = 0\)).} Proposition 3.1 identifies
\(\alpha\) throughout the open half-line \(\alpha > 0\), and the
log-odds identity \(\log[P(r)/P(s)] = \alpha(\eta_r - \eta_s)\) is
injective on the closed half-line \([0, \infty)\) whenever
\(\eta_r \neq \eta_s\), so \(\alpha = 0\) is reached continuously from
above and the genericity condition on \(\{\eta_r\}\) is \emph{not}
vacuous there. With \(\eta\) fixed the softmax log-likelihood
\(\ell = \alpha\eta_y
- \log\sum_j \exp(\alpha\eta_j)\) --- where \(\eta_y\) is the expected
utility of the chosen alternative and the sum runs over the menu --- is
in fact smooth at \(\alpha = 0\), with
\(\partial_\alpha \ell\big|_{\alpha=0} = \eta_y - \bar\eta\), where
\(\bar\eta\) is the menu-average expected utility. What is special about
\(\alpha = 0\) is therefore not non-differentiability but
\emph{non-regularity}: the choice rule is uniform (Theorem A.3) and
derivatives with respect to any parameter entering only through \(\eta\)
are multiplied by \(\alpha\), so \(\beta\) and \(\delta\) become locally
unidentified at the boundary. The proposition does not silently exclude
the near-random null; it simply does not assert differentiable
identification of the \(\eta\)-generating parameters \emph{at} the
boundary.

\end{minipage}%
\end{tcolorbox}

The structure of the claim matters: \(\alpha\) is identified \emph{from}
\(\eta\), not from \((\beta, \delta)\) directly. Whether
\((\beta, \delta)\) themselves are recoverable from uncertain choices is
a separate question --- the subject of §\hyperref[sec-bd-weak]{3.4}.

\subsection*{\texorpdfstring{3.4 Why \((\beta,\delta)\) are weakly
informed by uncertain
choices}{3.4 Why (\textbackslash beta,\textbackslash delta) are weakly informed by uncertain choices}}\label{sec-bd-weak}
\addcontentsline{toc}{subsection}{3.4 Why \((\beta,\delta)\) are weakly
informed by uncertain choices}

Proposition 3.1 identifies \(\alpha\) \emph{from} \(\eta\). The
remaining question is whether the data also pin down the two ingredients
that \emph{make up} \(\eta\) --- beliefs (\(\beta\)) and utilities
(\(\delta\)). They do not, at realistic sample sizes, and the reason is
structural. Expected utility composes the two multiplicatively, \[
\eta_r \;=\; \mathrm{softmax}(\beta\, w_r)^\top \bm{\upsilon}(\delta),
\] the product of a \(\beta\)-driven simplex and a \(\delta\)-driven
utility vector. Uncertain choices see \((\beta, \delta)\) only through
this scalar \(\eta_r\), so a family of \((\beta, \delta)\) pairs that
trades a change in beliefs against a compensating change in utilities
leaves the implied expected utilities --- and hence the choice
probabilities --- almost unchanged. The compensation is approximate
rather than exact: the data can pin down \(\eta\) (and through it
\(\alpha\)) sharply while still failing to separate the two factors
whose product forms it. The \(\bm{\upsilon}\)-endpoint convention
(§\hyperref[sec-notation]{2.6}) fixes the zero and unit of the utility
scale, but it does not touch this \emph{multiplicative}
\((\beta,\delta)\) coupling, which is not an indeterminacy that any
convention removes; the coupling is what leaves \(\beta\) and \(\delta\)
only weakly informed by uncertain-choice data.

This shows up directly in the Bayesian workflow rather than as a
separate formal claim. Parameter recovery in \texttt{m\_0}
(§\hyperref[sec-m0-recovery]{4.3}) returns wide marginal credible
intervals for both \(\beta\) and \(\delta\) --- in contrast to the
tight, well-calibrated \(\alpha\) intervals --- together with
\emph{correlated \(\beta\)--\(\delta\) estimation errors}: concretely,
across recovery replicates the posterior-mean estimation errors of a
representative weight entry \(\beta_{1,1}\) and a representative
increment \(\delta_1\) are correlated (Pearson;
§\hyperref[sec-m0-recovery]{4.3}). This is an \emph{across-replicate}
error correlation, not a within-posterior one; because raw \(\beta\)
entries carry the row-shift gauge we read this representative-component
diagnostic as an illustrative signature of the coupling rather than a
gauge-invariant statistic, and we do not attach significance to its
sign, which is not stable across simulation settings --- it shifts, for
instance, as the true \(\alpha\) used to generate the recovery data is
varied. The recovered posteriors concentrate not on \((\beta,\delta)\)
individually but on a compensating trade-off between them. That
correlated, barely-contracted joint posterior is the practical signature
of the multiplicative coupling, and it is what we mean throughout by
calling \((\beta,\delta)\) \emph{weakly informed} by uncertain choices.

We state this as an empirical feature of the posterior, not as a
theorem: no strict non-identifiability is claimed, and no gauge group on
\((\beta,\delta)\) beyond the \(\beta\) row-shift of
§\hyperref[sec-notation]{2.6} is asserted. The point is that
uncertain-choice data alone supply little information for separating
beliefs from utilities --- which is exactly what motivates the extended
model of §\hyperref[sec-m1-identifiability]{5}.

\subsection*{3.5 The two facts are one structural
picture}\label{sec-m0-link}
\addcontentsline{toc}{subsection}{3.5 The two facts are one structural
picture}

\(\alpha\) is identified \emph{from} \(\eta\) (Proposition 3.1), whereas
the two ingredients of \(\eta\) --- beliefs and utilities --- are only
weakly informed by uncertain choices (§\hyperref[sec-bd-weak]{3.4}).
These are two sides of one fact: the map \((\beta,\delta) \to \eta\)
collapses beliefs and utilities multiplicatively into a single scalar
per alternative, and choices act on those scalars through \(\alpha\).
Given the design and prior, the data therefore speak clearly about
\(\alpha\) --- the within-menu \(\eta\)-contrasts identify it up to the
scale fixed by the \(\bm{\upsilon}\)-endpoint convention --- while
saying little about how a given \(\eta\) splits into its
\((\beta,\delta)\) parts. Strictly, the observed log-odds identify the
products \(\alpha(\eta_r - \eta_s)\); it is the endpoint convention that
fixes the utility scale, together with the prior and design, that turns
this into a sharp statement about \(\alpha\) itself --- as the
posterior-contraction, parameter-recovery, and SBC diagnostics of
§\hyperref[sec-m0-implementation]{4} confirm. The dependence on the
design is not a formality: how sharply the data speak about \(\alpha\)
is governed by the within-menu \(\eta\)-gaps the design induces, and the
application reports these gaps, alongside the resulting
\(\alpha\)-posterior contraction, as explicit per-design diagnostics
(§\hyperref[sec-app-validation]{7.4}).

\textbf{Why \(\alpha\) recovers cleanly anyway.} Empirically, the two
effects do not interfere: in parameter recovery
(§\hyperref[sec-m0-recovery]{4.3}) \(\alpha\) is recovered with low bias
and well-calibrated intervals \emph{regardless} of the wide, correlated
spread in the \((\beta,\delta)\) posterior. Proposition 3.1 does not by
itself guarantee this: it identifies \(\alpha\) \emph{given} \(\eta\),
and since \((\beta,\delta)\) determine the \(\eta\)-spreads, a shift in
that spread can in principle trade off against \(\alpha\) inside the
products \(\alpha(\eta_r - \eta_s)\). What the recovery study shows is
that \emph{under this design and prior} the trade-off is not exercised:
the indeterminacy in how \(\eta\) splits into beliefs and utilities
leaves the \(\alpha\) estimate essentially untouched. We report this as
an observed, design- and prior-dependent feature of the recovered
posterior rather than derive it from a separability argument.

This is a \emph{structural} feature of decisions under uncertainty:
beliefs and utilities enter only through expected utility, and
uncertain-choice data alone supply little information for separating
them.

\subsection*{\texorpdfstring{3.6 From the weakly-informed
\((\beta,\delta)\) to the extended
model}{3.6 From the weakly-informed (\textbackslash beta,\textbackslash delta) to the extended model}}\label{sec-m0-to-m1}
\addcontentsline{toc}{subsection}{3.6 From the weakly-informed
\((\beta,\delta)\) to the extended model}

That uncertain choices barely inform \((\beta,\delta)\) motivates the
extended model of §\hyperref[sec-m1-identifiability]{5}. Before getting
there, §\hyperref[sec-m0-implementation]{4} shows that the practical
consequences --- wide credible intervals on \((\beta,\delta)\), tight
intervals on \(\alpha\) --- are directly visible in a Stan
implementation, and that marginal simulation-based calibration
nonetheless passes for all three parameters (the \emph{marginal-SBC
demarcation}).

\section*{\texorpdfstring{A Basic Computational Implementation: Model
\texttt{m\_0} in
Stan}{A Basic Computational Implementation: Model m\_0 in Stan}}\label{sec-m0-implementation}
\addcontentsline{toc}{section}{A Basic Computational Implementation:
Model \texttt{m\_0} in Stan}

This section turns the abstract \texttt{m\_0} model into an estimable
Bayesian model and shows that the identifiability picture of
§\hyperref[sec-m0-identifiability]{3} is directly visible in a Stan
implementation: \(\alpha\) recovers well, \((\beta,
\delta)\) recover poorly, and marginal simulation-based calibration
(SBC) nonetheless passes for all three --- the \emph{marginal-SBC
demarcation}.

\subsection*{4.1 Model specification}\label{sec-m0-spec}
\addcontentsline{toc}{subsection}{4.1 Model specification}

\textbf{Data.} \(M\) decision problems; \(R\) distinct alternatives with
\(D\)-dimensional feature vectors \(w_r\); availability indicators
\(I_{m,r} \in \{0,1\}\); observed choices \(y_m\). A problem \(m\)
presents \(N_m = \sum_r I_{m,r}\) alternatives.

\textbf{Parameters and priors.} \[
\alpha \sim \mathrm{Lognormal}(0, 1), \qquad
\beta_{k,d} \sim \mathcal{N}(0,1), \qquad
\delta \sim \mathrm{Dirichlet}(1, \dots, 1).
\] The \(\mathrm{Lognormal}(0,1)\) prior on \(\alpha\) (median \(1\),
roughly 90\% of mass in \((0.19, 5.18)\)) is a \emph{substantive}
choice: it spans the full near-uniform-to-near-deterministic sensitivity
range while concentrating mass near \(\alpha = 1\). The prior shape
co-determines which \(\alpha\) values are well-measured in finite
samples; we defend it via prior-predictive coverage in
§\hyperref[sec-m0-prior]{4.2}.

\begin{tcolorbox}[enhanced jigsaw, rightrule=.15mm, breakable, colback=white, arc=.35mm, leftrule=.75mm, colframe=quarto-callout-note-color-frame, toprule=.15mm, bottomrule=.15mm, left=2mm, opacityback=0]
\begin{minipage}[t]{5.5mm}
\textcolor{quarto-callout-note-color}{\faInfo}
\end{minipage}%
\begin{minipage}[t]{\textwidth - 5.5mm}

\textbf{Note for the reader (carried to §7).} This foundational
\(\mathrm{Lognormal}(0,1)\) prior on \(\alpha\) is \emph{not} the
application prior of §\hyperref[sec-application]{7}, which recalibrates
to \(\mathrm{Lognormal}(3.0, 0.75)\) for insurance and re-recalibrates
for Ellsberg. The \emph{likelihood} is unchanged across §4 and §7; only
the \(\alpha\) \emph{prior} is recalibrated.

\end{minipage}%
\end{tcolorbox}

\textbf{Transformed parameters.}
\(\bm{\psi}_r = \mathrm{softmax}(\beta\, w_r)\); the ordered utilities
\(\upsilon_k = \sum_{j<k}\delta_j\) via a cumulative sum (equivalently
\texttt{upsilon\ =\ cumulative\_sum(append\_row(0,\ delta))}), so
\(\upsilon_1 =
0\) and \(\upsilon_K = 1\); and
\(\eta_r = \bm{\psi}_r^\top \bm{\upsilon}\).

\textbf{Likelihood.}
\(y_m \sim \mathrm{Categorical}\big(\mathrm{softmax}(\alpha\,
\eta_{[m]})\big)\), where \(\eta_{[m]}\) collects the expected utilities
of the alternatives available in problem \(m\).

\begin{tcolorbox}[enhanced jigsaw, rightrule=.15mm, breakable, colback=white, arc=.35mm, leftrule=.75mm, colframe=quarto-callout-note-color-frame, toprule=.15mm, bottomrule=.15mm, left=2mm, opacityback=0]
\begin{minipage}[t]{5.5mm}
\textcolor{quarto-callout-note-color}{\faInfo}
\end{minipage}%
\begin{minipage}[t]{\textwidth - 5.5mm}

\textbf{The \(\beta\) gauge in the implementation.} Softmax invariance
to a common per-column row shift (§\hyperref[sec-notation]{2.6}) means
\(\beta\) has only \((K-1)\times
D\) effective degrees of freedom: adding the same shift to every row
leaves all choice probabilities unchanged, so the data can identify row
\emph{contrasts} but not absolute row levels. Rather than fix a
reference row, \texttt{m\_0} places an independent \(\mathcal{N}(0,1)\)
prior on all \(K\times D\) entries. The \emph{likelihood} is flat along
the gauge direction, but the Gaussian \emph{prior} is not: its density
decays as the common row shift moves \(\beta\) away from the origin, so
the prior softly pins the gauge --- effectively selecting the
minimum-norm representative of each gauge orbit --- and makes the
posterior proper in that otherwise likelihood-invariant direction,
without biasing the recovered \(\bm{\psi}\). Posterior summaries of
\(\beta\) should therefore target row \emph{contrasts} or the induced
\(\bm{\psi}\), never absolute row levels.

\end{minipage}%
\end{tcolorbox}

\subsection*{4.2 Prior predictive analysis}\label{sec-m0-prior}
\addcontentsline{toc}{subsection}{4.2 Prior predictive analysis}

The goal is to confirm that the priors permit a sensible \emph{range} of
choice behaviors before any data are seen. The summary statistic we
monitor is the per-dataset \textbf{SEU-maximizer rate}: the fraction of
choices that select the expected-utility-maximal available alternative
under the simulated parameters. Drawing \((\alpha, \beta, \delta)\) from
the prior, simulating data via \texttt{m\_0\_sim.stan}, and tabulating
this rate shows its prior-predictive distribution covering the full
range from near-uniform choice (rate \(\approx M^{-1}\sum_m 1/N_m\), the
chance baseline --- equal to \(1/\bar N\) only when menu sizes are
equal) to near-deterministic SEU maximization (rate \(\approx 1\)). The
\(\mathrm{Lognormal}(0,1)\) prior on \(\alpha\) places substantial mass
at both low (near-random) and high (near-deterministic) sensitivities,
so neither extreme of the sensitivity range is ruled out a priori.

\subsection*{4.3 Parameter recovery}\label{sec-m0-recovery}
\addcontentsline{toc}{subsection}{4.3 Parameter recovery}

\textbf{Paradigm.} Draw parameters from the prior via
\texttt{m\_0\_sim.stan}, simulate data, fit \texttt{m\_0.stan}, and
compare posterior summaries to the known true values. We evaluate bias
(\(\approx 0\) ideal), credible-interval coverage (\(\approx\) nominal
90\%), and interval width / RMSE as precision measures.

\begin{figure}

\centering{

\includegraphics[width=1\linewidth,height=\textheight,keepaspectratio]{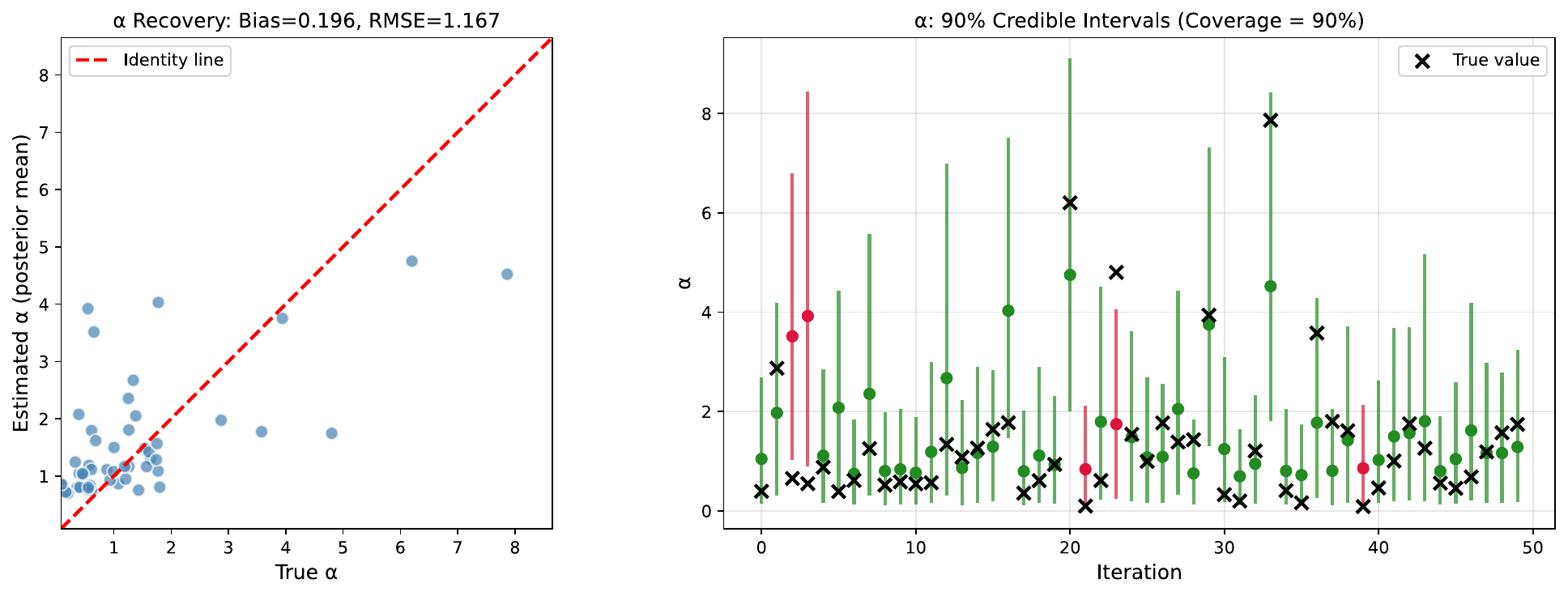}

}

\caption{\label{fig-alpha-recovery}Parameter recovery for \(\alpha\)
under \texttt{m\_0} on the canonical design (50 recovery replicates).
Left: posterior-mean \(\alpha\) against the true value with the identity
line; the aggregate signed bias is low, while the scatter shows the
expected prior shrinkage at large true \(\alpha\), which (together with
the wider posteriors there) also dominates the annotated RMSE --- a
point-estimate error summary, distinct from the posterior spread the
right panel's interval widths display. Right: per-replicate 90\%
credible intervals, one vertical bar per recovery replicate ordered by
iteration index, each with its true value marked by a \(\times\) and
colored green when the interval contains the true value and red when it
does not (5 of 50 miss, matching the nominal 90\% coverage). Because the
bars are closely spaced, a red (missed) interval can sit near a
\(\times\) that belongs to an adjacent replicate; each \(\times\) pairs
with the bar at its own iteration index.}

\end{figure}

\textbf{\(\alpha\) recovery.} \(\alpha\) recovers with low aggregate
bias, well-calibrated 90\% intervals, and useful precision, as
Figure~\ref{fig-alpha-recovery} shows: a true-versus-estimated scatter
with 90\% credible intervals, and per-replicate intervals whose
empirical coverage matches the nominal 90\%. Two qualifications keep
this from overstatement. First, the low bias is an \emph{aggregate}
(across-replicate) summary; the scatter shows the shrinkage a Bayesian
point estimate under a proper prior should show --- posterior means
pulled toward the prior center, visibly so for the largest true
\(\alpha\), so the signed error is \emph{conditionally} negative in the
upper range and positive at the bottom even though it is near zero on
average. Second, precision is not uniform: intervals widen materially
with \(\alpha\), so ``useful precision'' describes the bulk of the prior
range, not its upper tail. This is the computational counterpart of
Proposition 3.1 --- which identifies \(\alpha\) \emph{given} \(\eta\)
--- made empirical: under this design and prior the data supply enough
information to pin \(\alpha\) over the range the prior concentrates on.
The clean recovery is thus a design- and prior-dependent fact, not a
corollary of the proposition alone.

\textbf{\(\beta\) and \(\delta\) recovery.} By contrast
\((\beta, \delta)\) show detectably wider intervals and intervals that
narrow more slowly with sample size, together with correlated
posterior-mean estimation errors --- computed \emph{across recovery
iterations} between a representative weight entry \(\beta_{1,1}\) and a
representative increment \(\delta_1\). We report this last item as an
illustrative, non-gauge-invariant signature rather than a signed
quantity: its sign is not robust (it depends on the true \(\alpha\) used
to simulate the recovery data, since \(\beta_{1,1}\) carries the
row-shift gauge of §\hyperref[sec-notation]{2.6}), and it is the
\emph{presence} of across-replicate error coupling, not its direction,
that matters here. This is the \emph{computational manifestation} of the
weakly-informed \((\beta,\delta)\) of §\hyperref[sec-bd-weak]{3.4}: the
multiplicative \((\beta,\delta)\) coupling realized as a correlated
joint posterior and wide marginal intervals. Crucially, the poor
\((\beta, \delta)\) recovery does \emph{not} contaminate \(\alpha\) ---
exactly as the empirical separation noted in
§\hyperref[sec-m0-link]{3.5}.

\subsection*{4.4 Simulation-based calibration}\label{sec-m0-sbc}
\addcontentsline{toc}{subsection}{4.4 Simulation-based calibration}

Simulation-based calibration (SBC) checks whether the model-and-sampler
together return posteriors that are calibrated on average: if a true
parameter is drawn from the prior, data are simulated from it, and the
model is refit, then a well-calibrated 90\% credible interval should
contain that truth 90\% of the time. The rank test below checks this
across all quantiles at once.

\textbf{Method.} SBC draws a parameter from the prior, simulates data,
fits the model, and computes the rank of the true value within the
posterior draws; under correct calibration these ranks are uniform
(Talts et al. 2018) --- a truth drawn from the prior is equally likely
to fall at any rank within a correctly calibrated posterior, so
systematic departures from uniformity flag a mismatch among model, data,
and sampler. We diagnose uniformity with rank histograms and the
ECDF-difference plot with a simultaneous confidence band (Modrák et al.
2025), at \(N_{\mathrm{sbc}} = 999\), thinning \(4\), single chain, with
the \(L = N_{\mathrm{sbc}}\) convention (Appendix D.3).

\textbf{Result.} At \(N_{\mathrm{sbc}} = 999\) the marginal rank
distributions are consistent with uniformity for \(\alpha\), \(\beta\),
\emph{and} \(\delta\). This run does \textbf{not} independently
reproduce the weakly-informed \((\beta,\delta)\) finding of
§§\hyperref[sec-m0-identifiability]{3}--\hyperref[sec-m0-recovery]{4.3},
because that weakness is one of \emph{joint informativeness and
geometry} --- a correlated, barely-contracted \((\beta,\delta)\)
posterior that can itself be perfectly calibrated --- to which marginal
SBC is blind.

\begin{tcolorbox}[enhanced jigsaw, rightrule=.15mm, breakable, colback=white, arc=.35mm, leftrule=.75mm, colframe=quarto-callout-important-color-frame, toprule=.15mm, bottomrule=.15mm, left=2mm, opacityback=0]
\begin{minipage}[t]{5.5mm}
\textcolor{quarto-callout-important-color}{\faExclamation}
\end{minipage}%
\begin{minipage}[t]{\textwidth - 5.5mm}

\textbf{The marginal-SBC demarcation.} Three points must be kept apart.
\textbf{Marginal versus joint.} Marginal rank uniformity is necessary
but not sufficient for \emph{joint} posterior calibration: a joint
miscalibration that left every margin uniform would not be caught here.
\textbf{Weak informativeness is not miscalibration.} The correlated,
barely-contracted joint \((\beta,\delta)\) posterior is a statement
about weak \emph{joint informativeness and geometry}, not about
miscalibration: such a posterior can be exactly calibrated when the
model and sampler are correct. \textbf{What marginal SBC leaves unseen.}
What marginal SBC fails to reveal in our case is therefore the weak
joint identification and contraction of \((\beta,\delta)\), not a hidden
joint calibration failure --- we report no non-uniform joint SBC
statistic. The positive contribution of SBC here is threefold: (i)
\textbf{sampler/implementation validation} --- uniform marginal ranks at
\(N_{\mathrm{sbc}} = 999\) are strong evidence against a wide range of
implementation and sampler defects (priors, simplex and ordered
transforms, softmax curvature, HMC geometry) and establish
\emph{marginal} calibration; (ii) \textbf{marginal calibration} ---
per-parameter credible intervals are usable as marginal uncertainty
summaries; and (iii) the \textbf{demarcation} itself --- a precise
statement of what marginal calibration can and cannot diagnose, which
motivates a joint or projection-based rank statistic as future work
(§\hyperref[sec-discussion]{8.3}). We do \emph{not} claim that marginal
ranks certify correct sampling of the \emph{joint} posterior, nor
per-parameter calibration ``fingerprints'' or an
\texttt{m\_0}\(\to\)\texttt{m\_1} histogram improvement.

\end{minipage}%
\end{tcolorbox}

\subsection*{4.5 Section summary}\label{sec-m0-impl-summary}
\addcontentsline{toc}{subsection}{4.5 Section summary}

\texttt{m\_0} succeeds at recovering \(\alpha\) --- the parameter of
primary interest --- but the data leave \((\beta,\delta)\) only weakly
informed, leaving the expected-utility \emph{content} underdetermined at
realistic \(n\). This motivates the extended model of
§\hyperref[sec-m1-identifiability]{5}, whose implementation we take up
in §\hyperref[sec-m1-implementation]{6}.

\section*{The Extended Abstract Model with Risky
Choices}\label{sec-m1-identifiability}
\addcontentsline{toc}{section}{The Extended Abstract Model with Risky
Choices}

\subsection*{5.1 Enriching the choice domain --- a two-step
justification}\label{sec-m1-twostep}
\addcontentsline{toc}{subsection}{5.1 Enriching the choice domain --- a
two-step justification}

The extended model (``\texttt{m\_1}'') augments the uncertain-choice
setup with risky alternatives whose probabilities are objectively given.
Step 1 alone does not identify \(\delta\); the two steps below together
do.

\begin{tcolorbox}[enhanced jigsaw, rightrule=.15mm, breakable, colback=white, arc=.35mm, leftrule=.75mm, colframe=quarto-callout-note-color-frame, toprule=.15mm, bottomrule=.15mm, left=2mm, opacityback=0]
\begin{minipage}[t]{5.5mm}
\textcolor{quarto-callout-note-color}{\faInfo}
\end{minipage}%
\begin{minipage}[t]{\textwidth - 5.5mm}

\textbf{Step 1 (\(\beta\)-free risky expected utility).} A risky
alternative \(s\) carries an objective lottery
\(\pi_s \in \Delta^{K-1}\) over the \(K\) consequences, and its expected
utility is \(\eta^{(r)}_s = \pi_s^\top \bm{\upsilon}\), which depends
only on the utilities \(\bm{\upsilon}\) (hence \(\delta\)) and not on
the beliefs \(\beta\) --- it does not pass through the belief map
\(\bm{\psi}_r = \mathrm{softmax}(\beta w_r)\). The risky block therefore
supplies information about \(\bm{\upsilon}\) (hence \(\delta\)) that is
\emph{free of} \(\beta\): the gradient of the risky log-likelihood with
respect to \(\delta\) does not also depend on \(\beta\). By itself this
supplies unconfounded gradients but not invertibility.

\end{minipage}%
\end{tcolorbox}

\begin{tcolorbox}[enhanced jigsaw, rightrule=.15mm, breakable, colback=white, arc=.35mm, leftrule=.75mm, colframe=quarto-callout-note-color-frame, toprule=.15mm, bottomrule=.15mm, left=2mm, opacityback=0]
\begin{minipage}[t]{5.5mm}
\textcolor{quarto-callout-note-color}{\faInfo}
\end{minipage}%
\begin{minipage}[t]{\textwidth - 5.5mm}

\textbf{Step 2 (lottery diversity \(\Rightarrow\) identification).}
Recovering the \(K\) utilities from risky choices requires the lotteries
to point in enough different directions across the consequence space ---
intuitively, at least \(K\) lotteries whose payoff profiles differ
enough to solve for each utility. \emph{Given sufficient lottery
diversity} --- the presented lotteries' differences span the simplex
tangent space, equivalently a spanning affinely independent subset of
\(K\) lotteries with the menus linked into a connected contrast graph
(§\hyperref[sec-delta-id]{5.5}), which requires \(S \geq K\) --- the
\(\beta\)-free risky information \emph{identifies} \(\delta\).

\end{minipage}%
\end{tcolorbox}

\emph{History (brief).} Knight (1921) distinguished risk from
uncertainty; von Neumann and Morgenstern (1947) axiomatized expected
utility under risk; Savage (1954) gave SEU under uncertainty. Anscombe
and Aumann (1963) combined the two by considering acts mapping states to
lotteries, which lets one identify utilities from the risky margin and
beliefs from the state margin. Our \texttt{m\_1} uses a \emph{related
empirical strategy}: the connection is conceptual, not a direct
application of the Anscombe--Aumann representation theorem. The
important part is Steps 1+2, not the representation theorem.

\subsection*{5.2 The extended model}\label{sec-m1-model}
\addcontentsline{toc}{subsection}{5.2 The extended model}

Augment the \(M\) uncertain problems with \(N\) risky problems built
from \(S\) distinct lotteries \(\pi_s\) (objective simplexes over the
\(K\) consequences). The model shares the same \(\alpha\) and the same
\(\delta \to \bm{\upsilon}\) map across both blocks; risky expected
utilities \(\eta^{(r)}_s = \pi_s^\top \bm{\upsilon}\) depend only on
\(\delta\), not on \(\beta\). (In the Stan code the lottery simplices
enter the \texttt{data} block as \texttt{x} and are compacted per
problem into \texttt{x\_risky}; we use \(\pi\) in the body.)

\subsection*{5.3 Preservation of the three sensitivity
properties}\label{sec-m1-properties}
\addcontentsline{toc}{subsection}{5.3 Preservation of the three
sensitivity properties}

The softmax properties of §\hyperref[sec-three-properties]{2.3} are
properties of the choice rule given \emph{any} value function, so they
apply unchanged to both uncertain and risky sub-problems. Therefore the
interpretation of \(\alpha\) as sensitivity-to-SEU-maximization is
preserved in the extended model.

\begin{tcolorbox}[enhanced jigsaw, rightrule=.15mm, breakable, colback=white, arc=.35mm, leftrule=.75mm, colframe=quarto-callout-warning-color-frame, toprule=.15mm, bottomrule=.15mm, left=2mm, opacityback=0]
\begin{minipage}[t]{5.5mm}
\textcolor{quarto-callout-warning-color}{\faExclamationTriangle}
\end{minipage}%
\begin{minipage}[t]{\textwidth - 5.5mm}

\textbf{Single-\(\alpha\) assumption.} \texttt{m\_1} asserts a
\emph{single} \(\alpha\) governing both sub-problems --- i.e., the
agent's responsiveness to expected-utility differences does not depend
on whether probabilities are objective or subjective. The motivation is
that \(\alpha\) is meant to capture one disposition --- the care an
agent brings to acting on expected-utility differences --- which one
would expect to be a property of the agent rather than of whether the
probabilities it faces happen to be objective or subjective. This is a
substantive, testable assumption, and it is essential to the
§\hyperref[sec-m1-recovery]{6.4} interpretation of the matched recovery
contrasts: the quantity-versus-type reading of those contrasts requires
a \emph{common} \(\alpha\) scale across blocks. If \(\alpha\) differed
across blocks, the matched comparison would conflate a
choice-\emph{type} effect with a \emph{block-mix} effect on the
effective \(\alpha\)-precision. The generalization with separate
\(\alpha_{\text{unc}}, \alpha_{\text{risky}}\) (the companion model
\texttt{m\_2}) is named but not pursued here. \textbf{Consequently,
every conclusion that compares \(\alpha\) across the uncertain and risky
blocks is reported as explicitly conditional on this single-\(\alpha\)
assumption}; the assumption is not verified at the prior level (a
genuine test would be posterior-predictive,
§\hyperref[sec-m1-prior]{6.3}), so the conditional phrasing is
mandatory, not optional.

\end{minipage}%
\end{tcolorbox}

\subsection*{\texorpdfstring{5.4 Identifiability of \(\alpha\) in the
extended
model}{5.4 Identifiability of \textbackslash alpha in the extended model}}\label{sec-m1-alpha}
\addcontentsline{toc}{subsection}{5.4 Identifiability of \(\alpha\) in
the extended model}

\begin{tcolorbox}[enhanced jigsaw, rightrule=.15mm, breakable, colback=white, arc=.35mm, leftrule=.75mm, colframe=quarto-callout-note-color-frame, toprule=.15mm, bottomrule=.15mm, left=2mm, opacityback=0]
\begin{minipage}[t]{5.5mm}
\textcolor{quarto-callout-note-color}{\faInfo}
\end{minipage}%
\begin{minipage}[t]{\textwidth - 5.5mm}

\textbf{Proposition 5.1 (Identifiability of \(\alpha\)).} Proposition
3.1 holds unchanged when \(\eta\) is composed of both uncertain and
risky alternatives: if some menu of positive design probability has
non-constant \(\eta\), then \(\alpha > 0\) is identifiable from the
choice-probability function.

\end{minipage}%
\end{tcolorbox}

As in §\hyperref[sec-alpha-from-eta]{3.3}, this is identification of
\(\alpha\) \emph{given} \(\eta\). Jointly --- with \(\delta\) (and, for
uncertain alternatives, \(\beta\)) unknown --- non-constant values pin
down only the products \(\alpha(\eta_r - \eta_s)\); it is the additional
structure of \texttt{m\_1}, namely the lottery-diversity condition of
§\hyperref[sec-delta-id]{5.5} that recovers the utility scale from the
risky block, that turns the product into a statement about \(\alpha\)
itself. In \texttt{m\_0} the same role is played by the endpoint
convention together with the prior and design
(§\hyperref[sec-m0-link]{3.5}).

\subsection*{\texorpdfstring{5.5 Identifiability of \(\delta\) in the
extended
model}{5.5 Identifiability of \textbackslash delta in the extended model}}\label{sec-delta-id}
\addcontentsline{toc}{subsection}{5.5 Identifiability of \(\delta\) in
the extended model}

\begin{tcolorbox}[enhanced jigsaw, rightrule=.15mm, breakable, colback=white, arc=.35mm, leftrule=.75mm, colframe=quarto-callout-note-color-frame, toprule=.15mm, bottomrule=.15mm, left=2mm, opacityback=0]
\begin{minipage}[t]{5.5mm}
\textcolor{quarto-callout-note-color}{\faInfo}
\end{minipage}%
\begin{minipage}[t]{\textwidth - 5.5mm}

\textbf{Proposition 5.2 (Identifiability of \(\delta\)).} If the
presented lotteries' differences \(\{\pi_s - \pi_1\}\) span the
\((K-1)\)-dimensional simplex tangent space --- equivalently, the design
contains an affinely independent subset of \(K\) lotteries (requiring
\(S \geq K\) and a generic configuration) linked by a connected
menu-contrast graph --- then \(\alpha > 0\) and \(\delta\) are jointly
and globally identifiable from risky-choice probabilities alone.

\end{minipage}%
\end{tcolorbox}

\emph{Intuition (full proof in Appendix B.3).} A risky menu's softmax
depends on the linear functional \(\pi_s^\top \bm{\upsilon}\). Pinning
down \(K\) utilities requires \(K-1\) independent lottery directions
(one degree of freedom is fixed by the endpoint convention); with enough
diverse lotteries those directions are present. Formally, when the
lottery differences span the simplex tangent space, the map
\(\bm{\upsilon} \mapsto ((\pi_s - \pi_1)^\top
\bm{\upsilon})_s\) has rank \(K-1\), so the risky log-odds recover
\(\bm{\upsilon}\) --- and jointly \(\alpha\) --- once the endpoint
convention fixes the scale, hence \(\delta\). \(S \geq K\) is only the
necessary cardinality condition; the spanning condition itself concerns
the realized lotteries, so we verify it directly for the \(K = 3\),
\(S = 15\) lottery set of Appendix D.0: the \(14 \times 3\)
lottery-difference matrix (differences of the other 14 lotteries against
the first) has rank \(2 = K - 1\), with nonzero singular values \(1.08\)
and \(0.67\) (\texttt{spikes/report\_design\_diagnostics\_spike.py}) ---
the condition holds for the implemented design as a checked fact, not an
inference from the count. The proof is a linear-algebra inversion.

\subsection*{\texorpdfstring{5.6 What the extended model does \emph{not}
deliver:
\(\beta\)}{5.6 What the extended model does not deliver: \textbackslash beta}}\label{sec-m1-beta}
\addcontentsline{toc}{subsection}{5.6 What the extended model does
\emph{not} deliver: \(\beta\)}

\begin{tcolorbox}[enhanced jigsaw, rightrule=.15mm, breakable, colback=white, arc=.35mm, leftrule=.75mm, colframe=quarto-callout-note-color-frame, toprule=.15mm, bottomrule=.15mm, left=2mm, opacityback=0]
\begin{minipage}[t]{5.5mm}
\textcolor{quarto-callout-note-color}{\faInfo}
\end{minipage}%
\begin{minipage}[t]{\textwidth - 5.5mm}

\textbf{Proposition 5.3 (\(\beta\) constrained only through
\(\eta\)-contrasts; a design-rank condition).} With \(\delta\) pinned by
risky choices and \(\alpha\) pinned by either block, the
uncertain-choice probabilities constrain \(\beta\) only through the
per-menu contrasts of the expected utilities
\(\eta_r = \bm{\psi}_r^\top \bm{\upsilon}\) (Appendix B.4). Whether
those contrasts determine \(\beta\) --- modulo the row-shift gauge of
§\hyperref[sec-notation]{2.6} --- is then a property of the
\emph{design}: local identification requires the Jacobian of the
gauge-fixed map \(\beta \mapsto (\eta\text{-contrasts})\) to attain rank
\((K-1)D\), which is impossible whenever \((K-1)D\) exceeds the number
of independent \(\eta\)-contrasts the design supplies (at most \(R - 1\)
with \(R\) distinct alternatives).

\end{minipage}%
\end{tcolorbox}

The pointwise observation that a single scalar \(\eta_r\) does not
determine the \((K-1)\)-dimensional simplex \(\bm{\psi}_r\) when
\(K \geq 3\) does \emph{not} settle the question for \(\beta\), because
the belief vectors are linked across alternatives by the shared map
\(\bm{\psi}_r = \mathrm{softmax}(\beta w_r)\); what matters is the rank
of the composite map, and that is design-specific.

\emph{Some intuition for the rank condition.} The map
\(\beta \mapsto (\eta\text{-contrasts})\) sends the \((K-1)D\) free
coordinates of the gauge-fixed weight matrix to the list of
expected-utility contrasts the design produces. Its \emph{Jacobian} is
the matrix of first partial derivatives of those output contrasts with
respect to the input coordinates --- a local linear picture of how the
observable contrasts respond to a small change in \(\beta\). We call it
the \emph{contrast-Jacobian} to emphasize that it is assembled from the
\(\eta\)-\emph{contrasts} (which are gauge-invariant), not from the raw
expected utilities \(\eta_r\) (which are defined only up to the per-menu
additive constant of §\hyperref[sec-notation]{2.6}). By the inverse
function theorem, where this Jacobian has full column rank \((K-1)D\)
the map is locally invertible: two nearby \(\beta\) that differ modulo
the gauge cannot produce identical contrasts, so \(\beta\) is
\emph{locally identified}. Where the rank is deficient, some direction
in \(\beta\)-space changes no contrast --- and hence no choice
probability --- at all, so \(\beta\) can move undetected and is not
identified. Because the rank can in principle vary with \(\beta\), we do
not read it off at a single point but evaluate it numerically at a
spread of \(\beta\) drawn from the prior.

The two design regimes in this paper land on opposite sides of the
condition (\texttt{spikes/report\_design\_diagnostics\_spike.py}). At
the foundational design (\(K = 3\), \(D = 5\), \(R = 15\)) the
gauge-fixed \(\beta\) has \((K-1)D = 10\) degrees of freedom against up
to \(R - 1 = 14\) contrasts, and the computed contrast-Jacobian rank is
exactly \(10\) at every one of 20 prior draws checked --- so \(\beta\)
\emph{is} locally identified modulo the gauge there in principle, which
sharpens rather than undercuts the §\hyperref[sec-m0-recovery]{4.3}
finding: the weak \(\beta\) recovery at that design is a finite-sample
information problem, not a structural non-identification. At the
application designs (\(D = 32\), \(R = 30\);
§\hyperref[sec-app-design]{7.2}) the inequality reverses decisively ---
\((K-1)D = 64\) (\(K = 3\)) or \(96\) (\(K = 4\)) degrees of freedom
against a computed contrast rank of \(29 = R - 1\) --- so there
\(\beta\) is genuinely not identified, and no amount of per-design data
volume can change that. We accordingly reserve ``irreducible nuisance''
for designs where the rank condition fails, as it does in the
application; at designs where it holds, \(\beta\) is better described as
identified-but-weakly-informed. Either way, the contribution of the
risky block is to sharpen \(\bm{\upsilon}\), not to recover \(\beta\)
(Appendix B.4).

\subsection*{5.7 The conceptual upshot --- and its finite-sample
caveat}\label{sec-m1-upshot}
\addcontentsline{toc}{subsection}{5.7 The conceptual upshot --- and its
finite-sample caveat}

Enriching the choice domain with risky alternatives yields, \emph{in
principle}, identification of \(\alpha\) (from either block) and of the
utilities \(\delta\) (from the \(\beta\)-free risky block, Step 1, plus
lottery diversity, Step 2). It does \emph{not} deliver the beliefs:
\(\beta\) enters uncertain choices only through the per-menu
\(\eta\)-contrasts, so whether it is identified at all is a design-rank
question (Proposition 5.3) --- the condition fails outright at the
application designs, and even at the foundational design, where it
holds, the finite-sample information is meager
(§\hyperref[sec-m0-recovery]{4.3}). And even the utility gain is an
in-principle one --- it implies estimability only \emph{in the limit}.
§\hyperref[sec-m1-implementation]{6} asks whether the gain is realized
at realistic finite sample sizes, and finds that for \(\delta\) it
largely is not --- the cleanest illustration in the paper of the gap
between identifiability and precise estimability.

\section*{\texorpdfstring{A Basic Computational Implementation of the
Extended Model: \texttt{m\_1} in
Stan}{A Basic Computational Implementation of the Extended Model: m\_1 in Stan}}\label{sec-m1-implementation}
\addcontentsline{toc}{section}{A Basic Computational Implementation of
the Extended Model: \texttt{m\_1} in Stan}

\subsection*{6.1 Model specification}\label{sec-m1-spec}
\addcontentsline{toc}{subsection}{6.1 Model specification}

\texttt{m\_1} keeps the \texttt{m\_0} parameters and priors ---
\(\alpha \sim
\mathrm{Lognormal}(0,1)\), \(\beta_{k,d} \sim \mathcal{N}(0,1)\),
\(\delta \sim
\mathrm{Dirichlet}(1,\dots,1)\) --- and adds \(N\) risky problems built
from \(S\) distinct objective lotteries \(\pi_s \in \Delta^{K-1}\) (Stan
data: \texttt{x}). The log-likelihood is the sum of the \texttt{m\_0}
uncertain-choice term and a risky-choice term that uses
\(\eta^{(r)}_s = \pi_s^\top \bm{\upsilon}\) directly --- in Stan,
\texttt{eta\_risky{[}i{]}\ =\ dot\_product(x\_risky{[}i{]},\ upsilon)}.
The key structural contrast is that risky expected utilities depend only
on \(\bm{\upsilon}\) (hence \(\delta\)), not on \(\beta\)
(§\hyperref[sec-m1-twostep]{5.1}, Step 1).

\subsection*{6.2 Study design and the matched
comparison}\label{sec-m1-matched}
\addcontentsline{toc}{subsection}{6.2 Study design and the matched
comparison}

\textbf{Design.} We use the existing risky-choice configuration
(\(K = 3\), \(S = 15\) lotteries; full constants in Appendix D.0),
\emph{not} a \(\delta\)-optimal design. The matched design fixes one
study design across conditions and slices the same simulated choices
four ways, so that precision differences across conditions reflect the
estimand and the choice type rather than random variation between
separate simulations:

\begin{longtable}[]{@{}lllll@{}}
\toprule\noalign{}
Condition & Model & Uncertain \(M\) & Risky \(N\) & Total choices \\
\midrule\noalign{}
\endhead
\bottomrule\noalign{}
\endlastfoot
A & \texttt{m\_0} & 25 & --- & 25 \\
B & \texttt{m\_0} & 50 & --- & 50 \\
C & \texttt{m\_1} & 25 & 25 & 50 \\
D & \texttt{m\_1} & 50 & 50 & 100 \\
\end{longtable}

The central test is \textbf{B vs C}: same total choice count, same true
parameters per iteration, with only the model and the \emph{type} of
choice differing. The informative control is \textbf{A vs B}: doubling
the \emph{uncertain} block alone, holding the model fixed. Condition D
doubles both blocks and completes the factorial slicing; it plays no
role in the two named contrasts, so §\hyperref[sec-m1-recovery]{6.4}
reports no separate D comparison.

\begin{tcolorbox}[enhanced jigsaw, rightrule=.15mm, breakable, colback=white, arc=.35mm, leftrule=.75mm, colframe=quarto-callout-warning-color-frame, toprule=.15mm, bottomrule=.15mm, left=2mm, opacityback=0]
\begin{minipage}[t]{5.5mm}
\textcolor{quarto-callout-warning-color}{\faExclamationTriangle}
\end{minipage}%
\begin{minipage}[t]{\textwidth - 5.5mm}

\textbf{Statistical-power note.} A recovery contrast at this scale needs
enough iterations to pin effect \emph{magnitudes}, not merely a
\emph{direction} of effect: with too few iterations the matched-count
differences are too noisy to bound. We therefore report every §6.4
magnitude from a study of \textbf{\(n = 100\)} iterations, each
summarized with a bootstrap 90\% CI over iterations.

\end{minipage}%
\end{tcolorbox}

\begin{tcolorbox}[enhanced jigsaw, rightrule=.15mm, breakable, colback=white, arc=.35mm, leftrule=.75mm, colframe=quarto-callout-note-color-frame, toprule=.15mm, bottomrule=.15mm, left=2mm, opacityback=0]
\begin{minipage}[t]{5.5mm}
\textcolor{quarto-callout-note-color}{\faInfo}
\end{minipage}%
\begin{minipage}[t]{\textwidth - 5.5mm}

\textbf{Why a bootstrap rather than a posterior interval.} The quantity
summarized is a property of the \emph{estimator and design under
repeated simulation}, not a model parameter with a posterior --- so a
credible interval is not the right object. The Bayesian fit lives
\emph{inside} each iteration (every per-iteration RMSE or interval width
comes from a full posterior); the bootstrap only quantifies the
across-iteration variability of their paired median, distribution-free
because those per-iteration paired differences are non-normal.

\end{minipage}%
\end{tcolorbox}

\textbf{What each reported magnitude is.} Each of the \(n = 100\)
iterations runs the recovery loop of §\hyperref[sec-m0-recovery]{4.3} on
the matched design: draw one set of true parameters from the prior,
simulate the four sliced choice sets (conditions A--D) under those
parameters, fit the corresponding model in each condition, and record a
per-iteration recovery metric --- a posterior-mean RMSE against the
known truth, or a posterior credible-interval width --- for each
condition. Because every condition within an iteration shares the
\emph{same} true parameters, the conditions are \emph{paired}: a
contrast such as B vs C or A vs B is formed within each iteration and
then summarized across iterations by its paired-iteration median. The
interval we report is a \textbf{bootstrap 90\% CI over the \(n = 100\)
iterations} (10,000 resamples): we resample iterations with replacement,
recompute the paired median on each resample, and take its central 90\%
range, so the CI measures how stable the aggregate contrast is across
simulated datasets.

\textbf{Sign convention.} Each §6.4 magnitude is a \emph{signed
percentage change} in a precision measure (a posterior-mean RMSE or a
credible-interval width), signed so that a \emph{positive} value denotes
an \emph{improvement} --- a reduction in RMSE or in CI width --- and a
negative value denotes degraded precision. Its bootstrap 90\% CI is
expressed in these same signed-percentage units --- it is the interval
for that one signed quantity, not a separate scale --- so a CI lying
wholly above zero indicates a reliable improvement, whereas a CI
straddling zero leaves the direction undetermined.

\textbf{What we do not claim.} We do not engineer a design that makes
the \(\delta\) CI ``materially'' smaller than \texttt{m\_0}'s; at
\(n = 100\) the matched-count \(\delta\) CI-width gain is
\(\approx 0.8\%\) and \(\delta\) RMSE is statistically unchanged. A
\(\delta\)-information-optimal lottery design --- pitting, e.g., the
certain intermediate consequence against a 50/50 mix of the extremes to
triangulate each \(\delta_k\) --- is named as future work
(§\hyperref[sec-disc-limitations]{8.5}).

\subsection*{6.3 Prior predictive analysis}\label{sec-m1-prior}
\addcontentsline{toc}{subsection}{6.3 Prior predictive analysis}

On the combined (uncertain + risky) prior predictive, the SEU-maximizer
rate again covers the full sensitivity range. Because \texttt{m\_1}
assumes a \emph{shared} \(\alpha\) across blocks
(§\hyperref[sec-m1-properties]{5.3}), one might want to confirm that
risky and uncertain choices are mutually compatible with a single
\(\alpha\); but that is a claim about the \emph{posterior}, not the
prior. A genuine check would require a \emph{posterior}-predictive block
comparison, which we flag as the appropriate diagnostic for the
single-\(\alpha\) caveat (§\hyperref[sec-m1-alpha-recovery]{6.4.1})
rather than claim to have performed here.

\subsection*{6.4 Parameter recovery --- what the matched comparison
actually shows}\label{sec-m1-recovery}
\addcontentsline{toc}{subsection}{6.4 Parameter recovery --- what the
matched comparison actually shows}

\subsubsection*{\texorpdfstring{6.4.1 \(\alpha\): no per-choice
advantage from the risky
block}{6.4.1 \textbackslash alpha: no per-choice advantage from the risky block}}\label{sec-m1-alpha-recovery}
\addcontentsline{toc}{subsubsection}{6.4.1 \(\alpha\): no per-choice
advantage from the risky block}

At matched total choice count (B vs C), swapping in the \(\beta\)-free
risky block yields \textbf{no} \(\alpha\)-RMSE improvement: \(-8.1\%\),
bootstrap 90\% CI \([-22.0, +6.8]\) --- the point estimate is slightly
\emph{worse} (it favors \texttt{m\_0}) and the interval straddles zero.
The CI-width measure agrees (paired median \(-1.4\%\), improved in only
44\% of iterations; Wilcoxon signed-rank \(p \approx 0.39\)). Here and
below the Wilcoxon signed-rank test is a paired, distribution-free test
of whether the per-iteration paired differences between conditions are
centered at zero; we report it alongside the bootstrap CI because it
makes no normality assumption about those differences and a small \(p\)
would indicate a systematic shift in one direction across iterations.

The informative control is A\(\to\)B: doubling the \emph{uncertain}
block alone cuts \(\alpha\) RMSE by \textbf{27.1\%} (90\% CI
\([+15.0, +38.5]\); CI-width paired median \(+14.0\%\), improved in 78\%
of iterations). Read together, the two contrasts show that, \emph{in
this matched design}, \(\alpha\) precision at realistic \(n\) tracks
data \emph{quantity} rather than the \emph{type} of choice: the
\(\beta\)-free risky block confers no \emph{detected} per-choice
advantage --- the B\(\to\)C interval straddles zero, so a modest
risky-block advantage remains compatible with the data --- whereas more
uncertain choices sharpen \(\alpha\) directly. This is a statement about
the realized, unoptimized design, not a general claim that choice type
cannot matter. We draw out the methodological reading in
§\hyperref[sec-m1-lesson]{6.4.4}.

\begin{tcolorbox}[enhanced jigsaw, rightrule=.15mm, breakable, colback=white, arc=.35mm, leftrule=.75mm, colframe=quarto-callout-note-color-frame, toprule=.15mm, bottomrule=.15mm, left=2mm, opacityback=0]
\begin{minipage}[t]{5.5mm}
\textcolor{quarto-callout-note-color}{\faInfo}
\end{minipage}%
\begin{minipage}[t]{\textwidth - 5.5mm}

\textbf{Conditional on the single-\(\alpha\) assumption.} Comparing
\(\alpha\) precision across the uncertain and risky blocks presupposes a
common \(\alpha\) scale across blocks
(§\hyperref[sec-m1-properties]{5.3}). This shared-\(\alpha\) assumption
cannot be verified at the prior level (§\hyperref[sec-m1-prior]{6.3}); a
genuine data-level test would be posterior-predictive --- so the
quantity-versus-type reading is reported as explicitly conditional on
this assumption, not asserted outright. The block-specific
generalization (\texttt{m\_2}) is out of scope.

\end{minipage}%
\end{tcolorbox}

\subsubsection*{\texorpdfstring{6.4.2 \(\delta\): the payoff is
essentially
nil}{6.4.2 \textbackslash delta: the payoff is essentially nil}}\label{sec-m1-delta-recovery}
\addcontentsline{toc}{subsubsection}{6.4.2 \(\delta\): the payoff is
essentially nil}

At matched choice count (B vs C), \texttt{m\_1} narrows the \(\delta\)
CI width by only \(\approx 0.8\%\) (paired-iteration median; bootstrap
90\% CI \([0.6, 1.2]\); narrower in 72\% of iterations; Wilcoxon
signed-rank \(p \approx 1.2\times10^{-6}\)) and leaves \(\delta\) RMSE
statistically unchanged (\(0.3\%\), 90\% CI \([-4.0, +4.5]\)). The
CI-width effect is \emph{real but practically negligible}. The reading:
\(\delta\) is identifiable in \texttt{m\_1}
(§\hyperref[sec-delta-id]{5.5}, hence estimable \emph{in the limit}) but
\textbf{not precisely estimable at realistic sample sizes} in this
design.

\subsubsection*{\texorpdfstring{6.4.3
\(\beta\)}{6.4.3 \textbackslash beta}}\label{sec-m1-beta-recovery}
\addcontentsline{toc}{subsubsection}{6.4.3 \(\beta\)}

\(\beta\) recovery in \texttt{m\_1} is improved relative to
\texttt{m\_0} but remains subject to the additive row-shift gauge
(§\hyperref[sec-notation]{2.6}; Appendix B.4); summaries target
contrasts and induced \(\bm{\psi}\), not absolute row levels.

\subsubsection*{6.4.4 The methodological lesson --- two
phenomena}\label{sec-m1-lesson}
\addcontentsline{toc}{subsubsection}{6.4.4 The methodological lesson ---
two phenomena}

The matched comparison yields two separable substantive lessons, which
we keep apart.

\textbf{(i) For \(\delta\) --- the canonical lesson, sharpened.} The
principle is that \emph{identifiability (hence estimability in the
limit) does not entail precise estimability at realistic sample sizes.}
\(\delta\) is identifiable in \texttt{m\_1}
(§\hyperref[sec-delta-id]{5.5}), but the information about \(\delta\)
contributed by 25 risky choices at moderate \(\alpha\) is small: the
matched B\(\to\)C comparison shrinks the \(\delta\) CI width by
\(\approx 0.8\%\) and moves \(\delta\) RMSE not at all. With unboundedly
many risky choices \(\delta\) would be recovered exactly; with the
design's 25 it is not. In-principle identification buys
\emph{essentially nothing} at the design sample size --- the
identifiability-versus-estimability gap in its starkest form.

\textbf{(ii) For \(\alpha\) --- finite-\(n\) precision tracks quantity,
not type.} One might expect the \texttt{m\_1} \(\alpha\) gain to be a
per-choice advantage of the \(\beta\)-free risky block; at \(n = 100\)
there is no such advantage (B\(\to\)C null). What sharpens \(\alpha\) is
simply \emph{more data of the same kind} (A\(\to\)B: a \(27.1\%\) RMSE
\emph{reduction}). The correct lesson is \emph{not} ``a special
structure sharpens an already-identified parameter''; it is that
\textbf{in-principle identifiability tells you nothing about where
finite-\(n\) precision comes from} --- here, \emph{in this design}, it
comes from quantity, not type (the matched contrast detects no
risky-block advantage but does not rule out a smaller one). Phenomena
(i) and (ii) are distinct.

\begin{longtable}[]{@{}
  >{\raggedright\arraybackslash}p{(\linewidth - 4\tabcolsep) * \real{0.3333}}
  >{\raggedright\arraybackslash}p{(\linewidth - 4\tabcolsep) * \real{0.3333}}
  >{\raggedright\arraybackslash}p{(\linewidth - 4\tabcolsep) * \real{0.3333}}@{}}
\caption{The matched-design contrast. All percentages are
paired-iteration medians / bootstrap-over-iteration 90\% CIs from the
\(n = 100\) study (10,000 resamples).}\label{tbl-matched}\tabularnewline
\toprule\noalign{}
\begin{minipage}[b]{\linewidth}\raggedright
\end{minipage} & \begin{minipage}[b]{\linewidth}\raggedright
\(\delta\) in \texttt{m\_1}
\end{minipage} & \begin{minipage}[b]{\linewidth}\raggedright
\(\alpha\) in \texttt{m\_1}
\end{minipage} \\
\midrule\noalign{}
\endfirsthead
\toprule\noalign{}
\begin{minipage}[b]{\linewidth}\raggedright
\end{minipage} & \begin{minipage}[b]{\linewidth}\raggedright
\(\delta\) in \texttt{m\_1}
\end{minipage} & \begin{minipage}[b]{\linewidth}\raggedright
\(\alpha\) in \texttt{m\_1}
\end{minipage} \\
\midrule\noalign{}
\endhead
\bottomrule\noalign{}
\endlastfoot
Identifiable in \texttt{m\_0}? & not established --- weakly informed by
uncertain choices (§\hyperref[sec-bd-weak]{3.4}); the identifiability
result arrives with the risky block (§\hyperref[sec-delta-id]{5.5}) &
yes \\
What §5 promises & a \(\beta\)-free identification route that sharpens
\(\bm{\upsilon}\) & already identified \\
Matched-count payoff (B\(\to\)C) & CI-width \(+0.8\%\) \([0.6, 1.2]\)
(real but negligible); RMSE unchanged & no detected gain: RMSE
\(-8.1\%\) \([-22.0, +6.8]\) (straddles zero); CI-width unchanged \\
Where finite-\(n\) precision comes from & not realized at design \(n\) &
data \emph{quantity}, not choice \emph{type} (A\(\to\)B: \(+27.1\%\)
RMSE reduction \([+15.0, +38.5]\)) \\
Phenomenon & identifiability \(\neq\) precise estimability at realistic
\(n\) & in-principle identification \(\neq\) finite-\(n\) gain;
precision tracks quantity, not type \\
\end{longtable}

\begin{figure}

\centering{

\includegraphics[width=1\linewidth,height=\textheight,keepaspectratio]{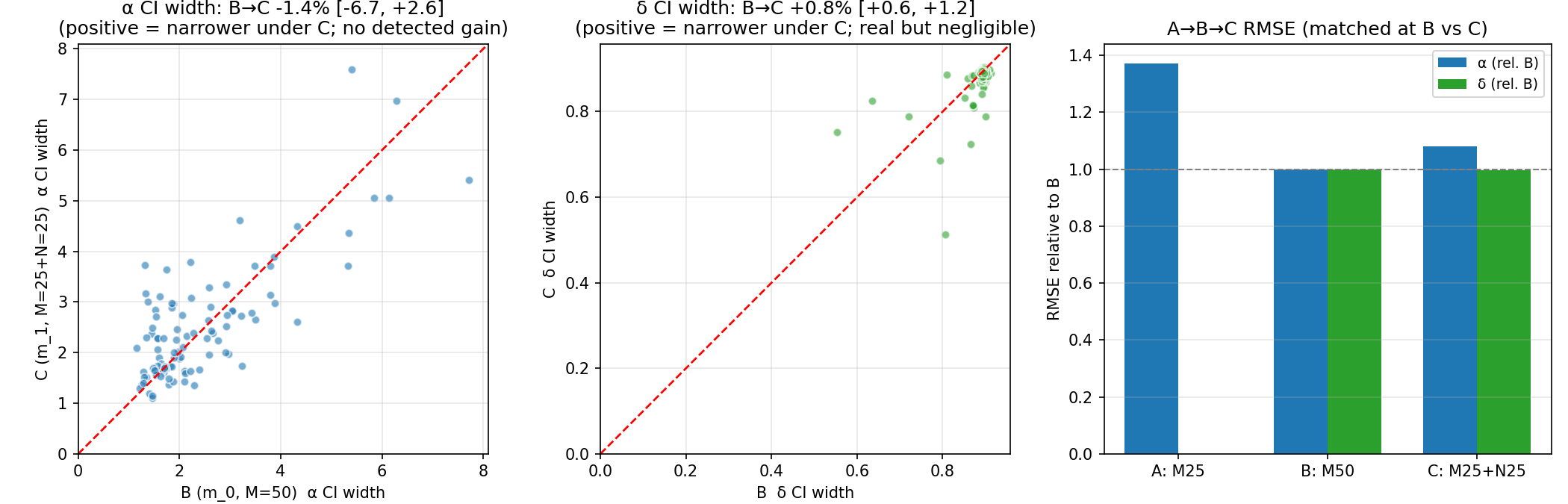}

}

\caption{\label{fig-matched}The matched-design comparison (conditions A:
\(M=25\); B: \(M=50\), \texttt{m\_0}; C: \(M=25\), \(N=25\),
\texttt{m\_1}), \(n = 100\) replicates. Left and centre: per-replicate
90\% credible-interval widths for \(\alpha\) and \(\delta\) at the
matched conditions B versus C, against the identity line. Right: RMSE
for \(\alpha\) and \(\delta\) relative to condition B. Doubling the
uncertain block (A\(\to\)B) sharpens \(\alpha\); swapping in the
\(\beta\)-free risky block at matched count (B\(\to\)C) does not, and
the \(\delta\) payoff is negligible.}

\end{figure}

Figure~\ref{fig-matched} summarizes the three matched conditions
visually.

\subsection*{\texorpdfstring{6.5 SBC for
\texttt{m\_1}}{6.5 SBC for m\_1}}\label{sec-m1-sbc}
\addcontentsline{toc}{subsection}{6.5 SBC for \texttt{m\_1}}

\begin{figure}

\centering{

\includegraphics[width=0.95\linewidth,height=\textheight,keepaspectratio]{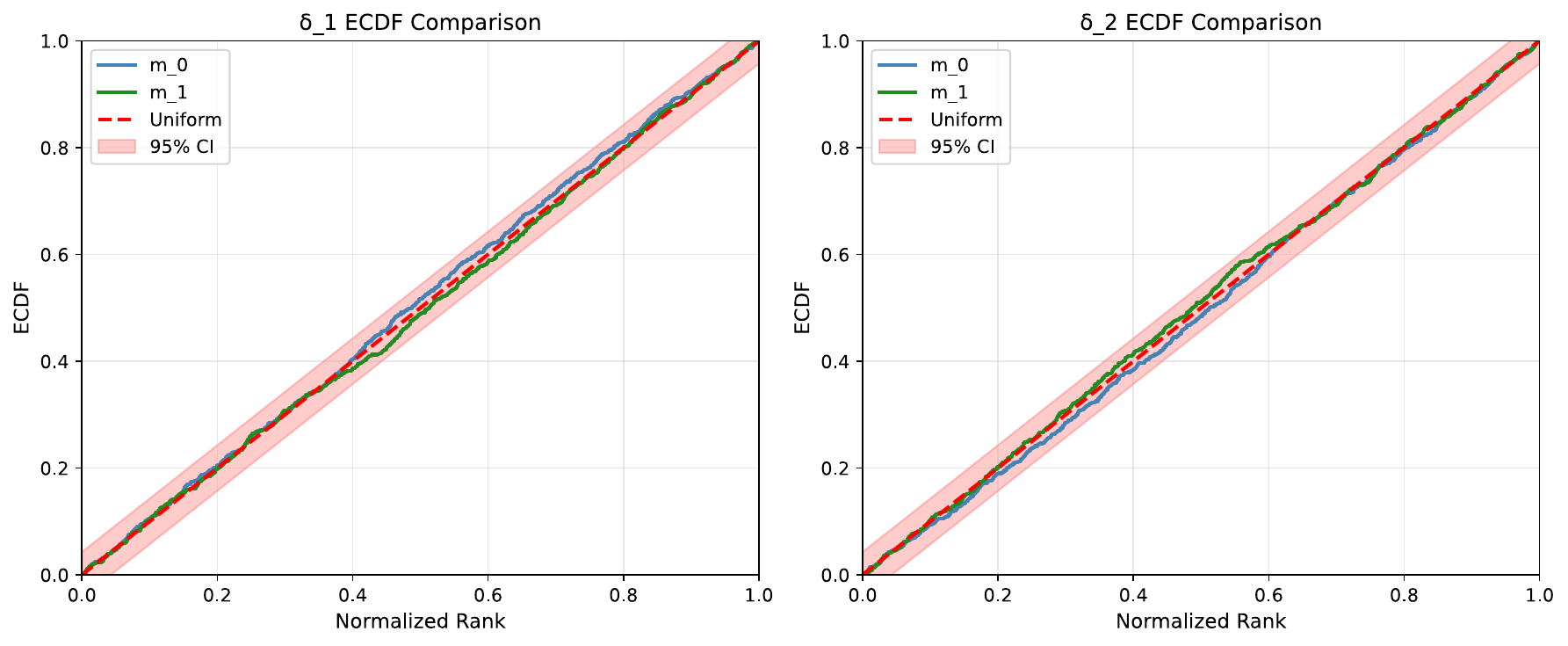}

}

\caption{\label{fig-sbc-ecdf}Empirical CDF of normalized SBC ranks for
\(\delta_1\) and \(\delta_2\) under \texttt{m\_0} and \texttt{m\_1},
with the 95\% simultaneous (Kolmogorov--Smirnov) band. Both models stay
inside the band, so marginal \(\delta\) calibration holds in each ---
the marginal-SBC demarcation: marginal rank uniformity coexists with a
jointly weakly-informed \((\beta,\delta)\) posterior that marginal SBC
cannot resolve.}

\end{figure}

Rank histograms and ECDF-difference plots for \(\alpha\), \(\beta\),
\(\delta\) are consistent with uniformity. We do \textbf{not} claim an
\texttt{m\_0}\(\to\)\texttt{m\_1} \(\delta\)-calibration improvement
(e.g., a flatter histogram): marginal SBC is uniform for \(\delta\) in
\emph{both} models, because the relevant \((\beta,\delta)\) weakness is
in the \emph{joint} posterior (the \textbf{marginal-SBC demarcation}
again, §\hyperref[sec-m0-sbc]{4.4}). Figure~\ref{fig-sbc-ecdf} makes
this concrete for \(\delta\): the empirical-CDF diagnostic keeps both
models inside the same simultaneous band. The appropriate diagnostic is
a joint or projection-based rank statistic
(§\hyperref[sec-discussion]{8.3}). The usual single-chain SBC caveats
apply: thinning, single-chain Monte-Carlo error, and sample-size driven
rank-resolution all bound what the histograms can detect.

\FloatBarrier

\section*{Illustrative Application: SEU Sensitivity in LLM
Decisions}\label{sec-application}
\addcontentsline{toc}{section}{Illustrative Application: SEU Sensitivity
in LLM Decisions}

\subsection*{7.1 Why include an application here}\label{sec-app-why}
\addcontentsline{toc}{subsection}{7.1 Why include an application here}

The methodology of
§§\hyperref[sec-abstract-model]{2}--\hyperref[sec-m1-implementation]{6}
is only useful insofar as it does real evaluative work on real choice
data. This section runs the full workflow end-to-end on a \(2\times2\)
factorial design --- two LLMs crossed with two task families --- to
demonstrate three things: (i) the workflow scales from the 25--100
simulated choices per condition of the
§§\hyperref[sec-m0-implementation]{4}/\hyperref[sec-m1-implementation]{6}
studies to \(N \approx 300\) real-LLM choices per condition without
redesign; (ii) prior recalibration is a routine, principled per-study
step rather than a redesign of the model; and (iii) the framework
registers a structured comparative effect when the posterior supports
one and withholds support when it does not --- where ``withholds
support'' means \emph{inconclusive at the achievable resolution}, not an
established zero (§\hyperref[sec-app-claude-insurance]{7.5.2}).

\begin{tcolorbox}[enhanced jigsaw, rightrule=.15mm, breakable, colback=white, arc=.35mm, leftrule=.75mm, colframe=quarto-callout-important-color-frame, toprule=.15mm, bottomrule=.15mm, left=2mm, opacityback=0]
\begin{minipage}[t]{5.5mm}
\textcolor{quarto-callout-important-color}{\faExclamation}
\end{minipage}%
\begin{minipage}[t]{\textwidth - 5.5mm}

\textbf{Reading guide (construct-validity layering).} Three distinct
claims should not be conflated, and this layering frames everything
below:

\begin{enumerate}
\def\labelenumi{\arabic{enumi}.}
\tightlist
\item
  \textbf{Model adequacy} --- the model fits the data adequately
  (validated in §\hyperref[sec-app-validation]{7.4}).
\item
  \textbf{Comparative claims under shared design} --- \(\alpha\) differs
  systematically across conditions that share item pool, prior, choice
  model, and feature-construction pipeline (the
  §\hyperref[sec-app-results]{7.5} reading). ``Shared'' here means the
  \emph{construction} is shared; the realized feature matrices are
  provider-generated and so differ across LLMs
  (§\hyperref[sec-app-design]{7.2}).
\item
  \textbf{Absolute claims about EU rationality} --- an LLM's \(\alpha\)
  value certifies a context-free rationality level.
\end{enumerate}

\textbf{Only layers 1 and 2 are supported by this paper.}
§\hyperref[sec-app-limits]{7.6.3} deepens this; it is stated here so the
reader has these limitations in mind when they reach the
§\hyperref[sec-app-results]{7.5} headline.

\end{minipage}%
\end{tcolorbox}

\begin{tcolorbox}[enhanced jigsaw, rightrule=.15mm, breakable, colback=white, arc=.35mm, leftrule=.75mm, colframe=quarto-callout-important-color-frame, toprule=.15mm, bottomrule=.15mm, left=2mm, opacityback=0]
\begin{minipage}[t]{5.5mm}
\textcolor{quarto-callout-important-color}{\faExclamation}
\end{minipage}%
\begin{minipage}[t]{\textwidth - 5.5mm}

\textbf{A recovered \(\alpha\) is model-conditional.} The \(\alpha\)
values reported below are parameters of the
§\hyperref[sec-m0-identifiability]{3} softmax-over-SEU likelihood, not
context-free rationality scores. The within-design comparisons of
§\hyperref[sec-app-results]{7.5} --- same model, same prior, same
feature-construction pipeline --- are supported; an \(\alpha\) read in
isolation is \emph{not} a model-free measure of an agent's rationality.
We return to this at §\hyperref[sec-disc-alpha]{8.4}.

\end{minipage}%
\end{tcolorbox}

\begin{tcolorbox}[enhanced jigsaw, rightrule=.15mm, breakable, colback=white, arc=.35mm, leftrule=.75mm, colframe=quarto-callout-note-color-frame, toprule=.15mm, bottomrule=.15mm, left=2mm, opacityback=0]
\begin{minipage}[t]{5.5mm}
\textcolor{quarto-callout-note-color}{\faInfo}
\end{minipage}%
\begin{minipage}[t]{\textwidth - 5.5mm}

\textbf{Reporting template (applied uniformly to all four cells).} Every
LLM \(\times\) task cell is reported with the \emph{same} four
quantities, defined and computed as spelled out in Appendix
\hyperref[sec-d6]{D.6}:

\begin{enumerate}
\def\labelenumi{\arabic{enumi}.}
\tightlist
\item
  \textbf{Per-condition \(\alpha\)} --- the posterior median at each
  temperature, with a 90\% credible interval.
\item
  \textbf{Global temperature slope \(\Delta\alpha/\Delta T\)} ---
  summarized by its posterior median, a 90\% credible interval
  \emph{around that median}, and \(P(\text{slope} < 0)\) (the posterior
  probability the slope is negative).
\item
  \textbf{\(P(\text{strict monotone decrease})\)} --- the posterior
  probability that \(\alpha\) decreases at every step across all five
  temperature levels.
\item
  \textbf{Cross-LLM comparison probability} --- the posterior
  probability that one LLM's temperature slope is more negative than the
  other's. Unlike quantities 1--3, this is a \emph{pairwise} summary
  computed once per LLM pair within a task family, not a per-cell
  quantity; it appears in the cross-LLM comparison
  (§\hyperref[sec-app-cross-llm]{7.5.2a}) rather than in each cell's
  report.
\end{enumerate}

Cross-\emph{task} differences are read descriptively (the \(2\times2\)
pattern of §\hyperref[sec-app-2x2]{7.5.5}) rather than as a single
formal contrast probability: because the two tasks differ in \(K\) and
in their separately calibrated \(\alpha\) priors, they do not share one
design, so a cross-task difference is informative about the broad
pattern but should be read with care rather than as a calibrated effect
size.

\end{minipage}%
\end{tcolorbox}

\subsection*{\texorpdfstring{7.2 The \(2\times2\)
design}{7.2 The 2\textbackslash times2 design}}\label{sec-app-design}
\addcontentsline{toc}{subsection}{7.2 The \(2\times2\) design}

\textbf{LLMs.} GPT-4o (OpenAI) and Claude 3.5 Sonnet (Anthropic).

\textbf{Task families.} (i) \emph{Insurance claims triage} over a
30-claim pool, with \(K = 3\) consequences --- the three
investigator-agreement outcomes (neither, one, or both investigators
agree the claim warrants investigation); and (ii) \emph{Ellsberg-style
urn gambles} over a 30-gamble pool, with \(K = 4\) consequences (\$0,
\$1, \$2, \$3 payouts), organized into three ambiguity tiers (Tier 1
unambiguous, Tier 2 moderately ambiguous, Tier 3 high-ambiguity in the
spirit of Ellsberg (1961)). In both tasks the \emph{alternatives}
presented in a problem are the pool items themselves (claims or
gambles). Each task has a single fixed design, built once and shared
across all temperature conditions; within that design the menu size of
each problem was drawn at random from \(N_m \in \{2, 3, 4\}\).

\textbf{Per cell.} Five sampling-temperature conditions;
\(M \approx 100\) base problems \(\times\) 3 position-counterbalanced
presentations \(= 300\) choices per condition (constants in Appendix
D.5). Temperature grids differ by provider: GPT-4o
\(\{0.0, 0.3, 0.7, 1.0, 1.5\}\); Claude \(\{0.0, 0.2, 0.5, 0.8, 1.0\}\)
(Anthropic-API range constraint). The unequal grids complicate direct
slope-\emph{magnitude} comparison across LLMs, though not the
qualitative reading; we flag this again at
§\hyperref[sec-app-results]{7.5.2a}.

\textbf{Feature pipeline.} Both task families use the \emph{same}
two-stage construction. \emph{Insurance:} a per-claim LLM assessment,
then choice over those assessments; assessment text embedded via
\texttt{text-embedding-3-small} (embedding turns each assessment's text
into a numeric vector), then reduced by pooled PCA (a linear projection
onto the leading directions of variation) to \(D = 32\) features --- a
compromise between explained variance and keeping the fit tractable.
\emph{Ellsberg:} the identical pipeline applied to gambles --- a
per-gamble LLM assessment of each urn (a short free-text analysis of the
payoff probabilities and ambiguity), then choice over those assessments,
embedded and pooled-PCA-projected to \(D = 32\). The objective urn
composition is \emph{not} entered into the model directly: in both tasks
the feature vector \(w_r\) is an embedding of an LLM-generated
assessment, and the choice is made over those assessments. Both
applications therefore measure the \emph{whole assessment-and-choice
pipeline}, not the choice stage in isolation; the two cells differ in
the decision domain (claims vs.~gambles) and in \(K\) (3 vs.~4), not in
which stages the temperature lever touches (we return to this at
§\hyperref[sec-app-limits]{7.6.5}). Position counterbalancing addresses
a position-bias problem identified in preliminary elicitation runs;
unparseable responses are recorded as missing rather than coerced to a
default, and missing rates are negligible across all 20 conditions.

One dimensional fact about this pipeline deserves explicit
acknowledgment: with \(D = 32\) features and only \(R = 30\) distinct
items, every realized feature matrix has full row rank (rank 30,
verified for all 20 conditions;
\texttt{spikes/report\_design\_diagnostics\_spike.py}). The linear map
\(\beta\) can therefore assign an arbitrary pattern of belief weights
across the 30 items --- the feature construction imposes no cross-item
restriction of its own, and the discipline on \(\beta\) comes from its
prior rather than from feature-space parsimony. None of the comparisons
below rest on \(\beta\) being identified (§\hyperref[sec-bd-weak]{3.4});
we state the rank fact so the \(D = 32\) choice is not misread as a
binding structural constraint.

\subsection*{\texorpdfstring{7.3 The model fit (\texttt{m\_01} /
\texttt{m\_02})}{7.3 The model fit (m\_01 / m\_02)}}\label{sec-app-model}
\addcontentsline{toc}{subsection}{7.3 The model fit (\texttt{m\_01} /
\texttt{m\_02})}

The fitted model is structurally identical to \texttt{m\_0} of
§\hyperref[sec-m0-implementation]{4} but with a
\(\mathrm{Lognormal}(\mu, \sigma)\) prior on \(\alpha\) calibrated to
each application's design and consequence space: \texttt{m\_01} for the
insurance task (\(K = 3\)) and \texttt{m\_02} for the Ellsberg task
(\(K = 4\)). The two programs differ in the consequence count \(K\) ---
and hence in the dimensions of \(\upsilon\), \(\delta\), and the
per-alternative belief objects \(\psi\) --- as well as in the \(\alpha\)
prior; the likelihood \emph{form} is identical.

\textbf{Prior calibration anchor.} Why does the \(K = 4\) Ellsberg task
need a different \(\alpha\) prior than the \(K = 3\) insurance task? Not
because random choice differs --- the menu size is the same --- but
because with four consequences a larger \(\alpha\) is empirically
required to reach the same prior-implied rate of good choice; the
calibration below determines it directly rather than from a closed-form
softmax argument. The calibration target is the prior predictive
SEU-maximizer selection rate --- the fraction of prior-simulated agents
that pick the SEU-maximal available alternative, obtained by drawing
choice problems from the design and parameters from the prior and
tabulating the outcome. At the application's \(R\) and \(K\) the
foundational \(\mathrm{Lognormal}(0,1)\) prior of
§\hyperref[sec-m0-spec]{4.1} places most of its mass on near-random
sensitivities, so the prior predictive concentrates well below the
SEU-max rates we take to be plausible for this application --- a range
we posit rather than derive, and whose influence on the findings is
examined in §\hyperref[sec-app-prior-sensitivity]{7.6.6}. (Numerical
overflow in \(\exp(\alpha\eta)\) is handled by the softmax
implementation and is \emph{not} the reason for recalibration --- indeed
the calibrated priors place \emph{more} mass at large \(\alpha\).) The
\(76\)--\(78\%\) target places the prior mode in the broad interior
between chance and determinism: informative enough to move mass off the
near-random end of the range that dominates under
\(\mathrm{Lognormal}(0,1)\), yet loose enough --- a wide 90\% prior
interval on \(\alpha\) --- to let the likelihood move the posterior. A
grid search over twelve Lognormal hyperparameter pairs scans
\((\mu, \sigma)\) to hit this interior target, selecting
\(\mathrm{Lognormal}(3.0, 0.75)\) for the insurance task (\(K = 3\)):
median \(\alpha \approx 20\), 90\% interval \(\approx [5.5, 67]\),
implied SEU-max rate \(\approx 78\%\). For the Ellsberg task (\(K = 4\))
the prior is \emph{recalibrated} by the same procedure, selecting
\(\mathrm{Lognormal}(3.5, 0.75)\) (median \(\alpha \approx 33\), 90\%
interval \(\approx [10, 124]\), implied SEU-max rate \(\approx 76\%\)).
(The 90\% intervals quoted here are the empirical 5\%--95\% quantiles of
the 200 \(\alpha\) draws used in the grid search, which is why they
differ slightly from the analytic lognormal quantiles.) The
recalibration is needed not because the random-choice baseline changes
--- that baseline is set by the menu size, not the consequence count
\(K\), and the menu size \(N_m \in \{2, 3, 4\}\) is identical across the
two tasks, so a uniform-random chooser selects the SEU-maximizer with
the same \(1/N_m\) probability either way --- but because the larger
consequence space (\(K = 4\)) changes the distribution of
expected-utility contrasts across alternatives, so the same grid search
selects a higher \(\alpha\) to reach the same prior-implied SEU-max rate
--- a direction we read off the calibration empirically rather than from
a closed-form softmax argument.

\begin{tcolorbox}[enhanced jigsaw, rightrule=.15mm, breakable, colback=white, arc=.35mm, leftrule=.75mm, colframe=quarto-callout-note-color-frame, toprule=.15mm, bottomrule=.15mm, left=2mm, opacityback=0]
\begin{minipage}[t]{5.5mm}
\textcolor{quarto-callout-note-color}{\faInfo}
\end{minipage}%
\begin{minipage}[t]{\textwidth - 5.5mm}

\textbf{Only the prior is recalibrated, not the likelihood.} The
\(\alpha\)-identifiability proposition of
§\hyperref[sec-alpha-from-eta]{3.3}/§\hyperref[sec-m1-alpha]{5.4} and
the weakly-informed \((\beta,\delta)\) finding of
§\hyperref[sec-bd-weak]{3.4} carry over unchanged from \texttt{m\_0} to
\texttt{m\_01}. The application inherits the identifiability story of
§§\hyperref[sec-m0-identifiability]{3} and
\hyperref[sec-m1-identifiability]{5} without re-derivation. This
foundational-vs-application prior distinction is the one flagged in
§\hyperref[sec-m0-spec]{4.1}: same likelihood, different \(\alpha\)
prior.

\end{minipage}%
\end{tcolorbox}

\subsection*{7.4 Validation at the application's
scale}\label{sec-app-validation}
\addcontentsline{toc}{subsection}{7.4 Validation at the application's
scale}

The workflow of
§§\hyperref[sec-m0-prior]{4.2}--\hyperref[sec-m0-sbc]{4.4} is re-run at
the application's scale (\(M = 300\), \(K = 3\) or \(4\), \(D = 32\),
\(R = 30\), 20 recovery iterations per cell).

\textbf{Parameter recovery for \(\alpha\).} Anchored on relative
metrics, because true \(\alpha\) lies on a wide multiplicative scale
under the calibrated prior: relative bias within \(\pm 10\%\) of the
mean true value, relative RMSE well below 25\%, and 90\% CI coverage at
nominal. With 20 iterations per cell these recovery checks are
reassurance against \emph{gross} miscalibration rather than precise
coverage estimates --- the binomial uncertainty on a coverage proportion
estimated from 20 draws is wide (a standard error of roughly 7
percentage points at nominal 90\%). As throughout, \((\beta, \delta)\)
recovery remains weak --- the §\hyperref[sec-m0-identifiability]{3}
picture is unchanged --- but the cross-condition comparison is a claim
about \(\alpha\), and \(\alpha\) is fit for purpose.

\textbf{SBC for \(\alpha\).} SBC (simulation-based calibration, defined
in §\hyperref[sec-m0-sbc]{4.4}) validates the \emph{(prior, likelihood,
sampler)} triple and is conditionally independent of which agent
generated the held-out empirical data.

\begin{tcolorbox}[enhanced jigsaw, rightrule=.15mm, breakable, colback=white, arc=.35mm, leftrule=.75mm, colframe=quarto-callout-note-color-frame, toprule=.15mm, bottomrule=.15mm, left=2mm, opacityback=0]
\begin{minipage}[t]{5.5mm}
\textcolor{quarto-callout-note-color}{\faInfo}
\end{minipage}%
\begin{minipage}[t]{\textwidth - 5.5mm}

\textbf{Remark (SBC reuse).} This is a workflow convention rather than a
proved proposition: if two fitted studies use the same Stan model, the
same prior hyperparameters, and the same feature pipeline, an SBC result
established for one can be reused for the other without re-running ---
which is what spares us a second, expensive calibration for the matched
cell. \emph{Application:} SBC for \(\alpha\) under
\(\mathrm{Lognormal}(3.0, 0.75)\) and the insurance likelihood is
established once in the GPT-4o \(\times\) insurance report and inherited
by the Claude \(\times\) insurance study. The Ellsberg studies differ in
\(K\) and in the \(\alpha\) prior and so receive \textbf{fresh} SBC
under the \(K = 4\) recalibrated prior.

\emph{Qualification.} SBC certifies the \emph{(prior, likelihood,
sampler)} triple at the model's dimensions \((K, D, R)\), but it fixes a
particular design when simulating, so strictly it speaks only to the
realized feature matrix used in that run --- the stacked per-item
feature vectors \(w_r\) of §\hyperref[sec-app-design]{7.2}. Because
GPT-4o and Claude generate different assessments, they yield different
feature matrices, so the insurance inheritance is an approximation
across two provider-specific designs of the same dimensions
\((K, D, R)\), not an identity; the per-cell parameter-recovery
diagnostics above provide design-specific reassurance, though recovery
checks point estimates, interval widths, and empirical coverage rather
than the full rank-uniformity property SBC certifies, so the inheritance
remains an approximation rather than a re-established calibration.

\end{minipage}%
\end{tcolorbox}

\textbf{MCMC diagnostics.} The split-\(\hat{R}\) statistic compares
between- and within-chain variance and should sit at or below \(1.01\)
for a well-mixed chain; the effective sample size (ESS) reports how many
independent draws the correlated chain is worth. Across all 20 factorial
fits, \(\alpha\) --- the parameter the §\hyperref[sec-app-results]{7.5}
temperature analysis rests on --- is \textbf{never} \(\hat{R}\)-flagged,
and ESS is satisfactory in every fit, so the per-condition \(\alpha\)
posteriors are a sound basis for the cross-condition comparison.
Divergent transitions are negligible (\(\leq 0.15\%\) per fit,
\(\leq 6/4000\); 34 total), though spread across 12/20 fits in both
tasks and both providers rather than confined to the highest GPT-4o
temperatures. \(\hat{R} > 1.01\) appears in 4/20 fits as marginal
exceedances on a minority of high-dimensional components,
\textbf{confined to the weakly-informed nuisance parameters} ---
\(\beta\), \(\delta\), and the global length-\(K\) utility vector
\(\upsilon\) (whose scale is fixed by the endpoint convention but whose
interior levels are only weakly informed, tracking \(\delta\)) ---
together with the per-trial latents \(\eta\) and \(\psi\) in the harder
\(K = 4\) GPT-4o \(\times\) Ellsberg fits (plus one Claude \(\times\)
insurance fit, \(\eta\) only). This \emph{corroborates} the
§\hyperref[sec-bd-weak]{3.4}/§\hyperref[sec-m1-recovery]{6.4}
weakly-informed \((\beta,\delta)\) finding --- the poorly-mixing
parameters are exactly the ones recovery flags as weakly informed ---
and is not a defect in \(\alpha\) inference. As a direct check, all 20
conditions were refit under stricter sampler settings
(\(\mathrm{adapt\_delta} = 0.99\), maximum tree depth \(12\), warmup
\(2000\); seed 42): the divergences vanish entirely (0 across all 20
fits, with no tree-depth saturations), the per-condition \(\alpha\)
posterior medians move by at most \(0.06\) committed-posterior standard
deviations, and the four §\hyperref[sec-app-results]{7.5} temperature
slopes are unchanged to within Monte-Carlo noise (e.g., GPT-4o
\(\times\) insurance \(-24.6 \to -24.8\), \(P(\beta_{\text{temp}} < 0)\)
\(0.99 \to 0.99\); Claude \(\times\) Ellsberg \(-15.0 \to -14.7\),
\(0.77 \to 0.77\)) --- confirming the divergences were benign step-size
artifacts rather than symptoms of a biased posterior (Appendix
\hyperref[sec-d2]{D.2}).

\textbf{Posterior predictive checks.} Three complementary summaries
(log-likelihood, modal-choice frequency, mean predicted probability of
the chosen alternative) yield posterior predictive \(p\)-values in
\([0.32, 0.66]\) across all 20 fits (60/60 in band, mean
\(\approx 0.46\)) --- no evidence of systematic misfit. PPC adequacy is
a \emph{necessary} (not sufficient) condition for the per-condition
\(\alpha\) posteriors to be a credible basis for the cross-condition
comparison; absent it, the comparison should be set aside.

\begin{tcolorbox}[enhanced jigsaw, rightrule=.15mm, breakable, colback=white, arc=.35mm, leftrule=.75mm, colframe=quarto-callout-note-color-frame, toprule=.15mm, bottomrule=.15mm, left=2mm, opacityback=0]
\begin{minipage}[t]{5.5mm}
\textcolor{quarto-callout-note-color}{\faInfo}
\end{minipage}%
\begin{minipage}[t]{\textwidth - 5.5mm}

\textbf{Design-conditional information diagnostics.} Because \(\alpha\)
is identified only through the products \(\alpha(\eta_r - \eta_s)\)
(§\hyperref[sec-alpha-from-eta]{3.3}), how much the data can say about
\(\alpha\) depends on the spread of utility indices the realized design
induces. Three inexpensive diagnostics, computed from the committed
per-condition draws and design matrices
(\texttt{spikes/report\_design\_diagnostics\_spike.py}), make this
design-dependence explicit at the application's scale. (i) \emph{Prior
\(\eta\)-gaps:} under the prior, the within-menu max--min gap in
\(\eta\) has pooled median \(\approx 0.18\)--\(0.20\) in the insurance
conditions and \(\approx 0.10\)--\(0.12\) in the Ellsberg conditions,
versus \(\approx 0.52\) for the foundational
§\hyperref[sec-m0-implementation]{4} design --- the application designs
separate alternatives less, which is exactly why their calibrated priors
select larger \(\alpha\) to reach the same prior-predictive SEU-max rate
(§\hyperref[sec-app-model]{7.3}). (ii) \emph{\(\alpha\) contraction:}
the posterior sd of \(\log \alpha\) is \(0.21\)--\(0.29\) against a
prior sd of \(0.75\) in every one of the 20 conditions --- contraction
to roughly a third of the prior scale --- so the per-condition
\(\alpha\) posteriors are data-driven rather than prior echoes. (iii)
\emph{Feature-matrix rank:} every realized feature matrix has full row
rank 30 (§\hyperref[sec-app-design]{7.2}). Together these quantify,
rather than merely assert, the sense in which the recovered \(\alpha\)
is a design-conditional quantity.

\end{minipage}%
\end{tcolorbox}

\subsection*{\texorpdfstring{7.5 Results --- the \(2\times2\) cross-LLM
\(\times\) cross-task
picture}{7.5 Results --- the 2\textbackslash times2 cross-LLM \textbackslash times cross-task picture}}\label{sec-app-results}
\addcontentsline{toc}{subsection}{7.5 Results --- the \(2\times2\)
cross-LLM \(\times\) cross-task picture}

\subsubsection*{\texorpdfstring{7.5.1 GPT-4o \(\times\)
insurance}{7.5.1 GPT-4o \textbackslash times insurance}}\label{sec-app-gpt4o-insurance}
\addcontentsline{toc}{subsubsection}{7.5.1 GPT-4o \(\times\) insurance}

Posterior medians of \(\alpha\) are highest at \(T = 0.0\) and lowest at
\(T = 1.5\), intermediate temperatures monotone-ish in between. The
global slope \(\Delta\alpha/\Delta T\) has median \(\approx -24.6\),
90\% CI \(\approx [-52.4, -6.7]\), \(P(\text{slope} < 0) \approx 0.99\).
The probability of strict monotonic decrease across all five levels is
only \(\approx 0.12\) (driven by overlap between \(T = 0.3\) and
\(T = 0.7\)); collapsing those two raises it to \(\approx 0.38\). The
headline \emph{directional} claim is well-supported; fine-grained
adjacent-step orderings are not.

\subsubsection*{\texorpdfstring{7.5.2 Claude \(\times\)
insurance}{7.5.2 Claude \textbackslash times insurance}}\label{sec-app-claude-insurance}
\addcontentsline{toc}{subsubsection}{7.5.2 Claude \(\times\) insurance}

The framework \textbf{does not detect} a monotonic
\(\alpha\)--temperature pattern (note the careful phrasing: \emph{does
not detect}, not \emph{establishes the absence of}). Posterior medians
are \(\approx \{71, 53, 73, 71, 55\}\) at \(\{0.0, 0.2, 0.5, 0.8,
1.0\}\). Global slope: median \(\approx -2.9\), 90\% CI
\(\approx [-42.9, 30.8]\), \(P(\text{slope} < 0) \approx 0.56\);
\(P(\text{strict monotone decrease}) < 0.01\). The oscillation in the
medians is consistent with posterior noise around a roughly flat
function, not a substantive non-monotonic response.

\begin{figure}

\centering{

\includegraphics[width=0.85\linewidth,height=\textheight,keepaspectratio]{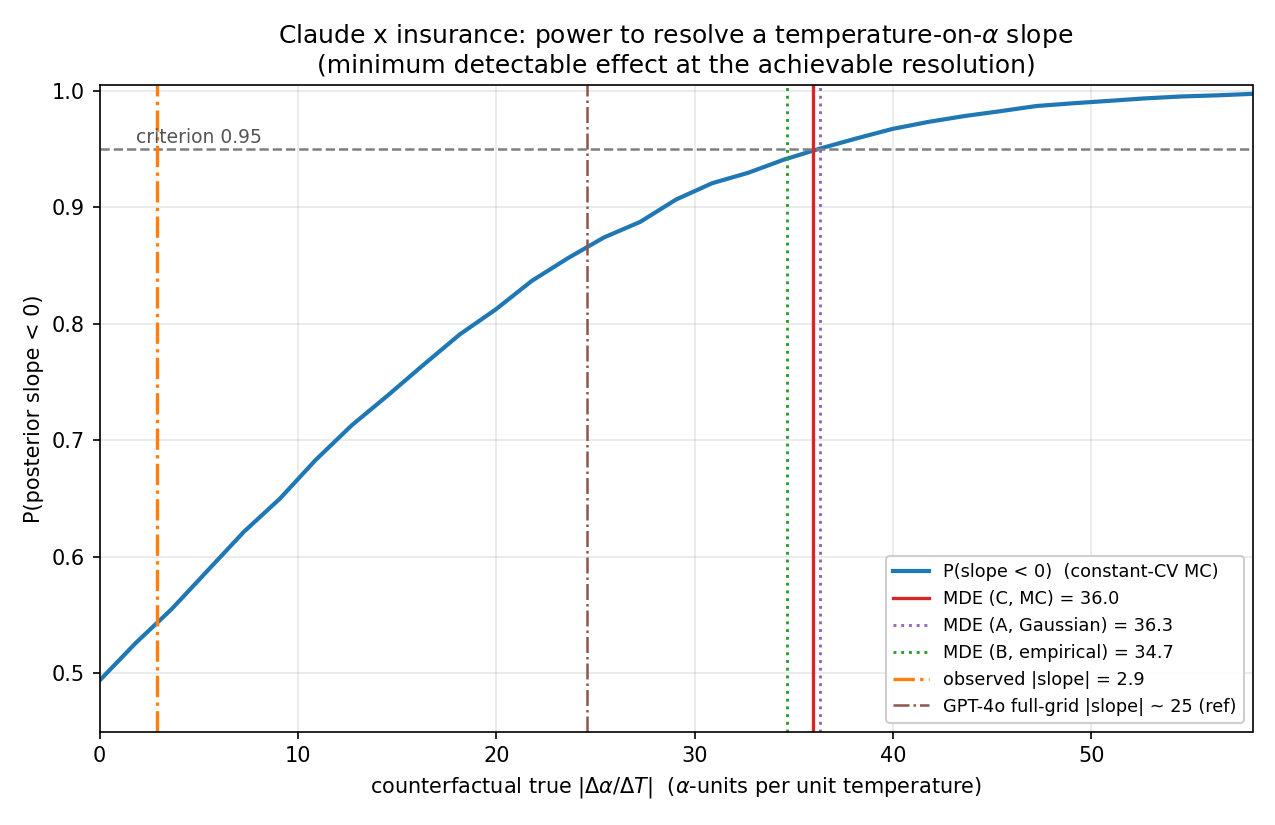}

}

\caption{\label{fig-mde-power}Claude \(\times\) insurance: power to
resolve a temperature-on-\(\alpha\) slope.
\(P(\text{posterior slope} < 0)\) as a function of the counterfactual
true slope magnitude \(|\Delta\alpha/\Delta T|\). The minimum detectable
effect at the \(P \geq 0.95\) criterion is \(\approx 36\)
\(\alpha\)-units per unit temperature (three estimators agree:
analytic-Gaussian, empirical-quantile, constant-CV Monte-Carlo), against
an observed \(|\text{slope}| \approx 2.9\) and a GPT-4o full-grid
reference \(\approx 25\) --- both far below the floor.}

\end{figure}

\begin{tcolorbox}[enhanced jigsaw, rightrule=.15mm, breakable, colback=white, arc=.35mm, leftrule=.75mm, colframe=quarto-callout-warning-color-frame, toprule=.15mm, bottomrule=.15mm, left=2mm, opacityback=0]
\begin{minipage}[t]{5.5mm}
\textcolor{quarto-callout-warning-color}{\faExclamationTriangle}
\end{minipage}%
\begin{minipage}[t]{\textwidth - 5.5mm}

\textbf{Minimum-detectable-effect (calibrating the Claude null).} To
read a null correctly we ask the design's \emph{resolution}: what is the
smallest temperature-on-\(\alpha\) slope this study could reliably
detect? The design could resolve a temperature-on-\(\alpha\) slope at
\(P(\text{slope} < 0) \geq 0.95\) only if its magnitude were
\(|\Delta\alpha/\Delta T| \gtrsim 36\) \(\alpha\)-units per unit
temperature (three independent estimators agree: analytic-Gaussian 36.3,
empirical-quantile 34.7, constant-CV Monte-Carlo 36.0;
\(\approx 41\)--\(43\) at the stricter \(P \geq 0.975\); Appendix
\hyperref[sec-d7]{D.7} defines each estimator and explains why they
agree). That floor is \(\approx 12\times\) the observed
\(|\text{slope}|\) (\(\approx 2.9\), the draw-level posterior median of
§\hyperref[sec-app-claude-insurance]{7.5.2}) and \(\approx 0.53\) of the
grand-mean \(\alpha\) (\(\approx 67.5\), the average of the five
per-temperature posterior means) --- an end-to-end \(\alpha\) change of
\(\approx 36\) over \(T \in [0, 1]\), roughly halving \(\alpha\). So the
Claude null is \emph{no effect at the achievable resolution},
\textbf{not} a measured zero; even a GPT-4o-sized slope (\(\approx 25\),
full-grid reference) sits below this floor (\(0.68\times\)). See the
power curve (Figure~\ref{fig-mde-power}). This calibration prevents the
§\hyperref[sec-app-limits]{7.6.1} ``declines to support'' framing from
being read as a positive no-effect claim.

\end{minipage}%
\end{tcolorbox}

\subsubsection*{7.5.2a Cross-LLM comparison}\label{sec-app-cross-llm}
\addcontentsline{toc}{subsubsection}{7.5.2a Cross-LLM comparison}

A formal cross-study comparison gives
\(P(\text{GPT-4o slope} < \text{Claude
slope}) = 0.817\) on the full grids; restricting GPT-4o to Claude's
\(T \leq 1.0\) ceiling (re-summarizing with \(T = 1.5\) dropped) gives
\(0.816\) --- the directional LLM contrast is \textbf{robust to the
unequal-grid confound} (GPT-4o spans \(T \in [0.0, 1.5]\); Claude spans
\([0.0, 1.0]\)). We report the Claude-grid-restricted number alongside
the full-grid one and make the \emph{qualitative pattern}
(monotone-decline vs.~no-monotone) the headline, not the numeric
inequality. The between-LLM probabilities (\(\sim 0.82\)) are lower than
GPT-4o's own within-cell \(P(\text{slope} < 0) > 0.98\) because they
fold in the independent uncertainty of both cells; they answer a
different question from either within-cell probability (Appendix
D.6(4)). ``Temperature'' is not the same instrument across providers
(§\hyperref[sec-app-limits]{7.6.2}(a)).

\subsubsection*{\texorpdfstring{7.5.3 GPT-4o \(\times\)
Ellsberg}{7.5.3 GPT-4o \textbackslash times Ellsberg}}\label{sec-app-gpt4o-ellsberg}
\addcontentsline{toc}{subsubsection}{7.5.3 GPT-4o \(\times\) Ellsberg}

The temperature--\(\alpha\) decline is reproduced on the Ellsberg task.
\(\alpha\) medians \(\{110.4, 106.9, 99.5, 84.0, 52.2\}\) at
\(T = \{0.0, 0.3,
0.7, 1.0, 1.5\}\) (90\% CIs \([74.4, 167.2]\) / \([72.6, 163.8]\) /
\([65.4, 154.6]\) / \([57.1, 126.1]\) / \([35.5, 80.3]\)); global slope
\(\Delta\alpha/\Delta T\) median \(-38.4\), 90\% CI \([-72.1, -10.0]\),
\(P(\text{slope} < 0) \, 0.984\);
\(P(\text{strict monotone} \downarrow) \, 0.090\) (a noisy decline, not
step-wise). In behavioral terms, the fitted decline means the model
selects the expected-utility-maximizing gamble progressively less often
as temperature rises, drifting toward the menu-uniform choice of the
§\hyperref[sec-three-properties]{2.3} low-\(\alpha\) limit without
reaching it.

\subsubsection*{\texorpdfstring{7.5.4 Claude \(\times\)
Ellsberg}{7.5.4 Claude \textbackslash times Ellsberg}}\label{sec-app-claude-ellsberg}
\addcontentsline{toc}{subsubsection}{7.5.4 Claude \(\times\) Ellsberg}

No monotonic \(\alpha\)--temperature pattern. \(\alpha\) medians
\(\{85.4, 56.1, 82.3,
53.0, 66.4\}\) at \(T = \{0.0, 0.2, 0.5, 0.8, 1.0\}\) (90\% CIs
\([58.9, 127.2]\) / \([38.2, 84.1]\) / \([52.3, 135.3]\) /
\([36.0, 80.5]\) / \([45.3, 99.8]\)) --- the medians oscillate rather
than decline; global slope \(\Delta\alpha/\Delta T\) median \(-15.0\),
90\% CI \([-52.2, 19.6]\), \(P(\text{slope} < 0) \, 0.766\);
\(P(\text{strict monotone} \downarrow) \, 0.0085\). As with Claude
\(\times\) insurance, this is \emph{does not detect}, not
\emph{establishes absence}. We do not transfer the insurance
minimum-detectable-effect number here --- the Ellsberg cell differs in
\(K\), in the \(\alpha\) prior, and in its wider per-condition
posteriors --- so we read this cell as an inconclusive null on its own
diagnostics (oscillating, non-monotone medians and a slope CI spanning
zero), the narrower claim the resolution discipline of
§\hyperref[sec-app-claude-insurance]{7.5.2} licenses.

\subsubsection*{\texorpdfstring{7.5.5 The \(2\times2\)
reading}{7.5.5 The 2\textbackslash times2 reading}}\label{sec-app-2x2}
\addcontentsline{toc}{subsubsection}{7.5.5 The \(2\times2\) reading}

\begin{longtable}[]{@{}
  >{\raggedright\arraybackslash}p{(\linewidth - 4\tabcolsep) * \real{0.3333}}
  >{\raggedright\arraybackslash}p{(\linewidth - 4\tabcolsep) * \real{0.3333}}
  >{\raggedright\arraybackslash}p{(\linewidth - 4\tabcolsep) * \real{0.3333}}@{}}
\caption{Per-cell global-slope \(\Delta\alpha/\Delta T\) posteriors
(median, 90\% CI, \(P(\text{slope}<0)\)). GPT-4o shows the
temperature--\(\alpha\) decline in both task families; Claude shows it
in neither.}\label{tbl-2x2}\tabularnewline
\toprule\noalign{}
\begin{minipage}[b]{\linewidth}\raggedright
\end{minipage} & \begin{minipage}[b]{\linewidth}\raggedright
\textbf{Insurance}
\end{minipage} & \begin{minipage}[b]{\linewidth}\raggedright
\textbf{Ellsberg}
\end{minipage} \\
\midrule\noalign{}
\endfirsthead
\toprule\noalign{}
\begin{minipage}[b]{\linewidth}\raggedright
\end{minipage} & \begin{minipage}[b]{\linewidth}\raggedright
\textbf{Insurance}
\end{minipage} & \begin{minipage}[b]{\linewidth}\raggedright
\textbf{Ellsberg}
\end{minipage} \\
\midrule\noalign{}
\endhead
\bottomrule\noalign{}
\endlastfoot
\textbf{GPT-4o} & slope \(-24.6\) \([-52.4, -6.7]\), \(P_{<0}\,0.99\) &
slope \(-38.4\) \([-72.1, -10.0]\), \(P_{<0}\,0.98\) \\
\textbf{Claude 3.5} & slope \(-2.9\) \([-42.9, 30.8]\), \(P_{<0}\,0.56\)
& slope \(-15.0\) \([-52.2, 19.6]\), \(P_{<0}\,0.77\) \\
\end{longtable}

\begin{figure}

\centering{

\includegraphics[width=0.85\linewidth,height=\textheight,keepaspectratio]{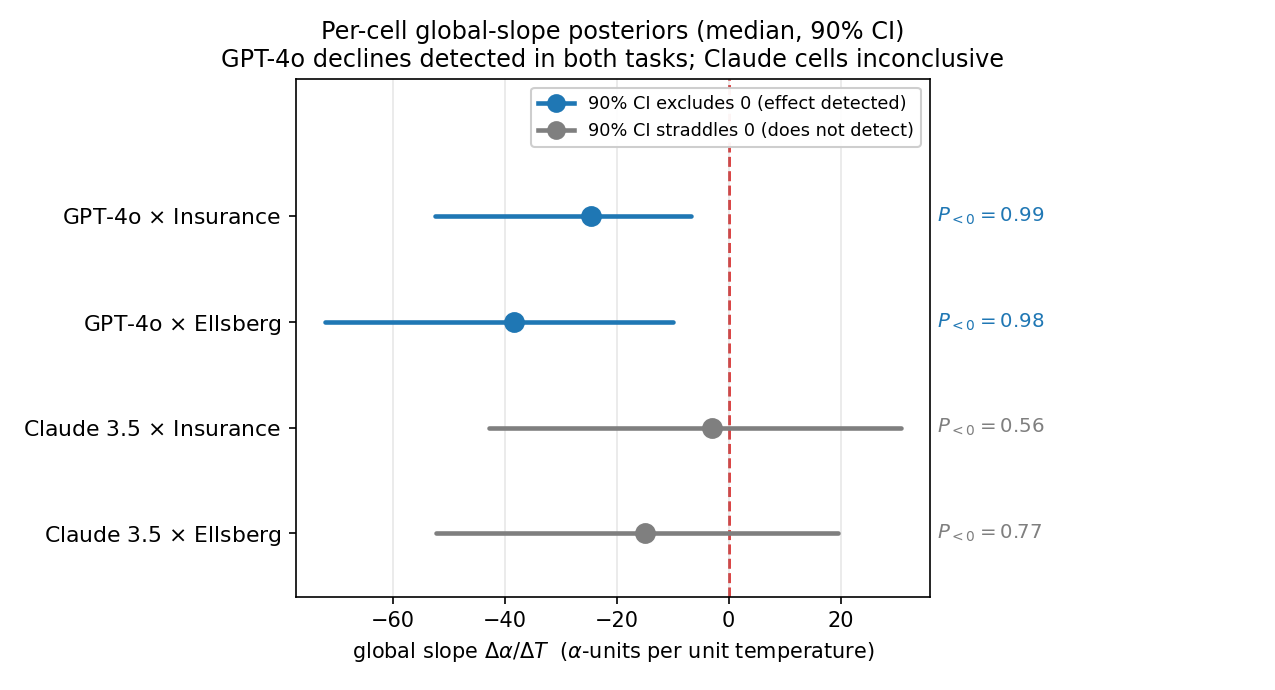}

}

\caption{\label{fig-forest}Per-cell global-slope
\(\Delta\alpha/\Delta T\) posteriors (median and 90\% credible interval)
for the \(2\times2\) LLM \(\times\) task design, with
\(P(\text{slope} < 0)\) annotated at right. Intervals that exclude zero
(blue) mark a detected temperature--\(\alpha\) decline --- GPT-4o in
both tasks; intervals that straddle zero (grey) mark a non-detection ---
Claude in neither. Visual companion to Table~\ref{tbl-2x2}.}

\end{figure}

In this \(2\times2\), the descriptive pattern lines up with the LLM axis
rather than the task axis: GPT-4o exhibits the temperature--\(\alpha\)
relationship across both task families, Claude exhibits it in neither.
But two of the four cells are inconclusive nulls rather than measured
zeros, so the pattern is \emph{suggestive of} an LLM-level difference,
not an established LLM effect. With only two task families and two LLMs
this is a \textbf{pattern statement, not a factorial generalization}
(§\hyperref[sec-app-limits]{7.6.2}(d)). Figure~\ref{fig-forest} presents
it as a \(2\times2\) forest plot of the per-cell global-slope
posteriors.

\subsection*{7.6 What the application demonstrates --- and what it does
not}\label{sec-app-limits}
\addcontentsline{toc}{subsection}{7.6 What the application demonstrates
--- and what it does not}

\subsubsection*{7.6.1 What it demonstrates}\label{sec-app-demonstrates}
\addcontentsline{toc}{subsubsection}{7.6.1 What it demonstrates}

\begin{enumerate}
\def\labelenumi{(\roman{enumi})}
\tightlist
\item
  The full workflow of §§\hyperref[sec-m0-implementation]{4} and
  \hyperref[sec-m1-implementation]{6} runs end-to-end on real LLM choice
  data, at the application's scale, with the \(\alpha\) inference and
  PPC summaries clean and the only flagged diagnostics confined to the
  weakly-informed nuisance parameters
  (§\hyperref[sec-app-validation]{7.4}). (ii) Prior recalibration via
  the prior predictive SEU-maximizer rate is a routine, principled
  per-study step. (iii) The framework detects a structured comparative
  effect in two of four LLM \(\times\) task cells (GPT-4o, both tasks)
  and \textbf{does not detect} one in the other two (Claude, both
  tasks). This last point is the methodological-value claim --- but
  stated carefully: the Claude cells are \emph{inconclusive nulls}
  (§\hyperref[sec-app-claude-insurance]{7.5.2}), so the supported claim
  is ``an instrument that does not manufacture an effect where its own
  diagnostics cannot resolve one,'' \textbf{not} ``an instrument that
  certifies the absence of an effect.'' The minimum-detectable-effect
  statement of §\hyperref[sec-app-claude-insurance]{7.5.2} is what keeps
  this claim resting on calibrated resolution rather than an unsupported
  null.
\end{enumerate}

\subsubsection*{7.6.2 What it does not
demonstrate}\label{sec-app-not-demonstrate}
\addcontentsline{toc}{subsubsection}{7.6.2 What it does not demonstrate}

\textbf{(a) That sampling temperature is a portable behavioral
instrument.} The Claude cells do not reproduce the GPT-4o pattern under
the same task and choice model; but the minimum-detectable-effect
analysis of §\hyperref[sec-app-claude-insurance]{7.5.2} shows that even
a GPT-4o-sized slope would sit below the resolution floor of the Claude
\(\times\) insurance cell --- the one cell for which an MDE was
computed; the Claude \(\times\) Ellsberg cell is read as an inconclusive
null on its own diagnostics --- so the non-reproduction is
\emph{inconclusive}: it does not establish that temperature behaves
differently across providers, only that any Claude-side effect, if
present, is below what these data can resolve. \textbf{(b) That GPT-4o
is ``more rational'' than Claude in any context-free sense.} The
supported reading of \(\alpha\) here is a \emph{within-design
comparative} one, not an absolute-rationality ranking. \textbf{(c)
Anything about ambiguity aversion in the Ellsberg study.} The
SEU+softmax model is, by assumption, of the wrong form to distinguish
ambiguity-driven choice from EU-driven choice --- but not because
ambiguity aversion must show up as a lower \(\alpha\). An
ambiguity-averse agent can be \emph{perfectly} accommodated by the model
with a high \(\alpha\) and pessimistically distorted beliefs: the belief
parameters \((\beta, \psi)\) are free to place extra weight on bad
outcomes of ambiguous gambles, in which case ambiguity aversion is
absorbed into the \emph{belief} side of the model rather than the
sensitivity side. A departure that the belief side cannot mimic is
instead absorbed into a lower \(\alpha\) (more randomness). Either way,
genuine ambiguity attitude, distorted beliefs, and plain noise are not
separable within this model class. Testing for ambiguity aversion
requires a model with explicit ambiguity-attitude parameters --- for
instance maxmin expected utility (Gilboa and Schmeidler 1989), which
adds a parameter governing how heavily the agent weights the worst-case
prior over an ambiguous event, or its \(\alpha\)-MEU generalization
(Ghirardato, Maccheroni, and Marinacci 2004) (whose mixing weight,
conventionally also written \(\alpha\), is unrelated to the sensitivity
parameter \(\alpha\) of this paper); and pooling across the three
ambiguity tiers as the application does would yield a mixture rather
than identify a mechanism. Tier-stratified \(\alpha\) is named as future
work, not a paper claim. \textbf{(d) That LLM identity is the only
relevant axis in general.} With only two task families and two LLMs, the
\(2\times2\) supports a pattern statement, not a factorial
generalization.

\subsubsection*{7.6.3 Construct-validity layering
(reaffirmed)}\label{sec-app-construct-validity}
\addcontentsline{toc}{subsubsection}{7.6.3 Construct-validity layering
(reaffirmed)}

The three-layer reading guide of §\hyperref[sec-app-why]{7.1} --- (1)
model adequacy, (2) comparative claims under shared design, (3) absolute
claims about EU rationality, with only (1) and (2) supported --- is the
single most important reading instruction for this section. The
§\hyperref[sec-app-results]{7.5} cross-condition contrast is a layer-(2)
claim; it is \emph{not} a layer-(3) certification of either LLM's
context-free rationality.

\subsubsection*{7.6.4 Ellsberg: do not adjudicate the normative
question}\label{sec-app-ellsberg-normative}
\addcontentsline{toc}{subsubsection}{7.6.4 Ellsberg: do not adjudicate
the normative question}

The Ellsberg stimuli carry historical weight that does not reduce to the
behavioral ``ambiguity aversion'' reading: alternative readings include
Ellsberg's own normative defense (Ellsberg 1961, 2001) and Levi's
reading as a violation of the completeness axiom rather than the
sure-thing principle (Levi 1980, 1986). A study that takes SEU as its
\emph{measurement device} cannot use the resulting estimates to
vindicate or refute SEU as a \emph{normative} standard. (See Camerer and
Weber (1992) and Trautmann and Kuilen (2015) for the broader
behavioral-ambiguity literature.) The paper's contribution lies
elsewhere.

\subsubsection*{7.6.5 Both applications measure the whole
pipeline}\label{sec-app-pipeline-caveat}
\addcontentsline{toc}{subsubsection}{7.6.5 Both applications measure the
whole pipeline}

Neither application isolates a single cognitive stage, and it is worth
being precise about what \(\alpha\) actually measures. Each trial ---
insurance or Ellsberg --- is produced by a \emph{composed system}: a
first LLM call assesses the stimulus (a free-text claim, or an urn
gamble) in natural language, an embedding model maps the resulting text
to a vector, a PCA step reduces it to the feature space the choice model
reads, and a second LLM call then chooses among the assessed
alternatives (§\hyperref[sec-app-design]{7.2}). The fitted sensitivity
\(\alpha\) is a property of that whole assembly --- the assessment and
choice calls together with the deterministic feature pipeline --- and
not of the choice stage taken on its own. The estimand is therefore the
SEU sensitivity \emph{of the pipeline as configured}, and the
temperature lever acts on every LLM-mediated component of it at once.
This is the appropriate object of study for the methodological claim ---
the workflow recovers a well-calibrated \(\alpha\) for a realistic
end-to-end system --- but it means that a within-cell change in
\(\alpha\) cannot, by itself, be attributed to any one stage. The GPT-4o
monotonicity is thus consistent with temperature affecting (i) the
assessment stage, (ii) the choice stage, or (iii) both. A further caveat
concerns the feature geometry itself: the belief-feature vectors \(w_r\)
are produced by \texttt{text-embedding-3-small} for \emph{both}
providers, so Claude's choices are read through an external embedding
space that need not preserve the distinctions driving its selections. A
within-cell change in \(\alpha\) could therefore reflect the assessment
stage, the choice stage, or upstream embedding/PCA mismatch.
Fixed-feature and alternative-embedding robustness checks --- holding
features constant across temperature, or re-embedding with a
provider-matched model --- are the natural way to separate these, and we
leave them to future work.

The Ellsberg cell does \emph{not} remove the assessment layer. It
applies the \emph{same} assess--embed--choose construction to a
different decision domain (\(K = 4\) monetary gambles in place of
\(K = 3\) claim-triage outcomes); the objective urn composition enters
only through the LLM's free-text assessment, never as a structural input
to the model. It is therefore a \emph{cross-domain replication} rather
than a stage-isolation design. What it establishes is correspondingly
more modest, but still useful: that the GPT-4o temperature pattern is
not an artifact of the particular insurance content or its assessment
prompts, since a comparable effect reappears when the same pipeline is
pointed at a structurally different task. It does \emph{not} decompose
the effect by stage --- because the domain and the assessment prompt
change together with everything else, the Ellsberg cell cannot say
whether the insurance effect originates in the assessment stage, the
choice stage, or both. The two cells together show that the effect
travels across domains; isolating its locus within the pipeline would
require dedicated stage-ablation designs (assessment-only versus
choice-only manipulations) that lie beyond this paper's scope.

\subsubsection*{\texorpdfstring{7.6.6 The cross-condition findings are
robust to the \(\alpha\)
prior}{7.6.6 The cross-condition findings are robust to the \textbackslash alpha prior}}\label{sec-app-prior-sensitivity}
\addcontentsline{toc}{subsubsection}{7.6.6 The cross-condition findings
are robust to the \(\alpha\) prior}

\begin{figure}

\centering{

\includegraphics[width=1\linewidth,height=\textheight,keepaspectratio]{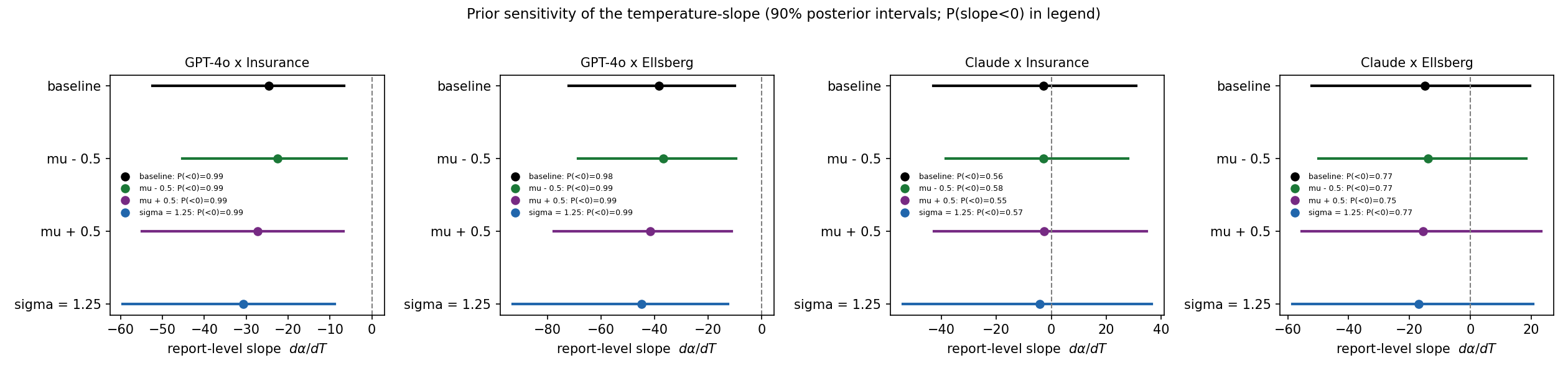}

}

\caption{\label{fig-prior-sensitivity}Prior sensitivity of the
temperature-slope \(\Delta\alpha/\Delta T\) for each of the four cells.
Each panel shows the slope posterior median and 90\% credible interval
under the calibrated baseline prior and three alternatives (location
shifted by \(\pm 0.5\); scale widened to \(1.25\)), with
\(P(\text{slope}<0)\) annotated. The two GPT-4o cells remain clearly
negative and the two Claude cells remain inconclusive nulls under every
prior: the qualitative conclusions do not depend on the \(\alpha\)
prior.}

\end{figure}

Because each cell's headline object is a temperature-slope of a
parameter carrying an informative, prior-predictive-calibrated prior
(§\hyperref[sec-app-model]{7.3}), it is fair to ask whether the slope
findings are an artifact of that prior. They are not. We re-estimated
every condition of all four cells under three alternative \(\alpha\)
priors --- shifting the lognormal location by \(\pm 0.5\)
(\(\mathrm{Lognormal}(\mu_{\text{base}} \mp 0.5,\, 0.75)\)) and widening
the scale to \(\mathrm{Lognormal}(\mu_{\text{base}},\, 1.25)\) ---
holding the choice data and the feature matrices fixed (the committed
per-condition Stan data), and recomputing each cell's slope and
\(P(\text{slope} < 0)\) with the same draw-wise population-OLS
functional. Across all four priors the qualitative reading of every cell
is unchanged (Figure~\ref{fig-prior-sensitivity}): both GPT-4o cells
retain a clearly negative slope --- insurance median in
\([-30.7, -22.6]\) with \(P(\text{slope}<0) \in [0.987, 0.991]\),
Ellsberg median in \([-44.8, -36.8]\) with
\(P(\text{slope}<0) \in [0.984, 0.994]\) --- while both Claude cells
remain inconclusive nulls, insurance median in \([-4.2, -2.7]\) with
\(P(\text{slope}<0) \in [0.55, 0.58]\) and Ellsberg median in
\([-17.0, -14.0]\) with \(P(\text{slope}<0) \in [0.75, 0.77]\). The
wider-prior variant moves the slope magnitudes the most, as expected,
but never flips a sign or crosses a resolution threshold. The prior
calibrates the \emph{scale} on which \(\alpha\) is read; it does not
manufacture the cross-condition contrasts. (Sweep code:
\texttt{spikes/report\_prior\_sensitivity\_spike.py}; claims-ledger row
C17.)

\subsection*{7.7 What the application motivates for follow-up
work}\label{sec-app-followup}
\addcontentsline{toc}{subsection}{7.7 What the application motivates for
follow-up work}

\textbf{Design-induced cross-condition correlation.} The five
temperature conditions in each cell draw from a single \emph{fixed set}
of \(R = 30\) alternatives --- the same claims (or gambles) at every
temperature, though their LLM assessments and embeddings are regenerated
independently at each temperature (§\hyperref[sec-app-design]{7.2}).
Whatever is idiosyncratic about that particular set --- its embedding
geometry, EU spread, or typical best-vs-second-best gap --- is a
property of \emph{this} pool rather than of the population of pools the
design might have drawn, so the cell's contrasts speak to the realized
pool and need not transport to another. Two concerns follow that
independent per-condition \texttt{m\_01} fits do not address: the
cross-condition contrasts may not generalize beyond the realized pool,
and the pool may interact with temperature --- a pool-by-temperature
effect that a single-pool design cannot separate from the main effect.
The principled fix is the hierarchical extension \texttt{h\_m01}:
\(\log \alpha_c = \gamma_0 + \gamma_1 T_c + \varepsilon_c\) with a small
cell-level random effect \(\varepsilon_c\), which estimates the
temperature--sensitivity relationship as a single regression slope with
one calibrated uncertainty statement and partially pools the five
conditions; its multi-pool generalization (the companion paper)
additionally lets the item pool be modeled as a source of variation
rather than treated as a fixed idiosyncrasy.

\textbf{Companion paper.} A planned alignment-study companion paper
(multi-LLM \(\times\) multi-prompt factorial) will introduce
\texttt{h\_m01} as the analysis vehicle and apply it to the
alignment-study data. That paper is out of scope here; the present paper
references it once, as the natural next step for which \texttt{h\_m01}
was developed.

\textbf{Tier-stratified \(\alpha\)} for the Ellsberg studies, and a
non-SEU comparator (e.g., a small \(\alpha\)-MEU instrument) that could
be paired with the SEU instrument to triangulate ambiguity-driven
vs.~EU-driven choice, are flagged as future work in the application
reports and remain out of scope here.

\section*{Discussion}\label{sec-discussion-top}
\addcontentsline{toc}{section}{Discussion}

\subsection*{8.1 What ``sensitivity to SEU maximization'' means,
restated}\label{sec-disc-meaning}
\addcontentsline{toc}{subsection}{8.1 What ``sensitivity to SEU
maximization'' means, restated}

Sensitivity to SEU maximization is the \emph{disposition} of an agent
committed to SEU maximization to act in accordance with that commitment.
The model decomposes choice behavior into three conceptually distinct
objects: (i) the agent's beliefs (\(\beta\), hence the subjective
probabilities \(\psi\)); (ii) the agent's utilities (\(\delta\), hence
the ordered utilities \(\bm{\upsilon}\)); and (iii) the agent's
sensitivity \(\alpha\) to the expected-utility ranking those beliefs and
utilities induce. What the agent is committed to is the \emph{SEU
standard} itself, and that standard places requirements on how the
agent's beliefs and values --- represented by a subjective probability
and a utility --- bear on choice; beliefs and utilities are those
inputs, while \(\alpha\) is \emph{how reliably} the agent's choices
track the expected-utility ranking the standard derives from them. The
three softmax-choice properties of §\hyperref[sec-three-properties]{2.3}
--- monotonicity in \(\alpha\), the optimization limit, and the
uniform-choice limit --- are exactly what license reading \(\alpha\) as
a graded measure that runs from near-uniform choice (\(\alpha \to 0\))
to deterministic SEU maximization (\(\alpha \to \infty\)).

This reading returns to Isaac Levi's distinction between an agent's
\emph{commitment} to a standard and the agent's \emph{performance}
relative to it (Levi 1980, chap. 1), introduced in
§\hyperref[sec-conceptual-payoff]{2.5}. One can be committed to a
standard one fails to live up to: most of us are committed to the laws
of arithmetic yet occasionally make calculation errors, and those errors
are lapses of performance rather than rejections of the standard. Taking
SEU theory to specify the agent's normative commitment, \(\alpha\) is
exactly the agent's \emph{tendency to perform in accordance with that
commitment}.

Two points are worth keeping in mind regarding our borrowing of Levi's
distinction. First, Levi was no defender of SEU as the standard of
rational choice; on the contrary, he developed an influential
generalization of it that admits indeterminate (imprecise) probabilities
and utilities (Levi 1986), and his substantive decision theory therefore
falls outside the real-valued template of this paper
(§\hyperref[sec-seu-standard]{1.4}). What we borrow is his
\emph{meta-level} distinction between commitment and performance, not
his account of the standard itself. Second, that distinction is of
interest even once SEU is granted as the relevant standard. Consider the
SEU violations documented by Kahneman and Tversky (Kahneman and Tversky
1979; Tversky and Kahneman 1992): do they reveal failures of
\emph{performance} by agents committed to SEU, or an \emph{absence of
commitment} to SEU as a standard? Our framework does not settle that
question, but it does make it tractable: \emph{under the assumption}
that the agent is committed to SEU, \(\alpha\) estimates how
\emph{reliably the agent's choices conform} to that commitment.

\subsection*{8.2 The methodological role of identifiability
analysis}\label{sec-disc-identifiability}
\addcontentsline{toc}{subsection}{8.2 The methodological role of
identifiability analysis}

Identifiability is a property of the \emph{likelihood}: it tells us
which parameters could in principle be recovered from choice data. It
does \emph{not} tell us how much data, or how good a design, is needed
to recover them precisely. The governing principle is:

\begin{tcolorbox}[enhanced jigsaw, rightrule=.15mm, breakable, colback=white, arc=.35mm, leftrule=.75mm, colframe=quarto-callout-important-color-frame, toprule=.15mm, bottomrule=.15mm, left=2mm, opacityback=0]
\begin{minipage}[t]{5.5mm}
\textcolor{quarto-callout-important-color}{\faExclamation}
\end{minipage}%
\begin{minipage}[t]{\textwidth - 5.5mm}

\emph{Identifiable (hence estimable in the limit) \(\neq\) precisely
estimable at realistic sample sizes.}

\end{minipage}%
\end{tcolorbox}

The \texttt{m\_0} / \texttt{m\_1} contrast illustrates \emph{both} sides
of this distinction, and the paper is careful to keep the two sides
apart (§\hyperref[sec-m1-lesson]{6.4.4}):

\begin{itemize}
\tightlist
\item
  \textbf{For \(\delta\) --- identifiability without precise
  estimability.} In \texttt{m\_1}, \(\delta\) is globally identifiable
  from the \(\beta\)-free risky block under lottery diversity
  (§\hyperref[sec-delta-id]{5.5}). Yet at the design's sample sizes the
  matched-count payoff is essentially nil: a \(\delta\) CI-width
  reduction of \(\approx 0.8\%\) (paired-iteration median; bootstrap
  90\% CI \([0.6, 1.2]\); Wilcoxon \(p \approx 1.2\times10^{-6}\)) and
  no change in \(\delta\) RMSE. The effect is statistically real but
  practically negligible --- the identifiability-versus-estimability gap
  in its starkest form.
\item
  \textbf{For \(\alpha\) --- in-principle identifiability is silent
  about where finite-\(n\) precision comes from.} \(\alpha\) is
  identifiable from \(\eta\) in both models
  (§\hyperref[sec-alpha-from-eta]{3.3}, §\hyperref[sec-m1-alpha]{5.4}).
  The matched recovery study shows that at realistic \(n\) its precision
  is governed by data \emph{quantity}, not by the \emph{type} of choice:
  doubling the uncertain block alone sharpens \(\alpha\) (\(A \to B\):
  RMSE reduced \(27.1\%\), 90\% CI \([+15.0, +38.5]\)), whereas swapping
  in an equal count of \(\beta\)-free risky choices does not
  (\(B \to C\): \(-8.1\%\), 90\% CI \([-22.0, +6.8]\), straddling zero).
  An in-principle identification argument predicts neither of these
  finite-\(n\) facts.
\end{itemize}

\subsection*{8.3 The methodological role of computational
validation}\label{sec-discussion}
\addcontentsline{toc}{subsection}{8.3 The methodological role of
computational validation}

Within the Bayesian workflow this paper adopts, three checks are
standard practice, and the paper applies all three. \emph{Prior
predictive checks} (§\hyperref[sec-m0-prior]{4.2},
§\hyperref[sec-m1-prior]{6.3}) expose what a prior implies before any
data are seen; in this setting the SEU-maximizer-rate statistic makes
the \(\alpha\) prior's behavioral content legible and turns per-study
prior recalibration into a principled, repeatable step
(§\hyperref[sec-app-model]{7.3}). \emph{Parameter recovery} and
\emph{simulation-based calibration} (SBC) together test whether the
posterior recovers known truths --- recovery for bias, precision, and
interval coverage; SBC for calibration of the whole \emph{(prior,
likelihood, sampler)} triple.

A specific caution emerges from our results, which we name so it is
citable rather than discursive:

\begin{tcolorbox}[enhanced jigsaw, rightrule=.15mm, breakable, colback=white, arc=.35mm, leftrule=.75mm, colframe=quarto-callout-warning-color-frame, toprule=.15mm, bottomrule=.15mm, left=2mm, opacityback=0]
\begin{minipage}[t]{5.5mm}
\textcolor{quarto-callout-warning-color}{\faExclamationTriangle}
\end{minipage}%
\begin{minipage}[t]{\textwidth - 5.5mm}

\textbf{The marginal-SBC demarcation (restated; full statement in
§\hyperref[sec-m0-sbc]{4.4}).} Marginal rank uniformity is
\emph{necessary but not sufficient} for \emph{joint} posterior
calibration; and --- distinctly --- the correlated, barely-contracted
joint \((\beta, \delta)\) posterior reflects \emph{weak joint
informativeness}, \textbf{not} a calibration failure: a correctly
specified model and sampler can be exactly calibrated and still return
such a posterior. Marginal SBC passes for \(\alpha\), \(\beta\),
\emph{and} \(\delta\) in both \texttt{m\_0} and \texttt{m\_1}
(§\hyperref[sec-m0-sbc]{4.4}, §\hyperref[sec-m1-sbc]{6.5}); what it does
\emph{not} do is reveal the weak joint identification of
\((\beta, \delta)\) at realistic \(n\) (§\hyperref[sec-bd-weak]{3.4}).
We report no non-uniform joint SBC statistic and claim none.

\end{minipage}%
\end{tcolorbox}

The positive contribution of SBC here --- sampler/implementation
validation, per-parameter marginal calibration, and the demarcation
itself --- is stated in full in §\hyperref[sec-m0-sbc]{4.4}. Two
follow-ups should be kept distinct. A \emph{joint} or
\emph{projection-based} rank statistic would extend SBC's reach from
marginal to joint \emph{calibration} --- a concrete next step for the
SBC literature. But no rank statistic, marginal or joint, measures
\emph{informativeness}: whether a posterior has contracted enough to be
useful is diagnosed by recovery, contraction, and interval-width
summaries (§\hyperref[sec-m0-recovery]{4.3},
§\hyperref[sec-app-validation]{7.4}), not by calibration checks. The
paper's \((\beta,\delta)\) finding is of the second kind, which is why
no SBC variant --- however joint --- would have surfaced it on its own.

\subsection*{\texorpdfstring{8.4 Why \(\alpha\) is the primary quantity
of
interest}{8.4 Why \textbackslash alpha is the primary quantity of interest}}\label{sec-disc-alpha}
\addcontentsline{toc}{subsection}{8.4 Why \(\alpha\) is the primary
quantity of interest}

A central methodological claim of the paper is that \(\alpha\) can be
measured precisely \emph{even when} \((\beta, \delta)\) cannot be --- a
separation the recovery study and application deliver
\emph{empirically}, under the paper's priors and designs, rather than
one guaranteed by the conditional identification proposition alone. The
mechanism is that beliefs and utilities enter the choice probabilities
only through the expected-utility vector \(\eta\), and the log-odds
scale with differences in \(\eta\) at rate \(\alpha\)
(§\hyperref[sec-m0-link]{3.5}). Proposition 3.1 identifies \(\alpha\)
only \emph{conditional} on \(\eta\); in the uncertain-choice model
\(\eta\) is itself generated by the unknown \((\beta, \delta)\), and
near the linear region of the belief softmax a change in the spread of
\(\eta\) can trade off with \(\alpha\) in the observed log-odds. What
makes \(\alpha\) sharply estimable in practice is therefore an empirical
fact about the calibrated prior and the realized designs --- the
posterior concentrates the \(\alpha\)-bearing log-odds scale while
leaving the \((\beta, \delta)\) trade-off broad
(§\hyperref[sec-bd-weak]{3.4}) --- not a structural guarantee read off
the conditional proposition. The §\hyperref[sec-application]{7}
application instantiates the claim: the cross-condition comparison is a
claim about \(\alpha\), \(\alpha\) is sharply and calibratedly recovered
in \texttt{m\_0} and the calibrated-prior variants
\texttt{m\_01}/\texttt{m\_02}, and the weak \emph{informativeness} of
\((\beta, \delta)\) does not undermine that comparison. This comparative
discipline is also the paper's answer to the cross-context comparability
critiques of softmax-precision parameters (Wilcox 2011; Apesteguia and
Ballester 2018): the \(\bm{\upsilon}\)-endpoint convention
(§\hyperref[sec-notation]{2.6}) fixes one global utility scale shared by
every problem in a design --- which supplies the within-context
normalization the contextual-utility critique asks for only because all
problems draw on the same consequence space
(§\hyperref[sec-softmax-rule]{2.2}) --- and the supported claims compare
\(\alpha\) only within a fixed design.

\begin{tcolorbox}[enhanced jigsaw, rightrule=.15mm, breakable, colback=white, arc=.35mm, leftrule=.75mm, colframe=quarto-callout-note-color-frame, toprule=.15mm, bottomrule=.15mm, left=2mm, opacityback=0]
\begin{minipage}[t]{5.5mm}
\textcolor{quarto-callout-note-color}{\faInfo}
\end{minipage}%
\begin{minipage}[t]{\textwidth - 5.5mm}

\textbf{Interpretive caveat (model-conditional reading of \(\alpha\)).}
A low recovered \(\alpha\) is \emph{not}, in a model-free sense, ``low
SEU sensitivity.'' The measured \(\alpha\) is a parameter of a specific
likelihood --- a softmax over SEU values under the parameterization of
§\hyperref[sec-m0-identifiability]{3}. A low \(\alpha\) is
observationally equivalent to several non-SEU alternatives (the agent
maximizes a different value functional, uses a non-softmax stochastic
rule, has heterogeneous within-session preferences, and so on). The
paper's claim is therefore conditional: \emph{under} the SEU + softmax
model, \(\alpha\) measures sensitivity to the EU ranking the model
attributes to the agent. The §\hyperref[sec-application]{7} application
makes this concrete --- the supported reading is the within-design
\emph{comparative} one (§\hyperref[sec-app-limits]{7.6.3}), not an
absolute-rationality ranking. Model misspecification is a substantive
interpretive risk for any procedural-evaluation instrument and is
acknowledged here rather than buried.

\end{minipage}%
\end{tcolorbox}

\subsection*{8.5 Limitations and extensions}\label{sec-disc-limitations}
\addcontentsline{toc}{subsection}{8.5 Limitations and extensions}

\textbf{Precise \(\delta\) estimation.} The modest finite-sample
\(\delta\) gain (§\hyperref[sec-m1-delta-recovery]{6.4.2}) is a
limitation, not a barrier in principle. Three approaches could improve
it --- a \(\delta\)-information-optimal lottery design (e.g.,
triangulating contrasts that pit the certain intermediate consequence
against a \(50/50\) mix of the extremes), substantially larger samples,
and prior regularization on \(\delta\) --- but they are \emph{not
independent}: a \(\delta\)-optimal design helps less when \(\alpha\) is
small (a shallow softmax at moderate \(\alpha\) dampens the same
contrasts), so precise \(\delta\) recovery likely requires improvements
on several of these fronts at once rather than any one of them alone. A
\(\delta\)-optimal lottery-design study is named here as future work
rather than attempted in this paper.

\textbf{The single-\(\alpha\) assumption.} A single \(\alpha\) governing
both the uncertain and risky blocks (§\hyperref[sec-m1-properties]{5.3})
is substantive and testable. It is essential to the
§\hyperref[sec-m1-recovery]{6.4} quantity-versus-type reading, which is
why that reading is stated as explicitly conditional on it. The natural
relaxation is a block-specific model fitting separate
\(\alpha_{\mathrm{unc}}, \alpha_{\mathrm{risky}}\) (our internal
\texttt{m\_2}); it is named here, not pursued.

\textbf{Menu size and sensitivity.} The model places no restriction on
the number of alternatives in a decision problem: the available set
\(\mathcal{A}_m\) may be of any size, the softmax of
§\hyperref[sec-softmax-rule]{2.2} is taken over it, and the Stan
implementation already accepts a per-problem alternative count (Appendix
C). The classical prospect-theory paradigm and its large-scale
replication (Kahneman and Tversky 1979; Tversky and Kahneman 1992;
Ruggeri et al. 2020) restricted attention to pairwise choice,
appropriately for their purpose; the present formulation does not, which
makes the \emph{relationship between sensitivity and menu size} an
estimable object rather than a fixed feature of the design. Treating
menu size \(N_m\) as a covariate on \(\log \alpha\) ---
e.g.~\(\log \alpha = \gamma_0 + \gamma_1 N_m\) in the hierarchical
vehicle above --- would let one ask whether sensitivity rises or falls
as alternatives are added; one illustrative, model-agnostic hypothesis
is that a larger menu raises the cognitive load on the decision maker
and thereby lowers \(\alpha\), but the sign is left open. This is named,
not pursued: the §\hyperref[sec-application]{7} designs randomize and
position-counterbalance \(N_m\) and fit a single \(\alpha\) pooled
across menu sizes, so estimating a menu-size effect cleanly would
require a purpose-built design that varies menu size as a factor rather
than a nuisance.

\textbf{Functional form.} Linear-softmax belief formation
(\(\psi_r = \mathrm{softmax}(\beta w_r)\)) is a strong functional form;
extensions to nonlinear belief formation and richer feature maps are
possible.

\textbf{Hierarchical extensions.} The §\hyperref[sec-application]{7}
application motivates the hierarchical extension \texttt{h\_m01} to
address the design-induced cross-condition structure of
§\hyperref[sec-app-followup]{7.7}: because the five conditions reuse a
single fixed pool of alternatives (with the LLM assessments and
embeddings regenerated independently at each temperature), fitting
\texttt{m\_01} independently per condition leaves two things on the
table --- partial pooling of information across the five temperature
conditions, and any treatment of the item pool as a modeled source of
variation. The two are distinct and should not be conflated: partial
pooling stabilizes the per-condition \(\alpha\) estimates and yields one
calibrated slope statement, but pooling across \emph{temperature
conditions} that all reuse the same pool cannot, by itself, generalize
the findings beyond that realized pool. Modelling
\(\log \alpha_c = \gamma_0 + \gamma_1 T_c + \varepsilon_c\) with
condition-level effects \(\varepsilon_c\) partially pools the five
temperature conditions and turns between-condition contrasts into
estimated regression effects on \(\log \alpha\). Because a \emph{single}
pool is reused across all conditions, however, \texttt{h\_m01} still
cannot separate a pool \emph{main} effect (confounded with the intercept
\(\gamma_0\)) or a pool-by-temperature interaction (absorbed into the
slope \(\gamma_1\)); doing so --- and with it any generalization over
item pools --- requires several distinct pools. That multi-pool
generalization is the analysis vehicle of a planned alignment-study
companion paper and is out of scope here.

\textbf{Further application studies.} A tier-stratified Ellsberg
analysis, a non-SEU comparator instrument (e.g., a small \(\alpha\)-MEU
model paired with the SEU instrument to triangulate ambiguity-driven
versus EU-driven choice), and the alignment study itself are pursued in
companion work and are out of scope here.

\subsection*{8.6 Closing}\label{sec-disc-closing}
\addcontentsline{toc}{subsection}{8.6 Closing}

A precise definition of sensitivity, a formal identifiability analysis,
a computational validation pipeline, and an end-to-end illustrative
application together provide a transparent, reusable framework for
\emph{procedurally} evaluating decision makers --- human or machine ---
against a stated rationality standard expressible as expected-value
maximization, with SEU as the canonical case. The framework reports
where a parameter is identifiable but not precisely estimable, where
in-principle arguments fail to predict finite-sample behavior, where
marginal calibration is blind to a jointly weakly-informed posterior,
and where an instrument declines to support an effect its own
diagnostics cannot resolve. Reporting these limits is part of what makes
the framework usable as a measurement instrument rather than a procedure
that returns an effect by construction.

\section*{Appendix A. Proofs of the Softmax-Choice
Properties}\label{sec-appendix-a}
\addcontentsline{toc}{section}{Appendix A. Proofs of the Softmax-Choice
Properties}

\begin{quote}
\emph{Proofs of the three properties stated in
§\hyperref[sec-three-properties]{2.3}, plus the scale-invariance result
underlying the \(\bm{\upsilon}\)-endpoint convention of
§\hyperref[sec-notation]{2.6}. These are the only results in the paper
that hold as closed-form proofs without a genericity or curvature
caveat.}
\end{quote}

Throughout, fix a decision problem \(m\) with available set
\(\mathcal{A} = \{r :
I_{m,r} = 1\}\) and a value function \(V\) on it, write
\(P(r) = \exp(\alpha V(r)) /
Z(\alpha)\) with partition function
\(Z(\alpha) = \sum_{j \in \mathcal{A}}
\exp(\alpha V(j))\), \(V^\ast = \max_{r \in \mathcal{A}} V(r)\),
\(\mathcal{R}^\ast = \{r \in \mathcal{A} : V(r) = V^\ast\}\), and
\(\mathcal{R}^- =
\mathcal{A} \setminus \mathcal{R}^\ast\). All maxima, minima, and sums
are taken over the available set \(\mathcal{A}\).

\subsection*{A.1 Monotonicity in sensitivity (Property
1)}\label{sec-a-monotonicity}
\addcontentsline{toc}{subsection}{A.1 Monotonicity in sensitivity
(Property 1)}

\begin{tcolorbox}[enhanced jigsaw, rightrule=.15mm, breakable, colback=white, arc=.35mm, leftrule=.75mm, colframe=quarto-callout-note-color-frame, toprule=.15mm, bottomrule=.15mm, left=2mm, opacityback=0]
\begin{minipage}[t]{5.5mm}
\textcolor{quarto-callout-note-color}{\faInfo}
\end{minipage}%
\begin{minipage}[t]{\textwidth - 5.5mm}

\textbf{Theorem A.1 (Monotonicity).} Holding \(V\) fixed: \emph{(i)} for
any two alternatives \(r, s \in \mathcal{A}\) the pairwise log-odds are
linear in \(\alpha\), \(\partial_\alpha \log[P(r)/P(s)] = V(r) - V(s)\),
so the odds move monotonically toward the higher-valued alternative;
\emph{(ii)} provided the values \(\{V(j)\}_{j \in \mathcal{A}}\) are not
all equal, the total probability
\(P(\mathcal{R}^\ast) = \sum_{r \in \mathcal{R}^\ast}
P(r)\) on the value-maximizing set is strictly increasing in \(\alpha\)
(if all values coincide then \(\mathcal{R}^\ast = \mathcal{A}\) and
every \(P(r)\) is constant in \(\alpha\)); \emph{(iii)} provided the
values \(\{V(j)\}_{j \in \mathcal{A}}\) are not all equal, each
\(r \in \mathcal{R}^\ast\) has \(P(r)\) strictly increasing in
\(\alpha\) and each value-\emph{minimizing} \(r\) (with
\(V(r) = \min_{j} V(j)\)) has \(P(r)\) strictly decreasing (in the
all-tied case every \(P(r)\) is constant, as in (ii)). A non-maximal
alternative with \(V(r) > \mathbb{E}_\alpha[V]\) has
\(\partial_\alpha P(r) > 0\), so individual non-maximal probabilities
need not be monotone in \(\alpha\).

\end{minipage}%
\end{tcolorbox}

\emph{Proof.} \emph{(i)} From \(P(r) = \exp(\alpha V(r))/Z(\alpha)\),
\(\log[P(r)/P(s)] = \alpha\,(V(r) - V(s))\), whose \(\alpha\)-derivative
is \(V(r) - V(s)\).

\emph{(iii)} Differentiating \(P(r)\) by the quotient rule with
\(Z'(\alpha) = \sum_j
V(j)\exp(\alpha V(j))\), \[
\frac{\partial P(r)}{\partial \alpha}
  = P(r)\Big[V(r) - \frac{Z'(\alpha)}{Z(\alpha)}\Big]
  = P(r)\big[V(r) - \mathbb{E}_\alpha[V]\big],
\] where \(\mathbb{E}_\alpha[V] = \sum_j P(j) V(j)\). With the values
not all equal, every alternative receives positive probability at any
finite \(\alpha\), so \(\min_j V(j) < \mathbb{E}_\alpha[V]
< V^\ast\) strictly (both inequalities require the nonconstant-values
case). Hence \(\partial P(r)/\partial\alpha > 0\) for
\(r \in \mathcal{R}^\ast\) and \(\partial P(r)/\partial\alpha < 0\) for
any \(r\) attaining \(\min_j V(j)\); for an intermediate \(r\) the sign
equals that of \(V(r) - \mathbb{E}_\alpha[V]\), which can be positive.

\emph{(ii)} Writing \(P^\ast(\alpha) = P(\mathcal{R}^\ast)\) and summing
the derivative identity over \(r \in \mathcal{R}^\ast\) (where
\(V(r) = V^\ast\)), \[
\frac{\partial P^\ast}{\partial \alpha}
  = \sum_{r \in \mathcal{R}^\ast} P(r)\big[V^\ast - \mathbb{E}_\alpha[V]\big]
  = P^\ast\big(V^\ast - \mathbb{E}_\alpha[V]\big) > 0,
\] since \(\mathbb{E}_\alpha[V] < V^\ast\) strictly.
\(\qquad\blacksquare\)

\subsection*{A.2 Optimization limit (Property
2)}\label{sec-a-optimization}
\addcontentsline{toc}{subsection}{A.2 Optimization limit (Property 2)}

\begin{tcolorbox}[enhanced jigsaw, rightrule=.15mm, breakable, colback=white, arc=.35mm, leftrule=.75mm, colframe=quarto-callout-note-color-frame, toprule=.15mm, bottomrule=.15mm, left=2mm, opacityback=0]
\begin{minipage}[t]{5.5mm}
\textcolor{quarto-callout-note-color}{\faInfo}
\end{minipage}%
\begin{minipage}[t]{\textwidth - 5.5mm}

\textbf{Theorem A.2.} As \(\alpha \to \infty\),
\(P(r) \to 1/|\mathcal{R}^\ast|\) for \(r \in \mathcal{R}^\ast\) and
\(P(r) \to 0\) for \(r \in \mathcal{R}^-\).

\end{minipage}%
\end{tcolorbox}

\emph{Proof.} Divide numerator and denominator of \(P(r)\) by
\(\exp(\alpha V^\ast)\). For \(r \in \mathcal{R}^\ast\), \[
P(r) = \frac{1}{|\mathcal{R}^\ast| + \sum_{j \in \mathcal{R}^-}\exp(\alpha[V(j)-V^\ast])}
  \;\longrightarrow\; \frac{1}{|\mathcal{R}^\ast|},
\] since \(V(j) - V^\ast < 0\) for \(j \in \mathcal{R}^-\). For
\(r \in \mathcal{R}^-\) the numerator is
\(\exp(\alpha[V(r)-V^\ast]) \to 0\) while the denominator is bounded
below by \(|\mathcal{R}^\ast| > 0\), so \(P(r) \to 0\).
\(\qquad\blacksquare\)

\emph{Remark.} Ties (\(|\mathcal{R}^\ast| > 1\)) have measure zero under
the continuous priors of §\hyperref[sec-m0-implementation]{4}; the
unique-maximizer case is generic and gives deterministic optimal choice
in the limit.

\subsection*{A.3 Uniform-choice limit (Property 3)}\label{sec-a-uniform}
\addcontentsline{toc}{subsection}{A.3 Uniform-choice limit (Property 3)}

\begin{tcolorbox}[enhanced jigsaw, rightrule=.15mm, breakable, colback=white, arc=.35mm, leftrule=.75mm, colframe=quarto-callout-note-color-frame, toprule=.15mm, bottomrule=.15mm, left=2mm, opacityback=0]
\begin{minipage}[t]{5.5mm}
\textcolor{quarto-callout-note-color}{\faInfo}
\end{minipage}%
\begin{minipage}[t]{\textwidth - 5.5mm}

\textbf{Theorem A.3.} As \(\alpha \to 0\), \(P(r) \to 1/|\mathcal{A}|\)
for all \(r \in \mathcal{A}\), independent of \(V\).

\end{minipage}%
\end{tcolorbox}

\emph{Proof.} By the expansion \(\exp(x) = 1 + x + O(x^2)\), \[
P(r) = \frac{1 + \alpha V(r) + O(\alpha^2)}{|\mathcal{A}| + \alpha\sum_j V(j) + O(\alpha^2)}
  \;\longrightarrow\; \frac{1}{|\mathcal{A}|}
\] as \(\alpha \to 0\). Equivalently,
\(\log P(r) = -\log|\mathcal{A}| + \alpha[V(r) -
\bar V] + O(\alpha^2)\) with \(\bar V = |\mathcal{A}|^{-1}\sum_j V(j)\),
so \(P(r) \to 1/|\mathcal{A}|\). \(\qquad\blacksquare\)

\subsection*{A.4 Scale invariance and the utility-endpoint
convention}\label{sec-a-scale}
\addcontentsline{toc}{subsection}{A.4 Scale invariance and the
utility-endpoint convention}

\begin{tcolorbox}[enhanced jigsaw, rightrule=.15mm, breakable, colback=white, arc=.35mm, leftrule=.75mm, colframe=quarto-callout-note-color-frame, toprule=.15mm, bottomrule=.15mm, left=2mm, opacityback=0]
\begin{minipage}[t]{5.5mm}
\textcolor{quarto-callout-note-color}{\faInfo}
\end{minipage}%
\begin{minipage}[t]{\textwidth - 5.5mm}

\textbf{Theorem A.4 (Scale invariance).} Let
\(\tilde{\upsilon}_k = a\,\upsilon_k + b\) with \(a > 0\). Then
\(\tilde{\eta}_r = a\,\eta_r + b\) for all \(r\), and \[
P(\text{choose } r \mid \alpha, \tilde{\bm{\upsilon}})
  = P(\text{choose } r \mid \alpha a,\, \bm{\upsilon}).
\] The pairs \((\alpha, \tilde{\bm{\upsilon}})\) and
\((\alpha a, \bm{\upsilon})\) generate identical choice probabilities.

\end{minipage}%
\end{tcolorbox}

\emph{Proof.} Since \(\sum_k \psi_{r,k} = 1\),
\(\tilde\eta_r = \sum_k \psi_{r,k}(a\upsilon_k
+ b) = a\eta_r + b\). Substituting into softmax, the additive constant
\(\alpha b\) cancels between numerator and denominator and the
multiplicative factor combines with \(\alpha\): \[
P(r \mid \alpha, \tilde{\bm{\upsilon}})
  = \frac{\exp(\alpha[a\eta_r + b])}{\sum_j \exp(\alpha[a\eta_j + b])}
  = \frac{\exp(\alpha a\,\eta_r)}{\sum_j \exp(\alpha a\,\eta_j)}
  = P(r \mid \alpha a, \bm{\upsilon}). \qquad\blacksquare
\]

\emph{Consequence.} Without fixing the utility scale, \(\alpha\) and the
scale of utility are confounded: scaling utilities by \(a\) is
equivalent to scaling sensitivity by \(1/a\). Fixing the endpoints
\(\upsilon_1 = 0\), \(\upsilon_K = 1\) (the \(\bm{\upsilon}\)-endpoint
convention of §\hyperref[sec-notation]{2.6}) selects a representative
from the affine equivalence class, which removes this particular
confound and makes \(\alpha\) a \emph{well-posed} target on a
standardized scale where the full utility range spans one unit, with the
interpretation \(\log[P(r)/P(s)] = \alpha(\eta_r - \eta_s)\). It does
not by itself \emph{identify} \(\alpha\): identification additionally
requires a positive-probability menu with non-constant expected
utilities (Proposition 3.1 / Appendix B.1), as
§\hyperref[sec-notation]{2.6} notes.

\section*{Appendix B. Identifiability Results}\label{sec-appendix-b}
\addcontentsline{toc}{section}{Appendix B. Identifiability Results}

\begin{quote}
\emph{B.1 and B.3 are proofs under explicit genericity conditions; B.4
records what uncertain choices do and do not pin down about \(\beta\)
--- all information flows through the per-menu \(\eta\)-contrasts, so
identification turns on a design-specific rank condition; B.2 is not a
proposition but a brief empirical note on the Bayesian workflow,
recording that uncertain-choice data weakly inform \((\beta,\delta)\) as
seen directly in parameter recovery.}
\end{quote}

\subsection*{\texorpdfstring{B.1 Identifiability of \(\alpha\) from
\(\eta\)}{B.1 Identifiability of \textbackslash alpha from \textbackslash eta}}\label{sec-b1}
\addcontentsline{toc}{subsection}{B.1 Identifiability of \(\alpha\) from
\(\eta\)}

\begin{tcolorbox}[enhanced jigsaw, rightrule=.15mm, breakable, colback=white, arc=.35mm, leftrule=.75mm, colframe=quarto-callout-note-color-frame, toprule=.15mm, bottomrule=.15mm, left=2mm, opacityback=0]
\begin{minipage}[t]{5.5mm}
\textcolor{quarto-callout-note-color}{\faInfo}
\end{minipage}%
\begin{minipage}[t]{\textwidth - 5.5mm}

\textbf{Proposition B.1.} Let \(\eta = (\eta_r)_{r \in \mathcal{R}}\) be
fixed. Suppose there is a menu \(\mathcal{M} \subseteq \mathcal{R}\) of
positive design probability and two alternatives
\(r, s \in \mathcal{M}\) with \(\eta_r \neq
\eta_s\). Then the map \(\alpha \mapsto P(\cdot \mid \alpha, \eta)\)
restricted to \(\mathcal{M}\) is injective on \((0, \infty)\); hence
\(\alpha\) is globally identifiable from choice probabilities on
\(\mathcal{M}\).

\end{minipage}%
\end{tcolorbox}

\emph{Proof.} On \(\mathcal{M}\) the softmax (Equation~\ref{eq-softmax})
gives, for the pair \(r, s\), \[
g(\alpha) \;:=\; \log\frac{P(r \mid \alpha, \eta)}{P(s \mid \alpha, \eta)}
  \;=\; \alpha\,(\eta_r - \eta_s).
\] Since \(\eta_r - \eta_s \neq 0\), \(g\) is strictly monotone (linear
with nonzero slope) in \(\alpha\), hence injective: distinct \(\alpha\)
values induce distinct log-odds, therefore distinct choice-probability
vectors on \(\mathcal{M}\). As the menu has positive design probability,
the choice distribution determines \(g(\alpha)\) and thus \(\alpha\).
\(\qquad\blacksquare\)

\emph{Remark (genericity).} The condition ``not all \(\eta_r\) equal on
some positive-probability menu'' replaces the vague ``sufficient feature
diversity'': it is exactly the condition under which softmax curvature
in \(\alpha\) is non-trivial. It fails only in the measure-zero event
that every presented menu has perfectly tied expected utilities.

\subsection*{\texorpdfstring{B.2 Weakly-informed \((\beta,\delta)\) in
\texttt{m\_0}: a workflow
note}{B.2 Weakly-informed (\textbackslash beta,\textbackslash delta) in m\_0: a workflow note}}\label{sec-b2}
\addcontentsline{toc}{subsection}{B.2 Weakly-informed \((\beta,\delta)\)
in \texttt{m\_0}: a workflow note}

This is not a proposition but a record of what the Bayesian workflow
shows about \((\beta,\delta)\) in \texttt{m\_0}. The structural reason
they are hard to separate is that expected utility composes beliefs and
utilities multiplicatively, \[
\eta_r \;=\; \mathrm{softmax}(\beta\, w_r)^\top \bm{\upsilon}(\delta),
\] so uncertain choices see \((\beta,\delta)\) only through the scalar
\(\eta_r\). A change in beliefs offset by a compensating change in
utilities leaves the implied expected utilities --- and hence the choice
probabilities --- almost unchanged. The \(\bm{\upsilon}\)-endpoint
convention (§\hyperref[sec-notation]{2.6}) fixes the zero and unit of
the utility scale, but the multiplicative \((\beta,\delta)\) coupling is
not an indeterminacy that any convention removes.

\textbf{What recovery shows.} Parameter recovery in \texttt{m\_0}
(§\hyperref[sec-m0-recovery]{4.3}) returns wide marginal credible
intervals for both \(\beta\) and \(\delta\) --- in sharp contrast to the
tight, well-calibrated \(\alpha\) intervals --- together with
\emph{correlated \(\beta\)--\(\delta\) estimation errors} across
recovery replicates: the posterior concentrates on a trade-off between
beliefs and utilities rather than on either separately. That correlated,
barely-contracted joint posterior is the practical signature of the
multiplicative coupling and is what we mean by calling
\((\beta,\delta)\) \emph{weakly informed} by uncertain choices. No
strict non-identifiability is claimed, and no gauge group on
\((\beta,\delta)\) beyond the \(\beta\) row-shift of
§\hyperref[sec-notation]{2.6} is asserted; the point is that
uncertain-choice data alone supply little information for separating
beliefs from utilities. The matched-design study of
§\hyperref[sec-m1-recovery]{6.4} quantifies how little the risky block
adds to this separation at realistic \(n\).

\subsection*{\texorpdfstring{B.3 Global identifiability of \(\delta\) in
\texttt{m\_1}}{B.3 Global identifiability of \textbackslash delta in m\_1}}\label{sec-b3}
\addcontentsline{toc}{subsection}{B.3 Global identifiability of
\(\delta\) in \texttt{m\_1}}

\begin{tcolorbox}[enhanced jigsaw, rightrule=.15mm, breakable, colback=white, arc=.35mm, leftrule=.75mm, colframe=quarto-callout-note-color-frame, toprule=.15mm, bottomrule=.15mm, left=2mm, opacityback=0]
\begin{minipage}[t]{5.5mm}
\textcolor{quarto-callout-note-color}{\faInfo}
\end{minipage}%
\begin{minipage}[t]{\textwidth - 5.5mm}

\textbf{Proposition B.3.} Let the risky problems present lotteries
\(\{\pi_s\}_{s=1}^{S} \subset \Delta^{K-1}\) whose differences span the
\((K-1)\)-dimensional tangent space of the simplex --- equivalently, the
set contains an affinely independent subset of \(K\) lotteries (which
requires \(S \geq K\) and a generic configuration) --- and let the
menu-contrast graph be connected. Then, with \(\alpha > 0\) known to be
positive and the endpoint convention \(\upsilon_1 = 0\),
\(\upsilon_K = 1\) in force, the utility vector \(\bm{\upsilon}\) ---
and hence \(\delta\) --- together with \(\alpha\) is globally identified
from risky-choice probabilities alone.

\end{minipage}%
\end{tcolorbox}

\emph{Proof.} For each risky menu and each co-menu pair \(s, s'\), the
softmax (Equation~\ref{eq-softmax}) gives the log-odds \[
\log\frac{P(s)}{P(s')} = \alpha\,(\pi_s - \pi_{s'})^\top \bm{\upsilon}.
\] The choice probabilities therefore identify the products
\(c_s := \alpha\,(\pi_s - \pi_1)^\top \bm{\upsilon}\) for every lottery
\(s\) sharing a menu with a common reference \(\pi_1\), and --- chaining
across the connected menu-contrast graph --- for all
\(s = 2, \dots, S\). We do \emph{not} invoke Proposition B.1 here:
\(\alpha\) and \(\bm{\upsilon}\) are identified jointly, as follows.
Because \(\alpha > 0\), the identified vector
\(c = \big(\alpha\,(\pi_s - \pi_1)^\top \bm{\upsilon}\big)_{s}\)
determines the linear functionals \((\pi_s - \pi_1)^\top \bm{\upsilon}\)
up to the single common positive factor \(\alpha\) --- no ratios are
needed (and ratios would discard sign information and be undefined
wherever a denominator vanishes). Because the differences
\(\{\pi_s - \pi_1\}\) span the \((K-1)\)-dimensional tangent space,
these functionals determine \(\bm{\upsilon}\) up to a positive affine
transformation; the endpoint convention \(\upsilon_1 = 0\),
\(\upsilon_K = 1\) fixes that transformation, so \(\bm{\upsilon}\) is
identified uniquely. Finally
\(\alpha = c_s / (\pi_s - \pi_1)^\top \bm{\upsilon}\) for any \(s\) with
non-zero denominator (one exists, as \(\bm{\upsilon}\) is non-constant),
identifying \(\alpha\). The cumulative-sum map
\(\delta \mapsto \bm{\upsilon}\) being a bijection onto the
ordered-utility set, \(\delta\) is identified. \(\qquad\blacksquare\)

\emph{Remark.} The substantive hypothesis is a design condition the
experimenter controls: the presented lotteries' differences must span
the simplex tangent space (equivalently, the design contains an affinely
independent subset of \(K\) lotteries), and the menus must be linked
into a connected contrast graph. Affine independence of \emph{all} \(S\)
lotteries is neither required nor possible when \(S > K\) (here
\(S = 15\), \(K = 3\)); what is needed is a spanning affinely
independent \emph{subset}.

\subsection*{\texorpdfstring{B.4 Uncertain choices sharpen utilities;
\(\beta\) is constrained only through
\(\eta\)}{B.4 Uncertain choices sharpen utilities; \textbackslash beta is constrained only through \textbackslash eta}}\label{sec-b4}
\addcontentsline{toc}{subsection}{B.4 Uncertain choices sharpen
utilities; \(\beta\) is constrained only through \(\eta\)}

\begin{tcolorbox}[enhanced jigsaw, rightrule=.15mm, breakable, colback=white, arc=.35mm, leftrule=.75mm, colframe=quarto-callout-note-color-frame, toprule=.15mm, bottomrule=.15mm, left=2mm, opacityback=0]
\begin{minipage}[t]{5.5mm}
\textcolor{quarto-callout-note-color}{\faInfo}
\end{minipage}%
\begin{minipage}[t]{\textwidth - 5.5mm}

\textbf{Proposition B.4.} Suppose \(\delta\) (hence \(\bm{\upsilon}\))
is identified by B.3 and \(\alpha > 0\) by B.1. Then the
uncertain-choice probabilities identify the expected utilities
\(\eta_r = \bm{\psi}_r^\top \bm{\upsilon}\) up to a per-menu additive
constant, and constrain \(\beta\) only through the composite map
\(\beta \mapsto \big(\mathrm{softmax}(\beta w_r)^\top \bm{\upsilon}\big)_r\).
For \(K \geq 3\) a scalar expected utility does not determine the
\((K-1)\)-dimensional belief simplex \(\bm{\psi}_r\), so whether
\(\beta\) is identified (modulo the additive row-shift gauge of
§\hyperref[sec-notation]{2.6}) turns entirely on a design-specific rank
condition on this composite map: local identification requires its
gauge-fixed Jacobian to attain rank \((K-1)D\), which is impossible
whenever \((K-1)D\) exceeds the number of independent \(\eta\)-contrasts
the design supplies (at most \(R-1\)). Where the condition fails,
\(\beta\) is an irreducible nuisance; the contribution of the risky
block (B.3) is in either case to sharpen \(\bm{\upsilon}\), not to
recover \(\beta\).

\end{minipage}%
\end{tcolorbox}

\emph{Proof.} With \(\alpha\) and \(\bm{\upsilon}\) fixed, Proposition
B.1 applied within each menu shows the choice probabilities determine
the within-menu contrasts of
\(\eta_r = \bm{\psi}_r^\top \bm{\upsilon}\), hence the \(\eta_r\) up to
a common per-menu constant. Each such value is a \emph{single} linear
equation in the \((K-1)\) free coordinates of \(\bm{\psi}_r\); for
\(K \geq 3\) the level set
\(\{\bm{\psi} : \bm{\psi}^\top \bm{\upsilon} = \eta_r\}\) is
\((K-2)\)-dimensional whenever \(\bm{\upsilon}\) is non-constant, so
\(\bm{\psi}_r\) is not recovered. The map
\(\bm{\psi} \mapsto \bm{\psi}^\top \bm{\upsilon}\) is therefore not
injective, and \(\beta\) enters the likelihood only through the
composite \(\beta \mapsto (\eta_r)_r\). Any two weight matrices inducing
the same expected utilities on every presented menu --- including those
related by the row-shift gauge, for which
\(\mathrm{softmax}(\beta w_r)\) itself is unchanged --- are
observationally indistinguishable. Point identification of \(\beta\)
would require the composite map to be injective modulo the gauge under
the design, an additional rank condition the scalar expected utilities
do not supply. \(\qquad\blacksquare\)

\emph{Remark (what is and is not claimed).} The proposition asserts that
identification of \(\beta\) is governed by a design-specific rank
condition, not a structural impossibility or a structural guarantee.
Crude dimension counts do not settle the matter: a nominal inequality
such as \(R \geq (K-1)D\) is necessary but not sufficient, since it
ignores the rank of the induced Jacobian. §\hyperref[sec-m1-beta]{5.6}
evaluates the condition numerically for the paper's two design regimes:
at the foundational design the gauge-fixed contrast-Jacobian rank equals
\((K-1)D = 10\), so \(\beta\) is locally identified modulo the gauge
(though weakly informed at finite \(n\)), while at the application
designs \((K-1)D = 64\) or \(96\) exceeds the maximal contrast count
\(R - 1 = 29\), so \(\beta\) is genuinely not identified there. The
narrow claim we rely on throughout is that each uncertain choice
supplies only the \emph{scalar} \(\eta_r\), so all information about
\(\beta\) flows through the \(\eta\)-contrasts.

\emph{Remark (how the rank is computed).} The gauge-fixed
contrast-Jacobian of \S\hyperref[sec-m1-beta]{5.6} is formed as follows.
Fixing the row-shift gauge by working with mean-centered utilities,
collect the design's independent \(\eta\)-contrasts into a single
vector-valued function of the \((K-1)D\) free entries of \(\beta\), and
evaluate the matrix of partial derivatives
\(\partial(\eta\text{-contrast})_j / \partial\beta_{k,d}\) at a given
\(\beta\) by automatic differentiation. Its numerical rank --- the
number of singular values above a small tolerance --- is the quantity
reported in \S\hyperref[sec-m1-beta]{5.6}; evaluating it at several
draws of \(\beta\) from the prior, rather than at a single point, guards
against mistaking a non-generic coincidence for identification
(\texttt{spikes/report\_design\_diagnostics\_spike.py}).

\section*{Appendix C. Stan Model Listings}\label{sec-appendix-c}
\addcontentsline{toc}{section}{Appendix C. Stan Model Listings}

\begin{quote}
\emph{Full listings of the Stan programs referenced by the brief code
excerpts in §\hyperref[sec-m0-implementation]{4} and
§\hyperref[sec-m1-implementation]{6}. Two model pairs are used
throughout. For each model there is an \textbf{inference} program
(\texttt{.stan}) that takes observed choices as data and samples the
posterior, and a matched \textbf{simulation} program
(\texttt{\_sim.stan}) that draws parameters from the prior and generates
synthetic choices in \texttt{generated\ quantities} --- the
data-generating half of every prior-predictive, parameter-recovery, and
SBC run. The calibrated-prior variant \texttt{m\_01} used in the
application (§\hyperref[sec-application]{7}) is structurally identical
to \texttt{m\_0} and is given as a one-line prior diff. Listings reflect
the pinned commit of Appendix E.0; the canonical files live at
\texttt{models/} in the repository.}
\end{quote}

The notation matches §\hyperref[sec-notation]{2.6}: \texttt{alpha} is
the sensitivity \(\alpha\), \texttt{beta} the belief-feature matrix
\(\beta\), \texttt{delta} the utility increments \(\delta\) (a
\texttt{simplex}, so \texttt{upsilon} \(= \bm{\upsilon}\) is ordered
with the endpoint convention \(\upsilon_1 = 0\) built in via
\texttt{cumulative\_sum(append\_row(0,\ delta))}), \texttt{psi} the
per-alternative belief simplices \(\bm{\psi}_r\), \texttt{eta} the
expected utilities \(\eta_r\), and \texttt{chi} the per-problem choice
probabilities. The \texttt{I} (uncertain) and \texttt{J} (risky)
indicator arrays encode which alternatives appear in each problem;
choices are \texttt{y} (uncertain) and \texttt{z} (risky).

\subsection*{\texorpdfstring{C.1 \texttt{m\_0.stan} ---
uncertain-choice-only inference
model}{C.1 m\_0.stan --- uncertain-choice-only inference model}}\label{sec-c-m0}
\addcontentsline{toc}{subsection}{C.1 \texttt{m\_0.stan} ---
uncertain-choice-only inference model}

\begin{Shaded}
\begin{Highlighting}[]
\KeywordTok{data}\NormalTok{\{}
  \DataTypeTok{int}\NormalTok{\textless{}}\KeywordTok{lower}\NormalTok{=}\DecValTok{1}\NormalTok{\textgreater{} M; }\CommentTok{// the number of decision problems}
  \DataTypeTok{int}\NormalTok{\textless{}}\KeywordTok{lower}\NormalTok{=}\DecValTok{2}\NormalTok{\textgreater{} K; }\CommentTok{// the number of possible consequences}
  \DataTypeTok{int}\NormalTok{\textless{}}\KeywordTok{lower}\NormalTok{=}\DecValTok{1}\NormalTok{\textgreater{} D; }\CommentTok{// the number of dimensions to describe an alternative}
  \DataTypeTok{int}\NormalTok{\textless{}}\KeywordTok{lower}\NormalTok{=}\DecValTok{2}\NormalTok{\textgreater{} R; }\CommentTok{// the number of distinct alternatives}
  \DataTypeTok{array}\NormalTok{[R] }\DataTypeTok{vector}\NormalTok{[D] w; }\CommentTok{// the descriptions of each distinct alternative}
  \DataTypeTok{array}\NormalTok{[M,R] }\DataTypeTok{int}\NormalTok{\textless{}}\KeywordTok{lower}\NormalTok{=}\DecValTok{0}\NormalTok{,}\KeywordTok{upper}\NormalTok{=}\DecValTok{1}\NormalTok{\textgreater{} I; }\CommentTok{// I[m,r] = 1 if alternative r is in problem m}
  \DataTypeTok{array}\NormalTok{[M] }\DataTypeTok{int}\NormalTok{\textless{}}\KeywordTok{lower}\NormalTok{=}\DecValTok{1}\NormalTok{\textgreater{} y; }\CommentTok{// selected alternative, as a LOCAL index into problem m\textquotesingle{}s}
                           \CommentTok{// menu: y[m] in 1..N[m] gives the position (in ascending}
                           \CommentTok{// global{-}r order, matching the compacted x rows) of the}
                           \CommentTok{// chosen alternative, NOT a global index in 1..R}
                           \CommentTok{// (bounds checked in transformed data)}
\NormalTok{\}}
\KeywordTok{transformed data}\NormalTok{ \{}
  \DataTypeTok{array}\NormalTok{[M] }\DataTypeTok{int}\NormalTok{\textless{}}\KeywordTok{lower}\NormalTok{=}\DecValTok{2}\NormalTok{\textgreater{} N; }\CommentTok{// the number of alternatives in each decision problem}
  \DataTypeTok{int}\NormalTok{ total\_alternatives = }\DecValTok{0}\NormalTok{;}
  \ControlFlowTok{for}\NormalTok{ (m }\ControlFlowTok{in} \DecValTok{1}\NormalTok{:M) \{}
\NormalTok{    N[m] = sum(I[m]);}
\NormalTok{    total\_alternatives += N[m];}
    \ControlFlowTok{if}\NormalTok{ (y[m] \textgreater{} N[m]) }\KeywordTok{reject}\NormalTok{(}\StringTok{"y["}\NormalTok{, m, }\StringTok{"] = "}\NormalTok{, y[m], }\StringTok{" must be \textless{}= N["}\NormalTok{, m, }\StringTok{"] = "}\NormalTok{, N[m]);}
\NormalTok{  \}}
  \CommentTok{// Construct x from w and I}
  \DataTypeTok{array}\NormalTok{[total\_alternatives] }\DataTypeTok{vector}\NormalTok{[D] x;}
\NormalTok{  \{}
    \DataTypeTok{int}\NormalTok{ pos = }\DecValTok{1}\NormalTok{;}
    \ControlFlowTok{for}\NormalTok{ (m }\ControlFlowTok{in} \DecValTok{1}\NormalTok{:M) \{}
      \ControlFlowTok{for}\NormalTok{ (r }\ControlFlowTok{in} \DecValTok{1}\NormalTok{:R) \{}
        \ControlFlowTok{if}\NormalTok{ (I[m, r] == }\DecValTok{1}\NormalTok{) \{ x[pos] = w[r]; pos += }\DecValTok{1}\NormalTok{; \}}
\NormalTok{      \}}
\NormalTok{    \}}
\NormalTok{  \}}
\NormalTok{\}}
\KeywordTok{parameters}\NormalTok{\{}
  \DataTypeTok{real}\NormalTok{\textless{}}\KeywordTok{lower}\NormalTok{=}\DecValTok{0}\NormalTok{\textgreater{} alpha; }\CommentTok{// sensitivity parameter (decision noise); strictly positive}
  \DataTypeTok{matrix}\NormalTok{[K,D] beta;    }\CommentTok{// coefficients mapping descriptions to subjective probabilities}
  \DataTypeTok{simplex}\NormalTok{[K {-} }\DecValTok{1}\NormalTok{] delta;}\CommentTok{// utility differences on unit scale (ensures ordered utilities)}
\NormalTok{\}}
\KeywordTok{transformed parameters}\NormalTok{\{}
  \DataTypeTok{array}\NormalTok{[sum(N)] }\DataTypeTok{simplex}\NormalTok{[K] psi; }\CommentTok{// subjective probability over outcomes per alternative}
  \DataTypeTok{ordered}\NormalTok{[K] upsilon;           }\CommentTok{// the subjective utility of each consequence}
  \DataTypeTok{vector}\NormalTok{[sum(N)] eta;           }\CommentTok{// expected utility of each alternative on unit scale}
  \DataTypeTok{array}\NormalTok{[M] }\DataTypeTok{simplex}\NormalTok{[max(N)] chi; }\CommentTok{// choice probs with padding}
  \ControlFlowTok{for}\NormalTok{(i }\ControlFlowTok{in} \DecValTok{1}\NormalTok{:sum(N))\{ psi[i] = softmax(beta*x[i]); \}}
\NormalTok{  upsilon = cumulative\_sum(append\_row(}\DecValTok{0}\NormalTok{, delta));}
  \ControlFlowTok{for}\NormalTok{(i }\ControlFlowTok{in} \DecValTok{1}\NormalTok{:sum(N))\{ eta[i] = dot\_product(psi[i],upsilon); \}}
\NormalTok{  \{}
    \DataTypeTok{int}\NormalTok{ pos = }\DecValTok{1}\NormalTok{;}
    \ControlFlowTok{for}\NormalTok{(i }\ControlFlowTok{in} \DecValTok{1}\NormalTok{:M)\{}
      \DataTypeTok{vector}\NormalTok{[N[i]] problem\_eta = segment(eta, pos, N[i]);}
\NormalTok{      chi[i] = append\_row(softmax(alpha * problem\_eta), rep\_vector(}\DecValTok{0}\NormalTok{, max(N) {-} N[i]));}
\NormalTok{      pos += N[i];}
\NormalTok{    \}}
\NormalTok{  \}}
\NormalTok{\}}
\KeywordTok{model}\NormalTok{\{}
\NormalTok{  alpha \textasciitilde{} lognormal(}\DecValTok{0}\NormalTok{, }\DecValTok{1}\NormalTok{);              }\CommentTok{// prior on choice sensitivity}
\NormalTok{  to\_vector(beta) \textasciitilde{} std\_normal();        }\CommentTok{// prior on belief{-}feature coefficients}
\NormalTok{  delta \textasciitilde{} dirichlet(rep\_vector(}\DecValTok{1}\NormalTok{,K}\DecValTok{{-}1}\NormalTok{));  }\CommentTok{// prior on ordered unit{-}scale utility increments}
  \ControlFlowTok{for}\NormalTok{(i }\ControlFlowTok{in} \DecValTok{1}\NormalTok{:M)\{ y[i] \textasciitilde{} categorical(chi[i]); \}}
\NormalTok{\}}
\KeywordTok{generated quantities}\NormalTok{ \{}
  \DataTypeTok{vector}\NormalTok{[M] log\_lik;}
  \ControlFlowTok{for}\NormalTok{ (i }\ControlFlowTok{in} \DecValTok{1}\NormalTok{:M) \{ log\_lik[i] = categorical\_lpmf(y[i] | chi[i]); \}}
  \DataTypeTok{array}\NormalTok{[M] }\DataTypeTok{int}\NormalTok{ y\_pred;}
  \ControlFlowTok{for}\NormalTok{ (i }\ControlFlowTok{in} \DecValTok{1}\NormalTok{:M) \{ y\_pred[i] = categorical\_rng(chi[i]); \}}

  \CommentTok{// Posterior{-}predictive check statistics (one{-}sided Bayesian p{-}values):}
  \CommentTok{// log{-}likelihood discrepancy, modal{-}choice accuracy, and sum of chosen}
  \CommentTok{// probabilities, each comparing T(y\_pred) to T(y\_obs).}
  \DataTypeTok{real}\NormalTok{ T\_obs\_ll = sum(log\_lik);}
  \DataTypeTok{real}\NormalTok{ T\_rep\_ll = }\DecValTok{0}\NormalTok{;}
  \ControlFlowTok{for}\NormalTok{ (m }\ControlFlowTok{in} \DecValTok{1}\NormalTok{:M) \{ T\_rep\_ll += categorical\_lpmf(y\_pred[m] | chi[m]); \}}
  \DataTypeTok{int}\NormalTok{\textless{}}\KeywordTok{lower}\NormalTok{=}\DecValTok{0}\NormalTok{,}\KeywordTok{upper}\NormalTok{=}\DecValTok{1}\NormalTok{\textgreater{} ppc\_ll = (T\_rep\_ll \textgreater{}= T\_obs\_ll) ? }\DecValTok{1}\NormalTok{ : }\DecValTok{0}\NormalTok{;}

  \DataTypeTok{int}\NormalTok{ T\_obs\_modal = }\DecValTok{0}\NormalTok{; }\DataTypeTok{int}\NormalTok{ T\_rep\_modal = }\DecValTok{0}\NormalTok{;}
  \ControlFlowTok{for}\NormalTok{ (m }\ControlFlowTok{in} \DecValTok{1}\NormalTok{:M) \{}
    \DataTypeTok{real}\NormalTok{ max\_prob = chi[m][}\DecValTok{1}\NormalTok{];}
    \ControlFlowTok{for}\NormalTok{ (j }\ControlFlowTok{in} \DecValTok{2}\NormalTok{:N[m]) \{ }\ControlFlowTok{if}\NormalTok{ (chi[m][j] \textgreater{} max\_prob) max\_prob = chi[m][j]; \}}
\NormalTok{    T\_obs\_modal += (chi[m][y[m]]      \textgreater{}= max\_prob {-} }\FloatTok{1e{-}9}\NormalTok{) ? }\DecValTok{1}\NormalTok{ : }\DecValTok{0}\NormalTok{;}
\NormalTok{    T\_rep\_modal += (chi[m][y\_pred[m]] \textgreater{}= max\_prob {-} }\FloatTok{1e{-}9}\NormalTok{) ? }\DecValTok{1}\NormalTok{ : }\DecValTok{0}\NormalTok{;}
\NormalTok{  \}}
  \DataTypeTok{int}\NormalTok{\textless{}}\KeywordTok{lower}\NormalTok{=}\DecValTok{0}\NormalTok{,}\KeywordTok{upper}\NormalTok{=}\DecValTok{1}\NormalTok{\textgreater{} ppc\_modal = (T\_rep\_modal \textgreater{}= T\_obs\_modal) ? }\DecValTok{1}\NormalTok{ : }\DecValTok{0}\NormalTok{;}

  \DataTypeTok{real}\NormalTok{ T\_obs\_prob = }\DecValTok{0}\NormalTok{; }\DataTypeTok{real}\NormalTok{ T\_rep\_prob = }\DecValTok{0}\NormalTok{;}
  \ControlFlowTok{for}\NormalTok{ (m }\ControlFlowTok{in} \DecValTok{1}\NormalTok{:M) \{ T\_obs\_prob += chi[m][y[m]]; T\_rep\_prob += chi[m][y\_pred[m]]; \}}
  \DataTypeTok{int}\NormalTok{\textless{}}\KeywordTok{lower}\NormalTok{=}\DecValTok{0}\NormalTok{,}\KeywordTok{upper}\NormalTok{=}\DecValTok{1}\NormalTok{\textgreater{} ppc\_prob = (T\_rep\_prob \textgreater{}= T\_obs\_prob) ? }\DecValTok{1}\NormalTok{ : }\DecValTok{0}\NormalTok{;}
\NormalTok{\}}
\end{Highlighting}
\end{Shaded}

\subsection*{\texorpdfstring{C.2 \texttt{m\_0\_sim.stan} ---
uncertain-choice data
generator}{C.2 m\_0\_sim.stan --- uncertain-choice data generator}}\label{sec-c-m0-sim}
\addcontentsline{toc}{subsection}{C.2 \texttt{m\_0\_sim.stan} ---
uncertain-choice data generator}

The simulation program shares \texttt{m\_0}'s
\texttt{data}/\texttt{transformed\ data} shape but draws
\((\alpha, \beta, \delta)\) from the prior inside
\texttt{generated\ quantities} (with the \(\alpha\)-prior
hyperparameters passed as data, so a single program serves both the
\texttt{Lognormal(0,1)} default and any calibrated prior), simulates
choices \texttt{y}, and reports \texttt{total\_seu\_max\_selected} ---
the count of problems in which the sampled choice coincided with the
SEU-maximizing alternative, the prior-predictive diagnostic of
§\hyperref[sec-m0-prior]{4.2}.

\begin{Shaded}
\begin{Highlighting}[]
\KeywordTok{data}\NormalTok{ \{}
  \DataTypeTok{int}\NormalTok{\textless{}}\KeywordTok{lower}\NormalTok{=}\DecValTok{1}\NormalTok{\textgreater{} M; }\DataTypeTok{int}\NormalTok{\textless{}}\KeywordTok{lower}\NormalTok{=}\DecValTok{2}\NormalTok{\textgreater{} K; }\DataTypeTok{int}\NormalTok{\textless{}}\KeywordTok{lower}\NormalTok{=}\DecValTok{1}\NormalTok{\textgreater{} D; }\DataTypeTok{int}\NormalTok{\textless{}}\KeywordTok{lower}\NormalTok{=}\DecValTok{2}\NormalTok{\textgreater{} R;}
  \DataTypeTok{array}\NormalTok{[R] }\DataTypeTok{vector}\NormalTok{[D] w;}
  \DataTypeTok{array}\NormalTok{[M,R] }\DataTypeTok{int}\NormalTok{\textless{}}\KeywordTok{lower}\NormalTok{=}\DecValTok{0}\NormalTok{,}\KeywordTok{upper}\NormalTok{=}\DecValTok{1}\NormalTok{\textgreater{} I;}
  \DataTypeTok{real}\NormalTok{ alpha\_mean;          }\CommentTok{// mean for log(alpha) (default 0); may be negative}
  \DataTypeTok{real}\NormalTok{\textless{}}\KeywordTok{lower}\NormalTok{=}\DecValTok{0}\NormalTok{\textgreater{} alpha\_sd;   }\CommentTok{// sd for log(alpha)   (default 1)}
  \DataTypeTok{real}\NormalTok{\textless{}}\KeywordTok{lower}\NormalTok{=}\DecValTok{0}\NormalTok{\textgreater{} beta\_sd;    }\CommentTok{// sd for beta coefficients (default 1)}
\NormalTok{\}}
\KeywordTok{transformed data}\NormalTok{ \{}
  \DataTypeTok{array}\NormalTok{[M] }\DataTypeTok{int}\NormalTok{\textless{}}\KeywordTok{lower}\NormalTok{=}\DecValTok{2}\NormalTok{\textgreater{} N; }\DataTypeTok{int}\NormalTok{ total\_alts = }\DecValTok{0}\NormalTok{;}
  \ControlFlowTok{for}\NormalTok{ (m }\ControlFlowTok{in} \DecValTok{1}\NormalTok{:M) \{ N[m] = sum(I[m]); total\_alts += N[m]; \}}
  \DataTypeTok{array}\NormalTok{[total\_alts] }\DataTypeTok{vector}\NormalTok{[D] x;}
\NormalTok{  \{}
    \DataTypeTok{int}\NormalTok{ pos = }\DecValTok{1}\NormalTok{;}
    \ControlFlowTok{for}\NormalTok{ (m }\ControlFlowTok{in} \DecValTok{1}\NormalTok{:M) }\ControlFlowTok{for}\NormalTok{ (r }\ControlFlowTok{in} \DecValTok{1}\NormalTok{:R) }\ControlFlowTok{if}\NormalTok{ (I[m, r] == }\DecValTok{1}\NormalTok{) \{ x[pos] = w[r]; pos += }\DecValTok{1}\NormalTok{; \}}
\NormalTok{  \}}
\NormalTok{\}}
\KeywordTok{generated quantities}\NormalTok{ \{}
  \DataTypeTok{real}\NormalTok{ alpha = lognormal\_rng(alpha\_mean, alpha\_sd);}
  \DataTypeTok{matrix}\NormalTok{[K,D] beta;}
  \ControlFlowTok{for}\NormalTok{ (k }\ControlFlowTok{in} \DecValTok{1}\NormalTok{:K) }\ControlFlowTok{for}\NormalTok{ (d }\ControlFlowTok{in} \DecValTok{1}\NormalTok{:D) beta[k,d] = normal\_rng(}\DecValTok{0}\NormalTok{, beta\_sd);}
  \DataTypeTok{simplex}\NormalTok{[K}\DecValTok{{-}1}\NormalTok{] delta = dirichlet\_rng(rep\_vector(}\FloatTok{1.0}\NormalTok{, K}\DecValTok{{-}1}\NormalTok{));}

  \DataTypeTok{array}\NormalTok{[total\_alts] }\DataTypeTok{vector}\NormalTok{[K] psi;}
  \ControlFlowTok{for}\NormalTok{ (i }\ControlFlowTok{in} \DecValTok{1}\NormalTok{:total\_alts) psi[i] = softmax(beta * x[i]);}
  \DataTypeTok{vector}\NormalTok{[K] upsilon = cumulative\_sum(append\_row(}\DecValTok{0}\NormalTok{, delta));}
  \DataTypeTok{vector}\NormalTok{[total\_alts] eta;}
  \ControlFlowTok{for}\NormalTok{ (i }\ControlFlowTok{in} \DecValTok{1}\NormalTok{:total\_alts) eta[i] = dot\_product(to\_vector(psi[i]), upsilon);}

  \DataTypeTok{array}\NormalTok{[M] }\DataTypeTok{int}\NormalTok{ y;}
  \DataTypeTok{array}\NormalTok{[M] }\DataTypeTok{int}\NormalTok{\textless{}}\KeywordTok{lower}\NormalTok{=}\DecValTok{0}\NormalTok{,}\KeywordTok{upper}\NormalTok{=}\DecValTok{1}\NormalTok{\textgreater{} selected\_seu\_max;}
\NormalTok{  \{}
    \DataTypeTok{int}\NormalTok{ pos = }\DecValTok{1}\NormalTok{;}
    \ControlFlowTok{for}\NormalTok{ (i }\ControlFlowTok{in} \DecValTok{1}\NormalTok{:M) \{}
      \DataTypeTok{vector}\NormalTok{[N[i]] problem\_eta = segment(eta, pos, N[i]);}
      \DataTypeTok{vector}\NormalTok{[N[i]] choice\_probs = softmax(alpha * problem\_eta);}
\NormalTok{      y[i] = categorical\_rng(choice\_probs);}
      \DataTypeTok{real}\NormalTok{ max\_eta = max(problem\_eta);}
\NormalTok{      selected\_seu\_max[i] = (abs(problem\_eta[y[i]] {-} max\_eta) \textless{} }\FloatTok{1e{-}10}\NormalTok{) ? }\DecValTok{1}\NormalTok{ : }\DecValTok{0}\NormalTok{;}
\NormalTok{      pos += N[i];}
\NormalTok{    \}}
\NormalTok{  \}}
  \DataTypeTok{int}\NormalTok{\textless{}}\KeywordTok{lower}\NormalTok{=}\DecValTok{0}\NormalTok{,}\KeywordTok{upper}\NormalTok{=M\textgreater{} total\_seu\_max\_selected = sum(selected\_seu\_max);}
\NormalTok{\}}
\end{Highlighting}
\end{Shaded}

\emph{(The full file additionally stores the per-problem
\texttt{problem\_etas} and \texttt{choice\_probabilities} for
inspection; those bookkeeping arrays are elided here.)}

\subsection*{\texorpdfstring{C.3 \texttt{m\_1.stan} --- combined
uncertain + risky inference
model}{C.3 m\_1.stan --- combined uncertain + risky inference model}}\label{sec-c-m1}
\addcontentsline{toc}{subsection}{C.3 \texttt{m\_1.stan} --- combined
uncertain + risky inference model}

\texttt{m\_1} adds a risky block: a second set of \texttt{N} problems
over \texttt{S} lotteries with \emph{objective} probability simplices
\texttt{x} supplied as data. The lotteries enter expected utility
directly (\texttt{eta\_risky\ =\ dot\_product(x\_risky,\ upsilon)}),
with no \(\beta\) --- this \(\beta\)-free channel is what identifies
\(\delta\) in principle (§\hyperref[sec-m1-identifiability]{5},
Proposition B.3). Crucially, \texttt{alpha} and the utility vector
\texttt{upsilon} are \textbf{shared} across the two blocks.

\begin{Shaded}
\begin{Highlighting}[]
\KeywordTok{data}\NormalTok{ \{}
  \CommentTok{// Uncertain problems}
  \DataTypeTok{int}\NormalTok{\textless{}}\KeywordTok{lower}\NormalTok{=}\DecValTok{1}\NormalTok{\textgreater{} M; }\DataTypeTok{int}\NormalTok{\textless{}}\KeywordTok{lower}\NormalTok{=}\DecValTok{2}\NormalTok{\textgreater{} K; }\DataTypeTok{int}\NormalTok{\textless{}}\KeywordTok{lower}\NormalTok{=}\DecValTok{1}\NormalTok{\textgreater{} D; }\DataTypeTok{int}\NormalTok{\textless{}}\KeywordTok{lower}\NormalTok{=}\DecValTok{2}\NormalTok{\textgreater{} R;}
  \DataTypeTok{array}\NormalTok{[R] }\DataTypeTok{vector}\NormalTok{[D] w;}
  \DataTypeTok{array}\NormalTok{[M,R] }\DataTypeTok{int}\NormalTok{\textless{}}\KeywordTok{lower}\NormalTok{=}\DecValTok{0}\NormalTok{,}\KeywordTok{upper}\NormalTok{=}\DecValTok{1}\NormalTok{\textgreater{} I;}
  \DataTypeTok{array}\NormalTok{[M] }\DataTypeTok{int}\NormalTok{\textless{}}\KeywordTok{lower}\NormalTok{=}\DecValTok{1}\NormalTok{\textgreater{} y;}
  \CommentTok{// Risky problems}
  \DataTypeTok{int}\NormalTok{\textless{}}\KeywordTok{lower}\NormalTok{=}\DecValTok{1}\NormalTok{\textgreater{} N; }\DataTypeTok{int}\NormalTok{\textless{}}\KeywordTok{lower}\NormalTok{=}\DecValTok{2}\NormalTok{\textgreater{} S;}
  \DataTypeTok{array}\NormalTok{[S] }\DataTypeTok{simplex}\NormalTok{[K] x;               }\CommentTok{// objective lottery simplices}
  \DataTypeTok{array}\NormalTok{[N,S] }\DataTypeTok{int}\NormalTok{\textless{}}\KeywordTok{lower}\NormalTok{=}\DecValTok{0}\NormalTok{,}\KeywordTok{upper}\NormalTok{=}\DecValTok{1}\NormalTok{\textgreater{} J;}
  \DataTypeTok{array}\NormalTok{[N] }\DataTypeTok{int}\NormalTok{\textless{}}\KeywordTok{lower}\NormalTok{=}\DecValTok{1}\NormalTok{\textgreater{} z;}
\NormalTok{\}}
\KeywordTok{transformed data}\NormalTok{ \{}
  \DataTypeTok{array}\NormalTok{[M] }\DataTypeTok{int}\NormalTok{\textless{}}\KeywordTok{lower}\NormalTok{=}\DecValTok{2}\NormalTok{\textgreater{} N\_uncertain; }\DataTypeTok{int}\NormalTok{ total\_uncertain\_alts = }\DecValTok{0}\NormalTok{;}
  \ControlFlowTok{for}\NormalTok{ (m }\ControlFlowTok{in} \DecValTok{1}\NormalTok{:M) \{}
\NormalTok{    N\_uncertain[m] = sum(I[m]); total\_uncertain\_alts += N\_uncertain[m];}
    \ControlFlowTok{if}\NormalTok{ (y[m] \textgreater{} N\_uncertain[m]) }\KeywordTok{reject}\NormalTok{(}\StringTok{"y["}\NormalTok{, m, }\StringTok{"] out of range"}\NormalTok{);}
\NormalTok{  \}}
  \DataTypeTok{array}\NormalTok{[N] }\DataTypeTok{int}\NormalTok{\textless{}}\KeywordTok{lower}\NormalTok{=}\DecValTok{2}\NormalTok{\textgreater{} N\_risky; }\DataTypeTok{int}\NormalTok{ total\_risky\_alts = }\DecValTok{0}\NormalTok{;}
  \ControlFlowTok{for}\NormalTok{ (n }\ControlFlowTok{in} \DecValTok{1}\NormalTok{:N) \{}
\NormalTok{    N\_risky[n] = sum(J[n]); total\_risky\_alts += N\_risky[n];}
    \ControlFlowTok{if}\NormalTok{ (z[n] \textgreater{} N\_risky[n]) }\KeywordTok{reject}\NormalTok{(}\StringTok{"z["}\NormalTok{, n, }\StringTok{"] out of range"}\NormalTok{);}
\NormalTok{  \}}
  \DataTypeTok{array}\NormalTok{[total\_uncertain\_alts] }\DataTypeTok{vector}\NormalTok{[D] x\_uncertain;}
\NormalTok{  \{}
    \DataTypeTok{int}\NormalTok{ pos = }\DecValTok{1}\NormalTok{;}
    \ControlFlowTok{for}\NormalTok{ (m }\ControlFlowTok{in} \DecValTok{1}\NormalTok{:M) }\ControlFlowTok{for}\NormalTok{ (r }\ControlFlowTok{in} \DecValTok{1}\NormalTok{:R) }\ControlFlowTok{if}\NormalTok{ (I[m,r]==}\DecValTok{1}\NormalTok{) \{ x\_uncertain[pos]=w[r]; pos+=}\DecValTok{1}\NormalTok{; \}}
\NormalTok{  \}}
  \DataTypeTok{array}\NormalTok{[total\_risky\_alts] }\DataTypeTok{simplex}\NormalTok{[K] x\_risky;}
\NormalTok{  \{}
    \DataTypeTok{int}\NormalTok{ pos = }\DecValTok{1}\NormalTok{;}
    \ControlFlowTok{for}\NormalTok{ (n }\ControlFlowTok{in} \DecValTok{1}\NormalTok{:N) }\ControlFlowTok{for}\NormalTok{ (s }\ControlFlowTok{in} \DecValTok{1}\NormalTok{:S) }\ControlFlowTok{if}\NormalTok{ (J[n,s]==}\DecValTok{1}\NormalTok{) \{ x\_risky[pos]=x[s]; pos+=}\DecValTok{1}\NormalTok{; \}}
\NormalTok{  \}}
\NormalTok{\}}
\KeywordTok{parameters}\NormalTok{ \{}
  \DataTypeTok{real}\NormalTok{\textless{}}\KeywordTok{lower}\NormalTok{=}\DecValTok{0}\NormalTok{\textgreater{} alpha;   }\CommentTok{// shared sensitivity}
  \DataTypeTok{matrix}\NormalTok{[K,D] beta;      }\CommentTok{// belief features (uncertain block only)}
  \DataTypeTok{simplex}\NormalTok{[K}\DecValTok{{-}1}\NormalTok{] delta;    }\CommentTok{// shared utility increments}
\NormalTok{\}}
\KeywordTok{transformed parameters}\NormalTok{ \{}
  \DataTypeTok{ordered}\NormalTok{[K] upsilon = cumulative\_sum(append\_row(}\DecValTok{0}\NormalTok{, delta)); }\CommentTok{// shared utilities}

  \CommentTok{// Uncertain block}
  \DataTypeTok{array}\NormalTok{[total\_uncertain\_alts] }\DataTypeTok{simplex}\NormalTok{[K] psi;}
  \DataTypeTok{vector}\NormalTok{[total\_uncertain\_alts] eta\_uncertain;}
  \DataTypeTok{array}\NormalTok{[M] }\DataTypeTok{simplex}\NormalTok{[max(N\_uncertain)] chi\_uncertain;}
  \ControlFlowTok{for}\NormalTok{ (i }\ControlFlowTok{in} \DecValTok{1}\NormalTok{:total\_uncertain\_alts) psi[i] = softmax(beta * x\_uncertain[i]);}
  \ControlFlowTok{for}\NormalTok{ (i }\ControlFlowTok{in} \DecValTok{1}\NormalTok{:total\_uncertain\_alts) eta\_uncertain[i] = dot\_product(psi[i], upsilon);}
\NormalTok{  \{}
    \DataTypeTok{int}\NormalTok{ pos = }\DecValTok{1}\NormalTok{;}
    \ControlFlowTok{for}\NormalTok{ (m }\ControlFlowTok{in} \DecValTok{1}\NormalTok{:M) \{}
      \DataTypeTok{vector}\NormalTok{[N\_uncertain[m]] pe = segment(eta\_uncertain, pos, N\_uncertain[m]);}
\NormalTok{      chi\_uncertain[m] = append\_row(softmax(alpha*pe),}
\NormalTok{                                    rep\_vector(}\DecValTok{0}\NormalTok{, max(N\_uncertain){-}N\_uncertain[m]));}
\NormalTok{      pos += N\_uncertain[m];}
\NormalTok{    \}}
\NormalTok{  \}}

  \CommentTok{// Risky block (beta{-}free: lotteries enter EU directly)}
  \DataTypeTok{vector}\NormalTok{[total\_risky\_alts] eta\_risky;}
  \DataTypeTok{array}\NormalTok{[N] }\DataTypeTok{simplex}\NormalTok{[max(N\_risky)] chi\_risky;}
  \ControlFlowTok{for}\NormalTok{ (i }\ControlFlowTok{in} \DecValTok{1}\NormalTok{:total\_risky\_alts) eta\_risky[i] = dot\_product(x\_risky[i], upsilon);}
\NormalTok{  \{}
    \DataTypeTok{int}\NormalTok{ pos = }\DecValTok{1}\NormalTok{;}
    \ControlFlowTok{for}\NormalTok{ (n }\ControlFlowTok{in} \DecValTok{1}\NormalTok{:N) \{}
      \DataTypeTok{vector}\NormalTok{[N\_risky[n]] pe = segment(eta\_risky, pos, N\_risky[n]);}
\NormalTok{      chi\_risky[n] = append\_row(softmax(alpha*pe),}
\NormalTok{                                rep\_vector(}\DecValTok{0}\NormalTok{, max(N\_risky){-}N\_risky[n]));}
\NormalTok{      pos += N\_risky[n];}
\NormalTok{    \}}
\NormalTok{  \}}
\NormalTok{\}}
\KeywordTok{model}\NormalTok{ \{}
\NormalTok{  alpha \textasciitilde{} lognormal(}\DecValTok{0}\NormalTok{, }\DecValTok{1}\NormalTok{);}
\NormalTok{  to\_vector(beta) \textasciitilde{} std\_normal();}
\NormalTok{  delta \textasciitilde{} dirichlet(rep\_vector(}\DecValTok{1}\NormalTok{, K}\DecValTok{{-}1}\NormalTok{));}
  \ControlFlowTok{for}\NormalTok{ (m }\ControlFlowTok{in} \DecValTok{1}\NormalTok{:M) y[m] \textasciitilde{} categorical(chi\_uncertain[m]);}
  \ControlFlowTok{for}\NormalTok{ (n }\ControlFlowTok{in} \DecValTok{1}\NormalTok{:N) z[n] \textasciitilde{} categorical(chi\_risky[n]);}
\NormalTok{\}}
\KeywordTok{generated quantities}\NormalTok{ \{}
  \DataTypeTok{vector}\NormalTok{[M] log\_lik\_uncertain;}
  \DataTypeTok{vector}\NormalTok{[N] log\_lik\_risky;}
  \ControlFlowTok{for}\NormalTok{ (m }\ControlFlowTok{in} \DecValTok{1}\NormalTok{:M) log\_lik\_uncertain[m] = categorical\_lpmf(y[m] | chi\_uncertain[m]);}
  \ControlFlowTok{for}\NormalTok{ (n }\ControlFlowTok{in} \DecValTok{1}\NormalTok{:N) log\_lik\_risky[n]     = categorical\_lpmf(z[n] | chi\_risky[n]);}
  \DataTypeTok{real}\NormalTok{ log\_lik\_total = sum(log\_lik\_uncertain) + sum(log\_lik\_risky);}
  \DataTypeTok{array}\NormalTok{[M] }\DataTypeTok{int}\NormalTok{ y\_pred; }\DataTypeTok{array}\NormalTok{[N] }\DataTypeTok{int}\NormalTok{ z\_pred;}
  \ControlFlowTok{for}\NormalTok{ (m }\ControlFlowTok{in} \DecValTok{1}\NormalTok{:M) y\_pred[m] = categorical\_rng(chi\_uncertain[m]);}
  \ControlFlowTok{for}\NormalTok{ (n }\ControlFlowTok{in} \DecValTok{1}\NormalTok{:N) z\_pred[n] = categorical\_rng(chi\_risky[n]);}
  \CommentTok{// Elided for space: the PPC indicator statistics. m\_1.stan computes, per}
  \CommentTok{// posterior draw, replicate{-}vs{-}observed indicators ppc\_ll\_*, ppc\_modal\_*,}
  \CommentTok{// ppc\_prob\_* (log{-}likelihood, modal{-}choice frequency, mean predicted}
  \CommentTok{// probability of the chosen alternative) built exactly as in m\_0.stan (C.1),}
  \CommentTok{// once for the uncertain block, once for the risky block, and combined}
  \CommentTok{// (e.g. ppc\_ll\_combined from the summed block log{-}likelihoods) {-}{-} seven}
  \CommentTok{// indicators in all; see models/m\_1.stan at the E.0 pin for the verbatim block.}
\NormalTok{\}}
\end{Highlighting}
\end{Shaded}

\subsection*{\texorpdfstring{C.4 \texttt{m\_1\_sim.stan} --- combined
data
generator}{C.4 m\_1\_sim.stan --- combined data generator}}\label{sec-c-m1-sim}
\addcontentsline{toc}{subsection}{C.4 \texttt{m\_1\_sim.stan} ---
combined data generator}

The matched simulator draws shared \((\alpha, \delta)\) and
uncertain-only \(\beta\) from the prior, then generates both \texttt{y}
and \texttt{z}. The four matched-design conditions of
§\hyperref[sec-m1-matched]{6.2} (Appendix D.4) are produced by slicing
the choices this program generates --- the same simulated parameters
feed every condition, which is what makes the A/B/C/D comparison a
\emph{matched} one.

\begin{Shaded}
\begin{Highlighting}[]
\KeywordTok{data}\NormalTok{ \{}
  \DataTypeTok{int}\NormalTok{\textless{}}\KeywordTok{lower}\NormalTok{=}\DecValTok{1}\NormalTok{\textgreater{} M; }\DataTypeTok{int}\NormalTok{\textless{}}\KeywordTok{lower}\NormalTok{=}\DecValTok{2}\NormalTok{\textgreater{} K; }\DataTypeTok{int}\NormalTok{\textless{}}\KeywordTok{lower}\NormalTok{=}\DecValTok{1}\NormalTok{\textgreater{} D; }\DataTypeTok{int}\NormalTok{\textless{}}\KeywordTok{lower}\NormalTok{=}\DecValTok{2}\NormalTok{\textgreater{} R;}
  \DataTypeTok{array}\NormalTok{[R] }\DataTypeTok{vector}\NormalTok{[D] w; }\DataTypeTok{array}\NormalTok{[M,R] }\DataTypeTok{int}\NormalTok{\textless{}}\KeywordTok{lower}\NormalTok{=}\DecValTok{0}\NormalTok{,}\KeywordTok{upper}\NormalTok{=}\DecValTok{1}\NormalTok{\textgreater{} I;}
  \DataTypeTok{int}\NormalTok{\textless{}}\KeywordTok{lower}\NormalTok{=}\DecValTok{1}\NormalTok{\textgreater{} N; }\DataTypeTok{int}\NormalTok{\textless{}}\KeywordTok{lower}\NormalTok{=}\DecValTok{2}\NormalTok{\textgreater{} S;}
  \DataTypeTok{array}\NormalTok{[S] }\DataTypeTok{simplex}\NormalTok{[K] x; }\DataTypeTok{array}\NormalTok{[N,S] }\DataTypeTok{int}\NormalTok{\textless{}}\KeywordTok{lower}\NormalTok{=}\DecValTok{0}\NormalTok{,}\KeywordTok{upper}\NormalTok{=}\DecValTok{1}\NormalTok{\textgreater{} J;}
  \DataTypeTok{real}\NormalTok{ alpha\_mean; }\DataTypeTok{real}\NormalTok{\textless{}}\KeywordTok{lower}\NormalTok{=}\DecValTok{0}\NormalTok{\textgreater{} alpha\_sd; }\DataTypeTok{real}\NormalTok{\textless{}}\KeywordTok{lower}\NormalTok{=}\DecValTok{0}\NormalTok{\textgreater{} beta\_sd;}
\NormalTok{\}}
\KeywordTok{transformed data}\NormalTok{ \{}
  \DataTypeTok{array}\NormalTok{[M] }\DataTypeTok{int}\NormalTok{\textless{}}\KeywordTok{lower}\NormalTok{=}\DecValTok{2}\NormalTok{\textgreater{} N\_uncertain; }\DataTypeTok{int}\NormalTok{ total\_uncertain\_alts = }\DecValTok{0}\NormalTok{;}
  \ControlFlowTok{for}\NormalTok{ (m }\ControlFlowTok{in} \DecValTok{1}\NormalTok{:M) \{ N\_uncertain[m] = sum(I[m]); total\_uncertain\_alts += N\_uncertain[m]; \}}
  \DataTypeTok{array}\NormalTok{[total\_uncertain\_alts] }\DataTypeTok{vector}\NormalTok{[D] x\_uncertain;}
\NormalTok{  \{ }\DataTypeTok{int}\NormalTok{ pos=}\DecValTok{1}\NormalTok{; }\ControlFlowTok{for}\NormalTok{ (m }\ControlFlowTok{in} \DecValTok{1}\NormalTok{:M) }\ControlFlowTok{for}\NormalTok{ (r }\ControlFlowTok{in} \DecValTok{1}\NormalTok{:R) }\ControlFlowTok{if}\NormalTok{ (I[m,r]==}\DecValTok{1}\NormalTok{)\{x\_uncertain[pos]=w[r];pos+=}\DecValTok{1}\NormalTok{;\} \}}
  \DataTypeTok{array}\NormalTok{[N] }\DataTypeTok{int}\NormalTok{\textless{}}\KeywordTok{lower}\NormalTok{=}\DecValTok{2}\NormalTok{\textgreater{} N\_risky; }\DataTypeTok{int}\NormalTok{ total\_risky\_alts = }\DecValTok{0}\NormalTok{;}
  \ControlFlowTok{for}\NormalTok{ (n }\ControlFlowTok{in} \DecValTok{1}\NormalTok{:N) \{ N\_risky[n] = sum(J[n]); total\_risky\_alts += N\_risky[n]; \}}
  \DataTypeTok{array}\NormalTok{[total\_risky\_alts] }\DataTypeTok{simplex}\NormalTok{[K] x\_risky;}
\NormalTok{  \{ }\DataTypeTok{int}\NormalTok{ pos=}\DecValTok{1}\NormalTok{; }\ControlFlowTok{for}\NormalTok{ (n }\ControlFlowTok{in} \DecValTok{1}\NormalTok{:N) }\ControlFlowTok{for}\NormalTok{ (s }\ControlFlowTok{in} \DecValTok{1}\NormalTok{:S) }\ControlFlowTok{if}\NormalTok{ (J[n,s]==}\DecValTok{1}\NormalTok{)\{x\_risky[pos]=x[s];pos+=}\DecValTok{1}\NormalTok{;\} \}}
\NormalTok{\}}
\KeywordTok{generated quantities}\NormalTok{ \{}
  \DataTypeTok{real}\NormalTok{ alpha = lognormal\_rng(alpha\_mean, alpha\_sd);}
  \DataTypeTok{matrix}\NormalTok{[K,D] beta;}
  \ControlFlowTok{for}\NormalTok{ (k }\ControlFlowTok{in} \DecValTok{1}\NormalTok{:K) }\ControlFlowTok{for}\NormalTok{ (d }\ControlFlowTok{in} \DecValTok{1}\NormalTok{:D) beta[k,d] = normal\_rng(}\DecValTok{0}\NormalTok{, beta\_sd);}
  \DataTypeTok{simplex}\NormalTok{[K}\DecValTok{{-}1}\NormalTok{] delta = dirichlet\_rng(rep\_vector(}\FloatTok{1.0}\NormalTok{, K}\DecValTok{{-}1}\NormalTok{));}
  \DataTypeTok{vector}\NormalTok{[K] upsilon = cumulative\_sum(append\_row(}\DecValTok{0}\NormalTok{, delta));}

  \CommentTok{// Uncertain choices}
  \DataTypeTok{array}\NormalTok{[total\_uncertain\_alts] }\DataTypeTok{vector}\NormalTok{[K] psi;}
  \ControlFlowTok{for}\NormalTok{ (i }\ControlFlowTok{in} \DecValTok{1}\NormalTok{:total\_uncertain\_alts) psi[i] = softmax(beta * x\_uncertain[i]);}
  \DataTypeTok{vector}\NormalTok{[total\_uncertain\_alts] eta\_uncertain;}
  \ControlFlowTok{for}\NormalTok{ (i }\ControlFlowTok{in} \DecValTok{1}\NormalTok{:total\_uncertain\_alts) eta\_uncertain[i] = dot\_product(to\_vector(psi[i]), upsilon);}
  \DataTypeTok{array}\NormalTok{[M] }\DataTypeTok{int}\NormalTok{ y; }\DataTypeTok{array}\NormalTok{[M] }\DataTypeTok{int}\NormalTok{\textless{}}\KeywordTok{lower}\NormalTok{=}\DecValTok{0}\NormalTok{,}\KeywordTok{upper}\NormalTok{=}\DecValTok{1}\NormalTok{\textgreater{} selected\_seu\_max\_uncertain;}
\NormalTok{  \{}
    \DataTypeTok{int}\NormalTok{ pos = }\DecValTok{1}\NormalTok{;}
    \ControlFlowTok{for}\NormalTok{ (m }\ControlFlowTok{in} \DecValTok{1}\NormalTok{:M) \{}
      \DataTypeTok{vector}\NormalTok{[N\_uncertain[m]] pe = segment(eta\_uncertain, pos, N\_uncertain[m]);}
      \DataTypeTok{vector}\NormalTok{[N\_uncertain[m]] cp = softmax(alpha * pe);}
\NormalTok{      y[m] = categorical\_rng(cp);}
\NormalTok{      selected\_seu\_max\_uncertain[m] = (abs(pe[y[m]] {-} max(pe)) \textless{} }\FloatTok{1e{-}10}\NormalTok{) ? }\DecValTok{1}\NormalTok{ : }\DecValTok{0}\NormalTok{;}
\NormalTok{      pos += N\_uncertain[m];}
\NormalTok{    \}}
\NormalTok{  \}}

  \CommentTok{// Risky choices (beta{-}free)}
  \DataTypeTok{vector}\NormalTok{[total\_risky\_alts] eta\_risky;}
  \ControlFlowTok{for}\NormalTok{ (i }\ControlFlowTok{in} \DecValTok{1}\NormalTok{:total\_risky\_alts) eta\_risky[i] = dot\_product(to\_vector(x\_risky[i]), upsilon);}
  \DataTypeTok{array}\NormalTok{[N] }\DataTypeTok{int}\NormalTok{ z; }\DataTypeTok{array}\NormalTok{[N] }\DataTypeTok{int}\NormalTok{\textless{}}\KeywordTok{lower}\NormalTok{=}\DecValTok{0}\NormalTok{,}\KeywordTok{upper}\NormalTok{=}\DecValTok{1}\NormalTok{\textgreater{} selected\_seu\_max\_risky;}
\NormalTok{  \{}
    \DataTypeTok{int}\NormalTok{ pos = }\DecValTok{1}\NormalTok{;}
    \ControlFlowTok{for}\NormalTok{ (n }\ControlFlowTok{in} \DecValTok{1}\NormalTok{:N) \{}
      \DataTypeTok{vector}\NormalTok{[N\_risky[n]] pe = segment(eta\_risky, pos, N\_risky[n]);}
      \DataTypeTok{vector}\NormalTok{[N\_risky[n]] cp = softmax(alpha * pe);}
\NormalTok{      z[n] = categorical\_rng(cp);}
\NormalTok{      selected\_seu\_max\_risky[n] = (abs(pe[z[n]] {-} max(pe)) \textless{} }\FloatTok{1e{-}10}\NormalTok{) ? }\DecValTok{1}\NormalTok{ : }\DecValTok{0}\NormalTok{;}
\NormalTok{      pos += N\_risky[n];}
\NormalTok{    \}}
\NormalTok{  \}}
  \DataTypeTok{int}\NormalTok{ total\_seu\_max\_selected =}
\NormalTok{      sum(selected\_seu\_max\_uncertain) + sum(selected\_seu\_max\_risky);}
\NormalTok{\}}
\end{Highlighting}
\end{Shaded}

\subsection*{\texorpdfstring{C.5 \texttt{m\_01.stan} ---
calibrated-prior variant
(application)}{C.5 m\_01.stan --- calibrated-prior variant (application)}}\label{sec-c-m01}
\addcontentsline{toc}{subsection}{C.5 \texttt{m\_01.stan} ---
calibrated-prior variant (application)}

The application model \texttt{m\_01} is \textbf{structurally identical}
to \texttt{m\_0} (§\hyperref[sec-c-m0]{C.1}) --- same \texttt{data},
\texttt{parameters}, \texttt{transformed\ parameters}, and
\texttt{generated\ quantities}. It differs only in the \(\alpha\) prior,
which is calibrated by the prior-predictive SEU-max-rate procedure of
§\hyperref[sec-app-model]{7.3} (claim C8) rather than left at the
convenience default. The single changed line in the \texttt{model} block
is:

\begin{Shaded}
\begin{Highlighting}[]
\CommentTok{// m\_0.stan:}
\NormalTok{alpha \textasciitilde{} lognormal(}\DecValTok{0}\NormalTok{, }\DecValTok{1}\NormalTok{);     }\CommentTok{// convenience default}
\CommentTok{// m\_01.stan:}
\NormalTok{alpha \textasciitilde{} lognormal(}\FloatTok{3.0}\NormalTok{, }\FloatTok{0.75}\NormalTok{); }\CommentTok{// calibrated: median \textasciitilde{} 20, 90\% CI \textasciitilde{} [5.5, 67],}
                              \CommentTok{// prior{-}implied SEU{-}max rate \textasciitilde{} 0.78 (insurance, K=3)}
\end{Highlighting}
\end{Shaded}

For the \(K = 4\) Ellsberg cells the same SEU-max-rate calibration is
re-run to set the \(\alpha\)-prior hyperparameters at that menu size
(Appendix D.5); the model code is otherwise unchanged.

\section*{Appendix D. Computational Details and Design
Constants}\label{sec-appendix-d}
\addcontentsline{toc}{section}{Appendix D. Computational Details and
Design Constants}

\begin{quote}
\emph{This appendix is the \textbf{canonical} location for the
study-design constants and sampler settings used throughout the paper;
the body (§§\hyperref[sec-m0-implementation]{4},
\hyperref[sec-m1-implementation]{6}, \hyperref[sec-application]{7})
references D.0 and D.5 rather than re-listing. D.0 fixes the
foundational (methodological) design; D.5 the application scale.
D.1--D.4 give software versions, sampler configuration, the SBC
protocol, and the matched-design comparison; D.6 defines and gives the
computation of the summary quantities reported per application cell in
§7.1; D.7 details the three minimum-detectable-effect estimators behind
the Claude-null calibration of §7.5.2. Exact, machine-readable values
live in \texttt{configs/*.json} and \texttt{applications/*/configs/} at
the pinned commit (Appendix E.0); the constants below are the
human-readable transcription.}
\end{quote}

\subsection*{D.0 Foundational study-design constants}\label{sec-d0}
\addcontentsline{toc}{subsection}{D.0 Foundational study-design
constants}

The methodological models \texttt{m\_0} and \texttt{m\_1}
(§§\hyperref[sec-m0-implementation]{4},
\hyperref[sec-m1-implementation]{6}) use a single small design, chosen
to be representative rather than power-optimized:

\begin{longtable}[]{@{}lll@{}}
\toprule\noalign{}
Constant & Symbol & Value \\
\midrule\noalign{}
\endhead
\bottomrule\noalign{}
\endlastfoot
Consequences per problem & \(K\) & 3 \\
Belief-feature dimension & \(D\) & 5 \\
Distinct uncertain alternatives & \(R\) & 15 \\
Distinct risky lotteries (\texttt{m\_1}) & \(S\) & 15 \\
Uncertain problems & \(M\) & \(\in \{25, 50\}\) \\
Risky problems (\texttt{m\_1}) & \(N\) & \(\in \{25, 50\}\) \\
Alternatives per problem & --- & 2--5 (uniform) \\
Feature distribution & --- & \(\mathcal{N}(0, 1)\), i.i.d. \\
Risky lottery simplices & \(\pi_s\) & fixed across iterations \\
\end{longtable}

The matched-design recovery study (D.4) instantiates these at the four
\((M, N)\) slices of the §\hyperref[sec-m1-matched]{6.2} table. The
simulation-based-calibration protocol (D.3) uses
\(N_{\mathrm{sbc}} = 999\) draws, thinning \(4\), a single chain, \(20\)
rank-histogram bins, and the \(L =
N_{\mathrm{sbc}}\) rank convention.

\subsection*{D.1 Software versions}\label{sec-d1}
\addcontentsline{toc}{subsection}{D.1 Software versions}

All fits use \textbf{CmdStan v2.37.0} driven by \textbf{CmdStanPy}
(\(\geq 1.2\)) under \textbf{Python 3.10} (conda environment
\texttt{seu-sensitivity}). Numerical and post-processing support: NumPy
(\(\geq 1.20\)), SciPy (\(\geq 1.7\)), pandas (\(\geq 1.3\)),
scikit-learn (\(\geq 1.0\), used for the \(D = 32\) PCA feature
reduction of the insurance application, D.5), and Matplotlib
(\(\geq 3.4\)). LLM choice data were collected via the OpenAI
(\(\geq 1.0\)) and Anthropic (\(\geq 0.18\)) Python clients. The exact,
pinned versions are recorded in \texttt{requirements.txt} and
\texttt{environment.yml} at the commit of Appendix E.0; the versions
above are the essential ones for reproducing the inference.

\subsection*{D.2 Sampler configuration and diagnostics
policy}\label{sec-d2}
\addcontentsline{toc}{subsection}{D.2 Sampler configuration and
diagnostics policy}

\textbf{Sampler.} Posterior fits use \textbf{4 chains} and \textbf{1000
sampling} iterations per chain (\textbf{4000 retained draws}), at
CmdStan's NUTS defaults --- target acceptance
\(\mathrm{adapt\_delta} = 0.8\) and maximum tree depth \(10\), not
overridden. Warmup is \(500\) iterations per chain in the
parameter-recovery driver and \(1000\) in the
§\hyperref[sec-application]{7} application fits; either way 4000
post-warmup draws are retained per fit. The 4000-draw count is the
denominator for the §\hyperref[sec-app-validation]{7.4}
\emph{divergence} fractions (\(\leq 6/4000\) per fit); the
\(\hat{R}\)-exceedance summary is denominated differently --- it counts
\emph{fits} (4/20) and, within a flagged fit, the monitored parameter
components that exceed \(1.01\), not draws. (The SBC fits of D.3 use a
different, single-chain configuration.) Random seeds are deterministic
per iteration: the simulation draw uses seed \(12345 + i\) and the
inference fit uses \(54321 + i\) for iteration \(i\), so every recovery
iteration is independently reproducible.

\textbf{Diagnostics policy.} A fit is considered well-behaved when, for
the key parameters, the potential-scale-reduction factor satisfies
\(\hat{R} \le 1.01\) and both bulk- and tail-effective sample size are
adequate for the reported quantiles; divergent transitions are tracked
per fit and reported as a fraction. In the application
(§\hyperref[sec-app-validation]{7.4}, claim C14) \(\alpha\) --- the only
parameter of primary interest for the temperature analysis --- is never
\(\hat{R}\)-flagged and has satisfactory ESS in all 20 fits; divergences
are \(\le 0.15\%\) per fit; the only \(\hat{R} > 1.01\) exceedances fall
on the weakly-informed \(\beta, \delta\) and the per-trial nuisance
latents (\(\eta, \psi, \upsilon\)) in the hardest \(K = 4\) fits ---
consistent with the §\hyperref[sec-bd-weak]{3.4} weakly-informed
\((\beta,\delta)\) finding rather than a defect in the \(\alpha\)
inference.

\textbf{Stricter-sampler refit check.} To verify that the residual
divergences are step-size artifacts rather than symptoms of posterior
bias, all 20 application conditions were refit at
\(\mathrm{adapt\_delta} = 0.99\), maximum tree depth \(12\), and warmup
\(2000\) (4 chains, 1000 sampling iterations, seed 42;
\texttt{spikes/report\_refit\_sensitivity\_spike.py}, Appendix
\hyperref[sec-e-manifest]{E.2}). Under the stricter settings the total
divergence count drops from 34 (spread across 12/20 fits) to \textbf{0},
with no tree-depth saturations; the largest shift in any per-condition
\(\alpha\) posterior median is \(0.06\) committed-posterior standard
deviations; and all four temperature slopes are reproduced to within
Monte-Carlo noise (GPT-4o \(\times\) insurance \(-24.6 \to -24.8\) with
\(P\) \(0.991 \to 0.986\); Claude \(\times\) insurance
\(-2.9 \to -2.0\), \(0.560 \to 0.537\); GPT-4o \(\times\) Ellsberg
\(-38.4 \to -37.5\), \(0.984 \to 0.992\); Claude \(\times\) Ellsberg
\(-15.0 \to -14.7\), \(0.766 \to 0.767\)). The committed draws are
retained as the paper's basis; the refit is a sensitivity check, not a
replacement.

\subsection*{D.3 SBC configuration and seed convention}\label{sec-d3}
\addcontentsline{toc}{subsection}{D.3 SBC configuration and seed
convention}

Simulation-based calibration (§\hyperref[sec-m0-sbc]{4.4},
§\hyperref[sec-m1-sbc]{6.5}, claim C5) draws \(N_{\mathrm{sbc}} = 999\)
prior samples from the matched \texttt{\_sbc} program (which embeds the
simulator of Appendix C.2/C.4: each replicate draws its true parameters
and generates its synthetic dataset in \texttt{generated\ quantities} of
the same program that is then fit, in a single Stan pass), fits the
inference model to each synthetic dataset, and ranks each true parameter
within its thinned posterior draws. Each SBC fit runs a \textbf{single
chain} with \(1998\) warmup and \(3996\) sampling iterations, thinned by
\(4\) to retain \(L = 999\) posterior draws --- the Stan User's Guide
convention of setting \(L = N_{\mathrm{sbc}}\), so that the rank
statistic takes one of \(L + 1 =
1000\) equally likely values under calibration. The rank histograms use
\(20\) bins (expected count \(\approx 50\) per bin). Calibration is read
off both the rank histograms and the ECDF-difference plot with
simultaneous KS bands (Modřák et al.~2023), which is more sensitive than
the rank histogram alone. Each replicate \(i\) uses a single
deterministic seed \(123 + i\): because the \texttt{\_sbc} program draws
the true parameters, simulates, and fits in one pass, one seed governs
all steps and each of the 999 replicates is reproducible in isolation.
(The study design itself is generated once under NumPy seed \(42\) and
held fixed across all replicates.)

\subsection*{D.4 The matched-design comparison (conditions
A/B/C/D)}\label{sec-d4}
\addcontentsline{toc}{subsection}{D.4 The matched-design comparison
(conditions A/B/C/D)}

The matched study underlying §\hyperref[sec-m1-recovery]{6.4} (claims
C1--C3) is \textbf{not} a \(\delta\)-optimal design study; it is a
controlled \emph{within-iteration} slicing. A single study design is
generated once \emph{externally} --- the feature draws \(w\) and the
lottery menus, which enter the simulator as data (\(w, I, x, J\);
Appendix C.4), not as quantities \texttt{m\_1\_sim} itself draws --- and
is \textbf{held fixed across all \(n = 100\) iterations}. Within each
iteration all four conditions share this one design and a single fresh
parameter draw (so the conditions are matched). Each iteration proceeds
as:

\begin{enumerate}
\def\labelenumi{\arabic{enumi}.}
\tightlist
\item
  draw a fresh true \((\alpha, \beta, \delta)\) from the prior;
\item
  with the fixed design supplied as data, \texttt{m\_1\_sim} (Appendix
  C.4) draws choices \emph{conditional on} the true parameters ---
  uncertain choices \texttt{y} (\(M = 50\)) and risky choices \texttt{z}
  (\(N = 50\));
\item
  fit four conditions to \textbf{slices of the same simulated choices},
  so the true parameters are matched across conditions within the
  iteration:
\end{enumerate}

\begin{longtable}[]{@{}lllll@{}}
\toprule\noalign{}
Condition & Model & Uncertain \(M\) & Risky \(N\) & Total \\
\midrule\noalign{}
\endhead
\bottomrule\noalign{}
\endlastfoot
A & \texttt{m\_0} & 25 & --- & 25 \\
B & \texttt{m\_0} & 50 & --- & 50 \\
C & \texttt{m\_1} & 25 & 25 & 50 \\
D & \texttt{m\_1} & 50 & 50 & 100 \\
\end{longtable}

The central comparison is \textbf{B vs C} (same total choice count; only
the model and the \emph{type} of choice differ) and the control is
\textbf{A vs B} (more uncertain data alone). All reported effects are
paired-iteration medians or bootstrap-over-iteration ratios with 90\%
CIs (10,000 resamples, bootstrap seed 20260617); see
\texttt{spikes/report14\_rerun\_analysis.py}. The headline finding ---
that the \(\beta\)-free risky block yields no matched-count
\(\alpha\)-precision gain (C1) while improving-\(\alpha\) comes from
data \emph{quantity} (A\(\to\)B, C2), and the \(\delta\) CI-width gain
is \(\approx 0.8\%\) (C3) --- is a property of \emph{this} matched
slicing, not of a \(\delta\)-optimized design.

\subsection*{D.5 Application-scale design constants}\label{sec-d5}
\addcontentsline{toc}{subsection}{D.5 Application-scale design
constants}

The \(2\times2\) application (§\hyperref[sec-application]{7}) uses the
calibrated-prior model \texttt{m\_01} (Appendix C.5) at a larger scale.
Per LLM \(\times\) task cell:

\begin{longtable}[]{@{}lll@{}}
\toprule\noalign{}
Constant & Insurance triage & Ellsberg urns \\
\midrule\noalign{}
\endhead
\bottomrule\noalign{}
\endlastfoot
Consequences \(K\) & 3 & 4 \\
Base problems \(M\) & 100 & 100 \\
Position-counterbalanced presentations & \(\times 3\) & \(\times 3\) \\
Choices per temperature condition & 300 & 300 \\
Temperature conditions per cell & 5 & 5 \\
Alternatives per pool \(R\) & 30 & 30 \\
Belief-feature dimension \(D\) & 32 (PCA) & 32 (PCA) \\
Parameter-recovery iterations & 20 & 20 \\
\end{longtable}

\textbf{Temperature grids} differ by provider: GPT-4o uses
\(T \in \{0.0, 0.3, 0.7,
1.0, 1.5\}\); Claude 3.5 Sonnet uses
\(T \in \{0.0, 0.2, 0.5, 0.8, 1.0\}\) (the provider-specific upper
bounds drive the unequal-grid caveat of
§\hyperref[sec-app-cross-llm]{7.5.2a} and the cross-LLM robustness
check, claim C11).

\textbf{Calibrated \(\alpha\) priors.} Insurance uses \(\alpha \sim
\mathrm{Lognormal}(3.0, 0.75)\) (median \(\approx 20\), 90\% CI
\(\approx [5.5, 67]\) --- the empirical quantiles of the grid-search
prior draws (§7.3) --- and prior-implied SEU-max rate \(\approx 0.78\);
claim C8). For the \(K = 4\) Ellsberg cells the prior is re-calibrated
by the same SEU-maximizer-rate prior-predictive procedure at that menu
size, since the SEU-max rate at fixed \(\alpha\) depends on \(K\) and
the number of alternatives. Across all four cells the fit count is
\(5 \text{ temperatures} \times 4 \text{ cells} = 20\) posterior fits,
plus \(20 \times 4 = 80\) recovery iterations.

\subsection*{D.6 Reported summary quantities: definitions and
computation}\label{sec-d6}
\addcontentsline{toc}{subsection}{D.6 Reported summary quantities:
definitions and computation}

The reporting template of §\hyperref[sec-app-why]{7.1} commits every LLM
\(\times\) task cell to the same four quantities. Each is a functional
of the per-condition \(\alpha\) posterior draws --- for a cell with
temperature levels \(T_1 < \dots < T_L\) (here \(L = 5\)), write
\(\alpha_\ell^{(s)}\) for the \(s\)-th posterior draw of \(\alpha\) at
temperature \(T_\ell\), \(s = 1, \dots, S\). All intervals are
equal-tailed posterior credible intervals (5th and 95th empirical
quantiles of the relevant draw set), and all probabilities are
Monte-Carlo averages over the \(S\) draws.

\textbf{(1) Per-condition \(\alpha\).} For each temperature \(T_\ell\)
we report the posterior median \(\mathrm{med}_s\{\alpha_\ell^{(s)}\}\)
and the 90\% credible interval \([\,q_{0.05}, q_{0.95}\,]\) of
\(\{\alpha_\ell^{(s)}\}_{s=1}^{S}\).

\textbf{(2) Global temperature slope \(\Delta\alpha/\Delta T\).} For
each posterior draw \(s\) we compute the population
ordinary-least-squares slope of \(\alpha\) on temperature across the
\(L\) levels, \[
b^{(s)} \;=\; \frac{\widehat{\mathrm{Cov}}\!\left(T,\, \alpha^{(s)}\right)}
                    {\widehat{\mathrm{Var}}(T)}
       \;=\; \frac{\sum_{\ell=1}^{L}\left(T_\ell - \bar{T}\right)
                    \left(\alpha_\ell^{(s)} - \bar{\alpha}^{(s)}\right)}
                   {\sum_{\ell=1}^{L}\left(T_\ell - \bar{T}\right)^2},
\] where \(\bar{T}\) and \(\bar{\alpha}^{(s)}\) are the across-level
means (the draw index \(s\) is held fixed, pairing the \(L\)
per-condition draws that share it). This yields a posterior distribution
\(\{b^{(s)}\}_{s=1}^{S}\), summarized by its median, its 90\% credible
interval, and the negativity probability
\(P(\text{slope} < 0) = \tfrac{1}{S}\sum_s \mathbf{1}\{b^{(s)} < 0\}\).
This \emph{draw-level population-OLS} slope is the single slope
convention used throughout the paper --- every slope in
§\hyperref[sec-app-results]{7.5}, the \(2\times2\) table of
§\hyperref[sec-app-2x2]{7.5.5}, and the claims ledger is this
functional. (An earlier draft mixed in a second estimator that inflated
slope magnitudes by exactly \(L/(L-1) = 1.25\) through mismatched
degrees-of-freedom conventions; the correction is recorded in Appendix
\hyperref[sec-e-manifest]{E.2}. \(P(\text{slope}<0)\) is invariant to
that scale factor and was unaffected.)

\textbf{(3) \(P(\text{strict monotone decrease})\).} The posterior
probability that \(\alpha\) falls at every step, \[
P(\text{strict monotone} \downarrow)
   \;=\; \frac{1}{S}\sum_{s=1}^{S}
         \mathbf{1}\!\left\{\alpha_1^{(s)} > \alpha_2^{(s)} > \dots > \alpha_L^{(s)}\right\}.
\] This is strictly stronger than a negative slope: a draw can have
\(b^{(s)} < 0\) while violating one adjacent-step ordering, which is why
the reported monotone-decrease probabilities are smaller than the
corresponding \(P(\text{slope} < 0)\)
(§\hyperref[sec-app-gpt4o-insurance]{7.5.1}).

\textbf{(4) Cross-LLM comparison probability.} For two independently
fitted cells with per-draw slopes \(\{b_A^{(s)}\}\) and
\(\{b_B^{(s)}\}\), we report
\(P(b_A < b_B) = \tfrac{1}{S}\sum_s \mathbf{1}\{b_A^{(s)} < b_B^{(s)}\}\),
pairing the two posteriors by the exchangeable draw index (the cells are
fit separately, so the pairing carries no cross-cell dependence). This
between-cell probability answers a different question from the
within-cell \(P(\text{slope}<0)\) values --- it measures the separation
of two independent slope posteriors, folding in both cells' uncertainty,
and is not bounded by either within-cell probability (in the insurance
comparison it sits between them: \(0.82\) against GPT-4o's \(0.99\) and
Claude's \(0.56\)). Because the providers use unequal temperature grids,
this comparison is reported both on the full grids and with the wider
grid restricted to the shared ceiling; see
§\hyperref[sec-app-cross-llm]{7.5.2a} and claim C11. The computation is
reproduced by \texttt{spikes/report11\_cross\_llm\_spike.py}.

\subsection*{D.7 Minimum-detectable effect: the three
estimators}\label{sec-d7}
\addcontentsline{toc}{subsection}{D.7 Minimum-detectable effect: the
three estimators}

The Claude \(\times\) insurance null
(§\hyperref[sec-app-claude-insurance]{7.5.2}) is read against the
design's \emph{resolution}: the smallest counterfactual
temperature-on-\(\alpha\) slope whose posterior would clear a detection
criterion. This appendix defines the minimum detectable effect (MDE) and
the three estimators that the body reports as agreeing; the computation
is reproduced by \texttt{spikes/report16\_mde\_spike.py} (seed
\texttt{20260618}).

\textbf{Setup.} The reported slope is the population-OLS functional of
§D.6(2). Writing \(\alpha_\ell^{(s)}\) for the \(s\)-th posterior draw
at temperature \(T_\ell\) (\(\ell = 1, \dots, L\), here \(L = 5\) at
\(T \in \{0.0, 0.2, 0.5, 0.8, 1.0\}\)), the per-draw slope is the fixed
linear combination \[
b^{(s)} \;=\; \sum_{\ell=1}^{L} w_\ell\, \alpha_\ell^{(s)},
\qquad
w_\ell \;=\; \frac{T_\ell - \bar{T}}{\sum_{j=1}^{L}\left(T_j - \bar{T}\right)^2},
\] so the slope posterior \(\{b^{(s)}\}\) is a weighted sum of the five
per-condition posteriors and
\(P(\text{slope} < 0) = \tfrac{1}{S}\sum_s \mathbf{1}\{b^{(s)} < 0\}\).
The \emph{resolving power} of the design is therefore fixed by the
per-condition posterior dispersions and the temperature weights
\(w_\ell\). The observed slope posterior has standard deviation
\(\sigma_{\text{slope}} \approx 22.1\) and
\(P(\text{slope} < 0) \approx 0.56\). Throughout this appendix \(\beta\)
denotes the \emph{signed} counterfactual true slope --- negative for a
temperature-driven decline in \(\alpha\) --- and the formulas below use
the signed value; MDEs are quoted as magnitudes \(|\beta|\). The MDE is
the smallest counterfactual decline magnitude for which
\(P(\text{slope} < 0)\) would reach the criterion; the primary criterion
is the one-sided \(P(\text{slope} < 0) \geq 0.95\)
(\(z_{0.95} = 1.645\)), the secondary \(P(\text{slope} < 0) \geq 0.975\)
(\(z_{0.975} = 1.960\)). The three estimators differ only in how they
model the \emph{counterfactual dispersion} of the slope under a shifted
mean; they share the per-condition spreads that dominate the
calculation, which is why they agree.

\textbf{(A) Analytic Gaussian.} Treat the slope posterior as
approximately normal with its observed dispersion held fixed, so a
signed counterfactual mean \(\beta < 0\) gives
\(P(\text{slope} < 0) = \Phi(-\beta / \sigma_{\text{slope}}) =
\Phi(|\beta| / \sigma_{\text{slope}})\). The criterion then fixes \[
\mathrm{MDE} \;=\; z\,\sigma_{\text{slope}},
\] giving \(\approx 36.3\) at \(P \geq 0.95\) and \(\approx 43\) at
\(P \geq 0.975\). This is the simplest estimator and is exact only to
the extent the slope posterior is Gaussian.

\textbf{(B) Empirical-quantile (additive-noise).} Drop the normality
assumption but keep the additive-shift model: re-center the actual slope
draws to a signed counterfactual mean \(\beta < 0\), so
\(\text{slope}^{\text{cf}}_s = (b^{(s)} - \bar{b}) + \beta\) with
residuals \(\varepsilon_s = b^{(s)} - \bar{b}\) carrying the observed
(non-Gaussian) shape. Then
\(P(\text{slope}^{\text{cf}} < 0) = P(\varepsilon < -\beta) =
P(\varepsilon < |\beta|)\), so the MDE is the empirical
criterion-quantile of \(\varepsilon\): the 95th percentile
(\(\approx 34.7\)) at \(P \geq 0.95\), the 97.5th at \(P \geq 0.975\).
This is faithful to the actual slope shape but assumes the per-condition
dispersion is unchanged by the counterfactual mean shift.

\textbf{(C) Constant-CV Monte-Carlo.} Allow the dispersion to scale with
the \(\alpha\) level, which the per-condition posteriors exhibit (their
coefficient of variation is near-constant, \(\mathrm{CV} \approx 0.28\),
across the five conditions). Impose a signed counterfactual mean line
\(\mu_\ell = \alpha_{\text{ref}} + \beta\,(T_\ell - \bar{T})\) about the
grand-mean level \(\alpha_{\text{ref}} \approx 67.5\), preserve each
condition's \emph{relative} residuals
\((\alpha_\ell^{(s)} - \bar\alpha_\ell)/\bar\alpha_\ell\), resample the
per-condition relative residuals, recompute the slope through the same
\(\{w_\ell\}\) functional, and root-find the \(\beta < 0\) where
\(P(\text{slope} < 0)\) hits the criterion (\(2 \times 10^{5}\)
Monte-Carlo resamples). This yields \(\approx 36.0\) at \(P \geq 0.95\)
and \(\approx 42\) at \(P \geq 0.975\), and is the estimator that
accounts for dispersion growing with the mean.

\textbf{Why they agree.} All three read off the same slope functional
and the same per-condition posterior spreads; they differ only in the
auxiliary assumption about how those spreads behave under a
counterfactual mean shift (fixed and Gaussian (A); fixed and empirical
(B); scaling with the mean at constant CV (C)). Because the resolving
power is dominated by the shared per-condition dispersions rather than
by that auxiliary assumption, the estimates cluster tightly ---
\(34.7\)--\(36.3\) at \(P \geq 0.95\) and \(41\)--\(43\) at
\(P \geq 0.975\) --- and the substantive conclusion (the observed
\(|\text{slope}| \approx 2.9\) sits an order of magnitude below the
floor) is insensitive to which one is used. The power curve of
Figure~\ref{fig-mde-power} is drawn from estimator (C).

\section*{Appendix E. Reproducibility}\label{sec-appendix-e}
\addcontentsline{toc}{section}{Appendix E. Reproducibility}

\begin{quote}
\emph{Provenance, compute budget, and a per-figure manifest. The goal is
that every key empirical claim in the paper can be regenerated from a
pinned state of the supporting repository, with no number depending on
an artifact that is not itself version-controlled.}
\end{quote}

\subsection*{E.0 Citation and provenance policy}\label{sec-e-provenance}
\addcontentsline{toc}{subsection}{E.0 Citation and provenance policy}

Every key empirical claim cites the supporting repository at a
\textbf{pinned commit or tag} --- and, at the point of archival deposit,
a DOI minted from that tagged commit (a Zenodo release) is substituted
for the commit pin in the citation --- never a homepage or blog URL. The
pin for this version is the Git tag \texttt{v1.0} (commit
\texttt{d1f56fd}) of the \texttt{seu-sensitivity} repository, archived
at Zenodo under DOI
\href{https://doi.org/10.5281/zenodo.21250951}{10.5281/zenodo.21250951}
--- the version-specific identifier for this exact snapshot; the
version-independent concept DOI
\href{https://doi.org/10.5281/zenodo.21250950}{10.5281/zenodo.21250950}
always resolves to the latest release. Supporting materials (data,
analysis scripts, configs, figures) are \textbf{referenced at that pin,
not copied} into the paper folder, so there is a single source of truth
and no drift: the Stan listings of Appendix C, the design constants of
Appendix D, and the figure manifest of E.2 all describe the repository
\emph{at the pin}. The dissemination pipeline (internal reports \(\to\)
homepage posts \(\to\) this working paper) exists for reach; it is not a
citation surface. The mapping from claim to artifact is maintained in
the repository's \texttt{claims\_ledger.md}, in which every row cited in
the paper carries status \texttt{computed} (finalized from a completed
run) before it appears in body text.

The analysis is driven by a small set of generic scripts parameterized
by JSON configs:

\begin{longtable}[]{@{}
  >{\raggedright\arraybackslash}p{(\linewidth - 2\tabcolsep) * \real{0.5000}}
  >{\raggedright\arraybackslash}p{(\linewidth - 2\tabcolsep) * \real{0.5000}}@{}}
\toprule\noalign{}
\begin{minipage}[b]{\linewidth}\raggedright
Script
\end{minipage} & \begin{minipage}[b]{\linewidth}\raggedright
Role
\end{minipage} \\
\midrule\noalign{}
\endhead
\bottomrule\noalign{}
\endlastfoot
\texttt{analysis/prior\_predictive.py} & prior-predictive SEU-max-rate
checks (§4.2, §6.3, §7.3) \\
\texttt{analysis/parameter\_recovery.py} & true-vs-estimated recovery
(§4.3, §6.4, §7.4) \\
\texttt{analysis/sbc.py} & simulation-based calibration (§4.4, §6.5) \\
\texttt{analysis/posterior\_predictive\_checks.py} & PPC test statistics
(§7.4) \\
\texttt{analysis/model\_estimation.py} & posterior fits on observed LLM
choices (§7.5) \\
\end{longtable}

driven by \texttt{configs/*.json} (methodological studies) and
\texttt{applications/*/configs/} (the four LLM \(\times\) task cells).
The Stan programs are those of Appendix C under \texttt{models/}.

\subsection*{E.1 Compute budget}\label{sec-e-compute}
\addcontentsline{toc}{subsection}{E.1 Compute budget}

All models are small --- a few hundred categorical choices, three
structural parameter \emph{blocks} (\(\alpha\) a scalar, \(\beta\) a
\(K\times D\) matrix, \(\delta\) a \((K-2)\)-simplex) plus per-trial
deterministic transforms --- so a single posterior fit (4 chains
\(\times\) 1000 draws) completes in seconds to a low number of minutes
on one core of a contemporary 8-core workstation. The end-to-end cost is
dominated not by any single fit but by the
\textbf{simulation-based-calibration} runs, which refit the model once
per SBC draw: at \(N_{\mathrm{sbc}} = 999\) draws per model (D.3) these
are the longest-running configs and account for the bulk of the
wall-clock budget; the \textbf{matched-design recovery study}
(\(n = 100\) iterations \(\times\) 4 conditions \(=\) 400 fits, D.4) and
the \textbf{application fits} (20 posterior fits plus
\(20 \times 4 = 80\) recovery iterations, D.5) are each substantially
cheaper. With the SBC runs parallelized across the 8 cores, a full
end-to-end reproduction of every figure and table is an overnight-scale
job rather than a multi-day one; the SBC configs are the ones to launch
first and the only ones for which a single-core reproduction is
uncomfortable. (These are order-of-magnitude figures from the
development runs and are reported as such; a timed clean-room
reproduction at the pinned commit will be the authoritative source and
will be recorded alongside the archival deposit.)

\subsection*{E.2 Per-figure manifest}\label{sec-e-manifest}
\addcontentsline{toc}{subsection}{E.2 Per-figure manifest}

Each headline figure and table maps to a script, a config, and a seed,
so it is reproducible in isolation. Every entry is \emph{computed} ---
its numbers are finalized from a completed run. The foundational-report
figures (§4.3 recovery and §6.5 SBC) carry their canonical design inline
rather than through a committed JSON config: the methodological reports
(\href{reports/foundations/04_parameter_recovery.qmd}{\texttt{04\_parameter\_recovery.qmd}},
\href{reports/foundations/06_sbc_validation.qmd}{\texttt{06\_sbc\_validation.qmd}})
fix the canonical design constants of D.0 in-script and reuse the
generic \texttt{parameter\_recovery.py} / \texttt{sbc.py} drivers, so
the committed \texttt{configs/*sbc*.json} and
\texttt{configs/*recovery*.json} files (smaller smoke-scale designs) are
not the pinned source for these three plots.

\begingroup\footnotesize\setlength{\tabcolsep}{3pt}

\begin{longtable}[]{@{}
  >{\raggedright\arraybackslash}p{(\linewidth - 12\tabcolsep) * \real{0.0700}}
  >{\raggedright\arraybackslash}p{(\linewidth - 12\tabcolsep) * \real{0.0400}}
  >{\raggedright\arraybackslash}p{(\linewidth - 12\tabcolsep) * \real{0.1800}}
  >{\raggedright\arraybackslash}p{(\linewidth - 12\tabcolsep) * \real{0.2200}}
  >{\raggedright\arraybackslash}p{(\linewidth - 12\tabcolsep) * \real{0.2400}}
  >{\raggedright\arraybackslash}p{(\linewidth - 12\tabcolsep) * \real{0.1600}}
  >{\raggedright\arraybackslash}p{(\linewidth - 12\tabcolsep) * \real{0.0900}}@{}}
\caption{Per-figure manifest: each headline figure or table mapped to
its script, config, and seed.}\label{tbl-manifest}\tabularnewline
\toprule\noalign{}
\begin{minipage}[b]{\linewidth}\raggedright
Fig/Tab
\end{minipage} & \begin{minipage}[b]{\linewidth}\raggedright
§
\end{minipage} & \begin{minipage}[b]{\linewidth}\raggedright
Description
\end{minipage} & \begin{minipage}[b]{\linewidth}\raggedright
Script
\end{minipage} & \begin{minipage}[b]{\linewidth}\raggedright
Config
\end{minipage} & \begin{minipage}[b]{\linewidth}\raggedright
Seed
\end{minipage} & \begin{minipage}[b]{\linewidth}\raggedright
Status
\end{minipage} \\
\midrule\noalign{}
\endfirsthead
\toprule\noalign{}
\begin{minipage}[b]{\linewidth}\raggedright
Fig/Tab
\end{minipage} & \begin{minipage}[b]{\linewidth}\raggedright
§
\end{minipage} & \begin{minipage}[b]{\linewidth}\raggedright
Description
\end{minipage} & \begin{minipage}[b]{\linewidth}\raggedright
Script
\end{minipage} & \begin{minipage}[b]{\linewidth}\raggedright
Config
\end{minipage} & \begin{minipage}[b]{\linewidth}\raggedright
Seed
\end{minipage} & \begin{minipage}[b]{\linewidth}\raggedright
Status
\end{minipage} \\
\midrule\noalign{}
\endhead
\bottomrule\noalign{}
\endlastfoot
Fig 6.4 & 6.4.1 & \(\alpha\)/\(\delta\) RMSE + CI-width, matched A/B/C/D
(\(n=100\)) & \texttt{spikes/report14\_rerun\_analysis.py} &
\texttt{configs/m1\_matched\_recovery\_n100\_config.json} & design
20260617 / boot 20260617 & computed \\
Tab (§7.5.5) & 7.5 & \(2\times2\) per-cell global-slope posteriors
(C9--C13) & \texttt{analysis/model\_estimation.py} &
\texttt{applications/*/configs/} & per-cell (see cell data) &
computed \\
Fig (§7.5.5) & 7.5.5 & \(2\times2\) forest plot of the per-cell
global-slope posteriors (companion to Table~\ref{tbl-2x2}) &
\texttt{spikes/report\_2x2\_forest\_spike.py} & committed per-condition
\(\alpha\) draws (computes C9/C10/C12/C13) & deterministic (no RNG) &
computed \\
Fig 7.5.2 & 7.5.2 / 7.6.1 & Claude-null MDE power curve; MDE
\(\approx 36\) marker (C16) & \texttt{spikes/report16\_mde\_spike.py} &
inline (data \texttt{alpha\_draws\_T*.npz}) & 20260618 & computed \\
Fig (§7.6.6) & 7.6.6 & Prior-sensitivity forest: per-cell slope
posteriors under baseline + 3 alternative \(\alpha\) priors (C17) &
\texttt{spikes/report\_prior\_sensitivity\_spike.py} & committed
per-condition Stan data + \texttt{models/m\_0\_prior\_sweep.stan} &
20260701 & computed \\
Fig (§4.3) & 4.3.2 & \(\alpha\) true-vs-estimated recovery scatter +
per-replicate 90\% CIs (\texttt{m\_0}) &
\texttt{analysis/parameter\_recovery.py} & inline (\texttt{m\_0}:
\(M25\) \(K3\) \(D5\) \(R15\); 50 iters, 4 chains \(\times\) 1000) &
design np \(42\); \(12345+i\) sim / \(54321+i\) fit (D.2) & computed \\
Fig (§6.5) & 6.5 & \(\delta\) ECDF comparison (\texttt{m\_0} and
\texttt{m\_1}) with 95\% KS band --- marginal-SBC demarcation &
\texttt{analysis/sbc.py} & inline (\texttt{m\_0} and \texttt{m\_1}
canonical SBC designs; \(N_{\mathrm{sbc}}999\) thin4 1 chain) & design
np \(42\); \(123+i\) per draw (D.3) & computed \\
\end{longtable}

\endgroup

The cross-LLM robustness number (C11) and the per-cell diagnostics/PPC
summaries (C14, C15) are reproduced by
\texttt{spikes/report11\_cross\_llm\_spike.py} and
\texttt{spikes/report1415\_diagnostics\_ppc\_spike.py} respectively
(both seed 20260618), reading the committed per-condition posterior
draws under \texttt{reports/applications/*/data/}. The prior-sensitivity
analysis (§7.6.6, C17) is reproduced by
\texttt{spikes/report\_prior\_sensitivity\_spike.py} (base seed
20260701), which refits each cell under the alternative \(\alpha\)
priors via \texttt{models/m\_0\_prior\_sweep.stan} on the committed
per-condition Stan data and caches the refits under
\texttt{reports/applications/*/data/prior\_sweep/}. The
design-conditional information diagnostics (§5.5 lottery-spanning check,
§5.6 contrast-Jacobian ranks, §7.2 feature-matrix rank, and the §7.4
\(\eta\)-gap and \(\alpha\)-contraction callout) are reproduced by
\texttt{spikes/report\_design\_diagnostics\_spike.py} (seed 20260705),
which reads the foundational design constants of D.0 and the committed
application Stan data. The stricter-sampler refit check reported in §7.4
and D.2 is reproduced by
\texttt{spikes/report\_refit\_sensitivity\_spike.py} (seed 42), which
refits all 20 application conditions at \(\mathrm{adapt\_delta} = 0.99\)
and maximum tree depth \(12\) on the committed per-condition Stan data.

\textbf{Correction note (slope estimator).} Earlier drafts of the
\(2\times2\) table and the application reports quoted per-cell slope
medians of \(-31\) (GPT-4o \(\times\) insurance), \(-3.6\) (Claude
\(\times\) insurance), and \(-18.8\) (Claude \(\times\) Ellsberg). Those
values came from a per-draw slope computed as
\texttt{np.cov(T,\ alpha){[}0,1{]}\ /\ np.var(T)}, which mixes a
ddof-\(1\) covariance with a ddof-\(0\) variance and therefore inflates
every draw's slope --- hence the posterior median \emph{and} the CI
endpoints --- by exactly \(L/(L-1) = 5/4 = 1.25\) for the five-point
temperature grids. The canonical draw-level population-OLS values
(D.6(2)) are \(-24.6\), \(-2.9\), and \(-15.0\) respectively; GPT-4o
\(\times\) Ellsberg always used the correct formula and is unchanged at
\(-38.4\). \(P(\text{slope} < 0)\) is invariant to a positive scale
factor, so no negativity probability, monotonicity probability, or
qualitative conclusion is affected. All slopes in the paper now use the
single canonical estimator;
\texttt{spikes/report\_2x2\_forest\_spike.py} records the superseded
values and the exact \(1.25\) factor in its results JSON.

\phantomsection\label{refs}

\end{document}